\RequirePackage{fix-cm}
\documentclass[smallextended]{svjour3}       
\smartqed  
\usepackage{graphicx}
\usepackage{mathptmx}      
\usepackage{amssymb}
\usepackage{amsfonts}
\usepackage{amsmath}
\usepackage{frcursive}
\usepackage{euscript}
\usepackage{color}
\usepackage{graphicx}
\usepackage{rotating}
\usepackage{bm}
\usepackage{curves}
\usepackage[svgnames,dvipsnames]{pstricks}
\usepackage{lettrine}
\usepackage{mathrsfs}
\usepackage{booktabs}
\DeclareGraphicsExtensions{{},.pdf}
\usepackage{scalefnt}
\providecommand{\tsn}[1]{{\text{\scalefont{0.80}#1}}}
\providecommand{\tabref}[1]{{\textup{(Tab.~\ref{#1})}}}
\providecommand{\figref}[1]{{\textup{(Fig.~\ref{#1})}}}

\providecommand{\tabrefnp}[1]{{\textup{Tab.~\ref{#1}}}}
\providecommand{\figrefnp}[1]{{\textup{Fig.~\ref{#1}}}}

\providecommand{\eqrefsatob} [2]{\textup{(\ref{#1}--\ref{#2})}}
\providecommand{\eqrefsab}   [2]{\textup{(\ref{#1}, \ref{#2})}}

\providecommand{\figrefsab}   [2]{{\textup{(Figs.~\ref{#1}, \ref{#2})}}}
\providecommand{\figrefsnpab} [2]{{\textup{Figs.~\ref{#1}, \ref{#2}}}}

\providecommand{\figrefsatob} [2]{{\textup{(Figs.~\ref{#1}--\ref{#2})}}}
\providecommand{\figrefsnpatob} [2]{{\textup{Figs.~\ref{#1}--\ref{#2}}}}
\providecommand{\parref}[1]{{\textup{(\S\ref{#1})}}}
\providecommand{\parrefsab}[2]{\textup{(\S\ref{#1}, \S\ref{#2})}}

\providecommand{\parrefnp}[1]{{\textup{\S\ref{#1}}}}
\providecommand{\etal}{et al. }
\providecommand{\cf}{{\em cf }}

\providecommand{\viz}{{\em viz }}

\newcommand{\abs}[1]{\left\lvert#1\right\rvert}

\providecommand{\const}{{\rm const}}

\begin{document}
\title{Reynolds-stress model prediction of 3-D duct flows}
\author{G. A. Gerolymos \and I. Vallet}
\institute{G. A. Gerolymos \at Sorbonne Universit\'es, Universit\'e Pierre-et-Marie-Curie (\tsn{UPMC}), 4 place Jussieu, 75005 Paris, France; \email{georges.gerolymos@upmc.fr}
          \and
           I. Vallet       \at Sorbonne Universit\'es, Universit\'e Pierre-et-Marie-Curie (\tsn{UPMC}), 4 place Jussieu, 75005 Paris, France; \email{isabelle.vallet@upmc.fr}}

\date{Received: date / Accepted: date}

\maketitle

\begin{abstract}
The paper examines the impact of different modelling choices in second-moment closures by assessing model performance in predicting 3-D duct flows.
The test-cases (developing flow in a square duct [Gessner F.B., Emery A.F.: {\em ASME J. Fluids Eng.} {\bf 103} (1981) 445--455],
circular-to-rectangular transition-duct [Davis D.O., Gessner F.B.: {\em AIAA J.} {\bf 30} (1992) 367--375],
and \tsn{S}-duct with large separation [Wellborn S.R., Reichert B.A., Okiishi T.H.: {\em J. Prop. Power} {\bf 10} (1994) 668--675])
include progressively more complex strains. Comparison of experimental data with selected 7-equation models (6 Reynolds-stress-transport and 1 scale-determining equations),
which differ in the closure of the velocity/pressure-gradient tensor $\Pi_{ij}$, suggests that rapid redistribution controls separation and secondary-flow prediction, whereas,
inclusion of pressure-diffusion modelling improves reattachment and relaxation behaviour.
\end{abstract}

%
%
%
%
%
%
%
%
%
\section{Introduction}\label{RSMP3DDF_s_I}
%
%
%
%
%
%
%
%
%

The accurate prediction of 3-D turbulent flow in geometrically complex ducts is important in many practical applications,
including aerospace \cite{Gerolymos_Joly_Mallet_Vallet_2010a}, process \cite{AnxionnazMinvielle_Cabassud_Gourdon_Tochon_2013a}
and nuclear \cite{Chang_Tavoularis_2007a} engineering,
and agrofood industry \cite{Aloui_Berrich_Pierrat_2011a}. These flows can be particularly complex, and turbulence structure may be influenced by
various mechanisms, including 3-D boundary-layer entrainment \cite{Kovasznay_Kibens_Blackwelder_1970a}, secondary flows \cite{Bradshaw_1987a},
flow separation \cite{Simpson_1989a}, especially 3-D \cite{Delery_2001a}, and important streamline curvature \cite{So_Mellor_1978a},
associated with the presence of convex and concave bends \cite{Smits_Young_Bradshaw_1979a}. Therefore, in a \tsn{RANS} (Reynolds-averaged Navier-Stokes) framework \cite{Rumsey_2010a},
differential full Reynolds-stress models (\tsn{RSM}s) are an appropriate choice \cite{Hanjalic_1994a}, in an effort to include terms in the model that
account for all these mechanisms, especially if one considers not only the prediction of the mean flow, but also of the detailed Reynolds-stress field \cite{Yakinthos_Vlahostergios_Goulas_2008a}.
In a recent study \cite{Gerolymos_Joly_Mallet_Vallet_2010a} of a double-\tsn{S}-shaped duct intake, typical of unmanned combat air vehicles (\tsn{UCAV}s), comparison of
\tsn{RSM} predictions with available measurements highlighted the importance of the closure for the rapid part of the velocity/pressure-gradient tensor \smash{$\Pi_{ij}:=-\overline{u_i'\partial_{x_j}p'}-\overline{u_j'\partial_{x_i}p'}$}
(where $u_i\in\{u,v,w\}$ are the velocity-components in the Cartesian frame $x_i\in\{x,y,z\}$, $p$ is the pressure, \smash{$\overline{(\cdot)}$} denotes Reynolds (ensemble) averaging,
and $(\cdot)'$ denotes Reynolds-fluctuations) in successfully predicting the complex 3-D flow structure dominated by 2 pairs of contrarotating streamwise vortices.

To improve our understanding of the predictive capability, but also of limitations, of \tsn{RSM}s applied to the computation of streamwise-developing 3-D duct flows,
it seemed worthwhile to study 3 configurations, where the effects of different
mechanisms could be assessed separately, or at least sequentially: (a) developing flow in a square duct \cite{Gessner_Emery_1981a}, (b) flow in a circular-to-rectangular (\tsn{C}-to-\tsn{R}) transition duct \cite{Davis_Gessner_1992a},
and (c) separated flow in a circular diffusing \tsn{S}-duct \cite{Wellborn_Reichert_Okiishi_1994a}.
These are highly anisotropic and inhomogeneous 3-D flows, driven by mechanisms that are not modelled in linear eddy-viscosity closures, and are therefore
well suited for the assessment of anisotropy-resolving closures \cite{Leschziner_2000a}.

In turbulent fully-developed (streamwise-invariant in the mean) flow in a straight square duct \cite{Gessner_Jones_1965a} the anisotropy of the diagonal stresses,
$\overline{v'^2}$ and $\overline{w'^2}$, in the crossflow plane $yz$ \cite{Bradshaw_1987a}, but also the inhomogeneity of the gradients of the secondary shear-stress $\overline{v'w'}$ \cite[(3), p. 378]{Brundrett_Baines_1964a},
trigger secondary ($\perp x$) flow, associated with streamwise vorticity \cite{Brundrett_Baines_1964a}. The Gessner and Emery \cite{Gessner_Emery_1981a} test-case is further complicated by the streamwise evolution of the very thin inflow boundary-layers,
on the duct walls, which grow streamwise, until they interact and fill the entire duct, resulting in fully-developed (streamwise-invariant in the mean) flow.
Previous \tsn{RSM} computations of this flow \cite{So_Yuan_1999a,Gerolymos_Sauret_Vallet_2004a,Vallet_2007a}
illustrated the difficulty to correctly predict the streamwise development of the centerline velocity $\bar u_\tsn{CL}$,
but also, near the duct's exit where the flow reaches a fully-developed state,
the underestimation of the secondary velocity along the corner bisector (diagonal); this underestimation of the secondary-flow velocities is also observed in fully-developed flow predictions \cite{PetterssonReif_Andersson_2002a}.
Notice that, in fully-developed turbulent square-duct flow, secondary "{\em velocities $\cdots$ are found to be smaller than the root-mean-square turbulent velocity}" \cite[p. 376]{Brundrett_Baines_1964a},
and, furthermore, "{\em secondary-flow velocities, when nondimensionalized with either the bulk velocity {\rm ($\bar u_\tsn{B}$)} or the axial mean-flow velocity at the channel centerline {\rm ($\bar u_\tsn{CL}$)} decrease for an increase
in Reynolds number}" \cite[p. 689]{Gessner_Jones_1965a}.
The So-Yuan \cite{So_Yuan_1999a} wall-normal-free (\tsn{WNF}) \tsn{RSM} slightly underestimates the centerline velocity peak \cite[Fig. 14, p. 51]{So_Yuan_1999a},
while results with different \tsn{WNF-RSM} variants \cite{Gerolymos_Sauret_Vallet_2004a,Vallet_2007a}
demonstrated the sensitivity of the prediction of the $x$-wise development of $\bar u_\tsn{CL}$ to the closures for both $\Pi_{ij}$ \cite{Gerolymos_Sauret_Vallet_2004a} and turbulent diffusion \cite{Vallet_2007a}.
Finally, the wall-geometry-dependent Launder-Shima \cite{Launder_Shima_1989a} \tsn{RSM} was found to perform poorly for this type of flows \cite{Gerolymos_Sauret_Vallet_2004a},
despite a slight improvement when using its \tsn{WNF} version.

Contrary to turbulence-driven secondary flows \cite[Prandtl's second kind]{Bradshaw_1987a}, pressure-driven secondary flows \cite[Prandtl's first kind]{Bradshaw_1987a}
are generally much stronger \cite{Demuren_1991a}. In the \tsn{C}-to-\tsn{R} transition duct studied by Davis and Gessner \cite{Davis_Gessner_1992a}, the curvature of the walls in the transition part of the duct induces pressure-gradients
in the crossflow plane $yz$ \cite[Fig. 14, p. 373]{Davis_Gessner_1992a}, driving relatively strong secondary flows that develop into 2 contrarotating vortex pairs.
The cross-sectional area of the duct varies in the divergent/convergent transition part of the duct \cite[Fig. 4, p. 242]{Lien_Leschziner_1996a},
and this further complicates the flow, although the diverging part of the duct was sufficiently long to exclude separation \cite{Davis_Gessner_1992a}. Previous \tsn{RSM} computations for this configuration
were reported by Sotiropoulos and Patel \cite{Sotiropoulos_Patel_1994a} with a variant of the Launder-Shima \cite{Launder_Shima_1989a} \tsn{RSM},
by Lien and Leschziner \cite{Lien_Leschziner_1996a} with a zonal Gibson-Launder \cite{Gibson_Launder_1978a} \tsn{RSM} coupled with a nonlinear $\mathrm{k}$-$\varepsilon$ model near the wall, and by
Craft and Launder \cite{Craft_Launder_2001a} with their two-component limit (\tsn{TCL}) \tsn{RSM}.
The detailed comparisons of model predictions with available experimental measurements presented in \cite{Sotiropoulos_Patel_1994a} showed quite satisfactory agreement, both for the mean-flow and for the Reynolds-stresses,
with the single exception of the Reynolds-stresses at the last measurement station, located 2 inlet diameters downstream of the end of the \tsn{C}-to-\tsn{R} transition,
where computations do not predict the measured increase of turbulence levels, compared to the previous measurement station located exactly at the end of the \tsn{C}-to-\tsn{R} transition.

The diffusing \tsn{S}-duct, that was experimentally investigated by Wellborn \etal\ \cite{Wellborn_Reichert_Okiishi_1994a}, combines centerline curvature and cross-sectional area increase,
both of which induce streamline curvature, with associated crossflow pressure-gradients which generate significant secondary flows. This configuration is further complicated by the strong adverse streamwise
pressure-gradient, related to the streamwise-diverging cross-sectional area of the duct, which induces a large separated-flow zone. The presence of several interacting mechanisms renders this test-case a difficult challenge,
even for the prediction of the mean-flow velocity and total-pressure fields \cite{Harloff_Smith_Bruns_DeBonis_1993a}. Previous \tsn{RSM} computations were reported by Vallet \cite{Vallet_2007a},
who found that the predictive quality of the models depended mainly on the ability of the redistribution closure to correctly predict separation.

The second-moment closure (\tsn{SMC}) that was assessed in the present work is the \tsn{GLVY} \tsn{RSM} \cite{Gerolymos_Lo_Vallet_Younis_2012a},
which is the final result of previous research \cite{Gerolymos_Vallet_2001a,Gerolymos_Sauret_Vallet_2004a,Sauret_Vallet_2007a,Vallet_2007a}
on the development of wall-normal-free (\tsn{WNF}) \tsn{RSM}s with quasi-linear closure for the rapid part of $\Pi_{ij}$.
To put the comparisons with measurements into perspective, results were also presented for the \tsn{GV} \tsn{RSM} \cite{Gerolymos_Vallet_2001a},
the \tsn{WNF--LSS} \tsn{RSM} \cite{Gerolymos_Sauret_Vallet_2004a}, and with the baseline Launder-Sharma $\mathrm{k}$-$\varepsilon$ model \cite{Launder_Sharma_1974a}.
All of the computations were run specifically for the present assessment, carefully adjusting the boundary-conditions separately for each model,
to obtain the best possible match with the experimental data at the first available measurement plane.

The \tsn{RSM}s used in the present work are briefly reviewed in \parrefnp{RSMP3DDF_s_TCsFS}, with particular emphasis on differences between modelling choices,
and their implications. In \parrefnp{RSMP3DDF_s_A} computational results using the various models are compared with available experimental measurements.
Conclusions from the present results, and recommendations for future research, are summarized in \parrefnp{RSMP3DDF_s_Cs}.


%
%
%
%
%
%
%
%
%
\section{Turbulence closures and flow solver}\label{RSMP3DDF_s_TCsFS}
%
%
%
%
%
%
%
%
%

All measurements were performed in airflow, and a compressible aerodynamic solver was used in the computations.
The square \cite[$\bar M_\tsn{CL}\sim0.05$]{Gessner_Emery_1981a} and \tsn{C}-to-\tsn{R} \cite[$\bar M_\tsn{CL}\sim0.1$]{Davis_Gessner_1992a} ducts test-cases were at sufficiently low Mach-number for the flow
to be essentially incompressible ($\bar M_\tsn{CL}$ is a typical centerline Mach number), whereas in the \tsn{S}-duct high-subsonic flow conditions prevail \cite[$\bar M_\tsn{CL}\sim0.6$]{Wellborn_Reichert_Okiishi_1994a}.
Obviously in all of the previous cases, density fluctuations have negligible influence \cite{Bradshaw_1977a}, so that Favre (used in \parrefnp{RSMP3DDF_s_TCsFS_ss_TCs}) or Reynolds averages are, for practical purposes, equivalent.
The flow is modelled by the Reynolds-averaged Navier-Stokes (\tsn{RANS}) equations \cite{Gerolymos_Vallet_2001a,Vallet_2008a},
coupled with the appropriate modelled turbulence-transport equations \parrefsab{RSMP3DDF_s_TCsFS_ss_TCs}
                                                                               {RSMP3DDF_s_TCsFS_ss_FS}.
All computations were performed for air thermodynamics \cite{Vallet_2008a}.

%
%
%
%
%
\subsection{Turbulence closures}\label{RSMP3DDF_s_TCsFS_ss_TCs}
%
%
%
%
%

Details on the development of the \tsn{RSM}s used in the present work, can be found in the original papers \cite{Gerolymos_Vallet_2001a,Gerolymos_Sauret_Vallet_2004a,Gerolymos_Lo_Vallet_Younis_2012a}.
They are summarized below for completeness, in a common representation which highlights differences in the closure choices between different models.
Define
\begin{subequations}
                                                                                                                                    \label{Eq_RSMP3DDF_s_TCsFS_ss_TCs_001}
\begin{align}
r_{ij}:=\dfrac{1}{\bar\rho}\overline{\rho u_i''u_j''}     \;\; ; \;\;
\mathrm{k}:=\tfrac{1}{2}r_{\ell\ell}                      \;\; ; \;\;
a_{ij}:=\dfrac{r_{ij}}{\mathrm{k}}-\tfrac{2}{3}\delta_{ij}
                                                                                                                                    \label{Eq_RSMP3DDF_s_TCsFS_ss_TCs_001a}\\
A_2:=a_{ik}a_{ki}                                         \;\; ; \;\;
A_3:=a_{ik}a_{kj}a_{ji}                                   \;\; ; \;\;
A:=1-\tfrac{9}{8}(A_2-A_3)
                                                                                                                                    \label{Eq_RSMP3DDF_s_TCsFS_ss_TCs_001b}\\
\varepsilon=:\varepsilon^*+2\breve\nu\dfrac{\sqrt{\mathrm{k}}}{\partial x_\ell}
                                     \dfrac{\sqrt{\mathrm{k}}}{\partial x_\ell}\;\; ; \;\;
\ell_\tsn{T}  :=\dfrac{\mathrm{k}^\frac{3}{2}}{\varepsilon}                             \;\; ; \;\;
\ell_\tsn{T}^*:=\dfrac{\mathrm{k}^\frac{3}{2}}{\varepsilon^*}                           \;\; ; \;\;
Re_\tsn{T}    :=\dfrac{\mathrm{k}^2        }
                      {\breve\nu\varepsilon}                                   \;\; ; \;\;
Re_\tsn{T}^*  :=\dfrac{\mathrm{k}^2          }
                      {\breve\nu\varepsilon^*}
                                                                                                                                    \label{Eq_RSMP3DDF_s_TCsFS_ss_TCs_001c}\\
\breve\mu:=\mu_{\rm Sutherland}(\tilde T)                                      \;\; ; \;\;
\breve\nu:=\frac{\breve\mu}{\bar\rho}                                          \;\; ; \;\;
\breve S_{ij}:=\tfrac{1}{2}\left(\dfrac{\partial\tilde u_i}{\partial x_j}
                                +\dfrac{\partial\tilde u_j}{\partial x_i}\right)
                                                                                                                                    \label{Eq_RSMP3DDF_s_TCsFS_ss_TCs_001d}
\end{align}
\end{subequations}
where $\rho$ is the density, $r_{ij}$ are the 2-moments of velocity-fluctuations, $\mathrm{k}$ is the turbulent kinetic energy, $\delta_{ij}$ is the identity tensor, $a_{ij}$ is the deviatoric Reynolds-stress anisotropy-tensor,
with invariants $A_2$ and $A_3$, $A$ is Lumley's \cite{Lumley_1978a} flatness parameter,
$\varepsilon$ is the dissipation-rate of $\mathrm{k}$, $\varepsilon^*$ is the modified dissipation-rate \cite{Launder_Sharma_1974a},
$\ell_\tsn{T}$ ($\ell_\tsn{T}^*$) is the turbulent lengthscale and $Re_\tsn{T}$ ($Re_\tsn{T}^*$) is the turbulent Reynolds-number,
(defined using either $\varepsilon$ or $\varepsilon^*$), $\breve\mu$ is the dynamic viscosity evaluated from
Sutherland's law \cite[(6), p. 528]{Vallet_2008a} at mean temperature $\tilde T$, $\breve\nu$ is the kinematic viscosity at mean-flow conditions, $\breve S_{ij}$ is the deformation-rate tensor of the mean-velocity field,
\smash{$\widetilde{(\cdot)}$} denotes Favre averaging, \smash{$(\cdot)''$} are Favre fluctuations, and \smash{$\breve{(\cdot)}$} denotes a function of averaged quantities that
cannot be identified with a Reynolds or a Favre average \cite{Gerolymos_Vallet_1996a,Gerolymos_Vallet_2014a}.
Recall that $\varepsilon$ and $\varepsilon^*$ are significantly different only very near the wall \cite{Launder_Sharma_1974a,Gerolymos_Vallet_1996a,Gerolymos_Vallet_1997a}.

All of the 3 \tsn{RSM}s \cite{Gerolymos_Vallet_2001a,Gerolymos_Sauret_Vallet_2004a,Gerolymos_Lo_Vallet_Younis_2012a}
use the same scale-determining equation, solving for the modified dissipation-rate $\varepsilon^*$ \cite{Launder_Sharma_1974a,Gerolymos_Vallet_1997a}
\begin{subequations}
                                                                                                                                    \label{Eq_RSMP3DDF_s_TCsFS_ss_TCs_002}
\begin{alignat}{6}
\dfrac{\partial\bar\rho\varepsilon^*}
      {\partial t                   }+\dfrac{\partial\left(\bar\rho\varepsilon^*\tilde u_\ell\right)}
                                            {\partial x_\ell                                        }=&
\dfrac{\partial       }
      {\partial x_\ell}\left[C_\varepsilon\dfrac{\mathrm{k}   }
                                                {\varepsilon^*}\bar\rho r_{m\ell}\dfrac{\partial\varepsilon^*}
                                                                                       {\partial x_m         }
                           +              \breve\mu\dfrac{\partial\varepsilon^*}
                                                         {\partial x_\ell      }
                     \right]
                                                                                                                                    \notag\\
                            +&C_{\varepsilon_1} P_\mathrm{k}\dfrac{\varepsilon^*}
                                                                 {\mathrm{k}   }
                            -C_{\varepsilon_2}\bar\rho\dfrac{{\varepsilon^*}^2}{\mathrm{k}}
                            +2 \breve\mu C_{\mu} \dfrac{{\rm k}^2}{\varepsilon^*} \dfrac{\partial^2 \tilde{u}_i}{\partial x_\ell \partial x_\ell} \dfrac{\partial^2 \tilde{u}_i}{\partial x_m \partial x_m}
                                                                                                                                    \label{Eq_RSMP3DDF_s_TCsFS_ss_TCs_002a}
\end{alignat}
\begin{align}
P_\mathrm{k}:=\tfrac{1}{2} P_{\ell\ell}                   \;\; ;\;\;
C_\varepsilon=0.18                                        \;\; ;\;\;
C_{\varepsilon1}=1.44
                                                                                                                                    \label{Eq_RSMP3DDF_s_TCsFS_ss_TCs_002b}\\
C_{\varepsilon2}=1.92(1-0.3\mathrm{e}^{-{Re^*_\tsn{T}}^2})\;\; ;\;\;
C_\mu=0.09\mathrm{e}^{-\frac{3.4                    }
                             {(1+0.02 Re^*_\tsn{T})^2}}
                                                                                                                                    \label{Eq_RSMP3DDF_s_TCsFS_ss_TCs_002c}
\end{align}
\end{subequations}
where $t$ is the time, $P_{ij}$ is the Reynolds-stress production-tensor \eqref{Eq_RSMP3DDF_s_TCsFS_ss_TCs_003} and $P_\mathrm{k}$ is the production-rate of turbulent kinetic energy $\mathrm{k}$.
The scale-determining equation \eqref{Eq_RSMP3DDF_s_TCsFS_ss_TCs_002} is solved along with the 6 transport equations for the components of the symmetric tensor $r_{ij}$ \cite[(1), p. 2849]{Gerolymos_Lo_Vallet_Younis_2012a}
\begin{align}
\underbrace{\dfrac{\partial}{\partial t}\left(\overline{\rho}r_{ij}\right)
           +\dfrac{\partial}{\partial x_\ell}\left(\overline{\rho}r_{ij}\tilde{u}_\ell\right)}_{\displaystyle C_{ij}}=&
\underbrace{-\overline{\rho}r_{i\ell}\dfrac{\partial \tilde{u}_j}{\partial x_\ell}
            -\overline{\rho}r_{j\ell}\dfrac{\partial \tilde{u}_i}{\partial x_\ell}}_{\displaystyle P_{ij}}+
\underbrace{\dfrac{\partial}{\partial x_\ell}\left(\breve{\mu}\dfrac{\partial r_{ij}}{\partial x_\ell}\right)}_{\displaystyle d_{ij}^{(\mu)}}
                                                                                                                                    \notag\\
+&d_{ij}^{(u)}+\Pi_{ij}-\overline{\rho}\varepsilon_{ij}+K_{ij}
                                                                                                                                    \label{Eq_RSMP3DDF_s_TCsFS_ss_TCs_003}
\end{align}
where convection \smash{$C_{ij}$}, production \smash{$P_{ij}$} and viscous diffusion \smash{$d_{ij}^{(\mu)}$} are computable terms, and diffusion by the fluctuating velocity field \smash{$d_{ij}^{(u)}$},
the velocity/pressure-gradient correlation \smash{$\Pi_{ij}$}, the dissipation-tensor \smash{$\varepsilon_{ij}$} and the fluctuating-density terms \smash{$K_{ij}$} require closure.
For all of the 3 \tsn{RSM}s \cite{Gerolymos_Vallet_2001a,Gerolymos_Sauret_Vallet_2004a,Gerolymos_Lo_Vallet_Younis_2012a}
the fluctuating-density terms \smash{$K_{ij}$} and the pressure-dilatation correlation \smash{$\phi_p$} \cite[(1), p. 2849]{Gerolymos_Lo_Vallet_Younis_2012a} were neglected
\begin{align}
K_{ij}=0\qquad;\qquad
\phi_p=0
                                                                                                                                    \label{Eq_RSMP3DDF_s_TCsFS_ss_TCs_004}
\end{align}
this being a safe assumption for the subsonic flows that were investigated \cite{Gessner_Emery_1981a,Davis_Gessner_1992a,Wellborn_Reichert_Okiishi_1994a}.
The closure for the remaining terms ($d_{ij}^{(u)}$, $\Pi_{ij}$, $\varepsilon_{ij}$) differs between the 3 \tsn{RSM}s \cite{Gerolymos_Vallet_2001a,Gerolymos_Sauret_Vallet_2004a,Gerolymos_Lo_Vallet_Younis_2012a},
either in the functional dependence of the model coefficients on the local turbulent scales, or in the tensorial representation that was used \tabref{Tab_RSMP3DDF_s_TCsFS_ss_TCs_001}.

Diffusion by the triple velocity correlation
\begin{align}
d_{ij}^{(u)}:=\dfrac{\partial}{\partial x_\ell}\left(-\overline{\rho u_i'' u_j'' u_\ell''}\right)
                                                                                                                                    \label{Eq_RSMP3DDF_s_TCsFS_ss_TCs_005}
\end{align}
is modelled \tabref{Tab_RSMP3DDF_s_TCsFS_ss_TCs_001} using either the Daly-Harlow \cite{Daly_Harlow_1970a} closure in the \tsn{WNF--LSS} \tsn{RSM} \cite{Gerolymos_Sauret_Vallet_2004a},
or the Hanjali\'c-Launder \cite{Hanjalic_Launder_1976a} closure in  the \tsn{GV} \cite{Gerolymos_Vallet_2001a} and \tsn{GLVY} \cite{Gerolymos_Lo_Vallet_Younis_2012a} \tsn{RSM}s.

The dissipation-rate tensor is modelled as
\begin{align}
\bar\rho\varepsilon_{ij}=\tfrac{2}{3}\bar\rho\varepsilon\left(1-f_\varepsilon\right)\delta_{ij}+f_\varepsilon\dfrac{\varepsilon}{\mathrm{k}}\bar\rho r_{ij}
                                                                                                                                    \label{Eq_RSMP3DDF_s_TCsFS_ss_TCs_006}
\end{align}
The anisotropic part modelled via $f_\varepsilon$ \tabref{Tab_RSMP3DDF_s_TCsFS_ss_TCs_001} is only present in the \tsn{GLVY} \tsn{RSM},
the \tsn{GV} and \tsn{WNF--LSS} \tsn{RSM}s following Lumley's \cite{Lumley_1978a} suggestion to include the anisotropy of $\varepsilon_{ij}$ in the closure for the slow-redistribution terms \cite{Gerolymos_Lo_Vallet_2012a}.
\begin{sidewaystable}[h!]
\vspace{4.7in}
\begin{center}
\caption{Coefficients in the closure relations \eqrefsatob{Eq_RSMP3DDF_s_TCsFS_ss_TCs_005}
                                                          {Eq_RSMP3DDF_s_TCsFS_ss_TCs_007},
for the \tsn{GLVY} \cite{Gerolymos_Lo_Vallet_Younis_2012a},
the \tsn{GV} \cite{Gerolymos_Vallet_2001a},
and the \tsn{WNF--LSS} \cite{Gerolymos_Sauret_Vallet_2004a} \tsn{RSM}s.}
\label{Tab_RSMP3DDF_s_TCsFS_ss_TCs_001}
\scalebox{.8}{
\begin{tabular}{c|l|c|c|c}\hline\hline
Eq.                                    &  term                              & \tsn{GLVY}                                                        & \tsn{GV}      & \tsn{WNF--LSS} \\\hline
                                       &                                    &                                                                   &               &                \\
\eqref{Eq_RSMP3DDF_s_TCsFS_ss_TCs_005} &$-\overline{\rho u_i'' u_j'' u_k''}$&\multicolumn{2}{|c|}{$\displaystyle C^{(\mathrm{\tiny S}u)}\dfrac{\mathrm{k} }
                                                                                                                                              {\varepsilon}\left(\bar\rho r_{im}\dfrac{\partial r_{jk}}{\partial x_m}
                                                                                                                                                                +\bar\rho r_{jm}\dfrac{\partial r_{ki}}{\partial x_m}
                                                                                                                                                                +\bar\rho r_{km}\dfrac{\partial r_{ij}}{\partial x_m}\right)$}
                                                                            &$\displaystyle C^{(\tsn{S}u)}\dfrac{\mathrm{k} }
                                                                                                                {\varepsilon}\bar\rho r_{km}\dfrac{\partial r_{ij}}{\partial x_m}$\\
                                       &$C^{(\tsn{\tiny S}u)}$              & 0.11                                                              & 0.11          & 0.22           \\
\eqref{Eq_RSMP3DDF_s_TCsFS_ss_TCs_006} &$f_\varepsilon$                     &$1-A^{[1+A^2]}\left[1-\mathrm{e}^{-\frac{Re_{\tsn{T}}}{10}}\right]$& 0             & 0              \\
\eqref{Eq_RSMP3DDF_s_TCsFS_ss_TCs_007b}&$C^{(\tsn{Sp1})}$                   &-0.005                                                             & 0             & 0              \\
\eqref{Eq_RSMP3DDF_s_TCsFS_ss_TCs_007b}&$C^{(\tsn{Sp2})}$                   &+0.022                                                             & 0             & 0              \\
\eqref{Eq_RSMP3DDF_s_TCsFS_ss_TCs_007b}&$C^{(\tsn{Rp })}$                   &-0.005                                                             & 0             & 0              \\
\eqref{Eq_RSMP3DDF_s_TCsFS_ss_TCs_007c}&$C_\phi^{(\tsn{RH})}$               &\multicolumn{2}{|c|}{$\displaystyle\min{\left[1,0.75+1.3\max{[0,A-0.55]}\right]}\;A^{[\max(0.25,0.5-1.3\max{[0,A-0.55]})]}[1-\max(0,1-{\frac{Re_\tsn{T}}
                                                                                                                                                                                                                         {50        }})]$}
                                                                            &$\displaystyle0.75\sqrt{A}$\\
\eqref{Eq_RSMP3DDF_s_TCsFS_ss_TCs_007c}&$C_\phi^{(\tsn{RI})}$               &$\displaystyle\max{\left[{\tfrac{2}{3}-\frac{1}{6C_\phi^{(\tsn{RH})}},0}\right]}
                                                                                           \abs{\mathrm{grad}\Biggl(\dfrac{\ell_\tsn{T}[1-\mathrm{e}^{-\frac{Re^*_\tsn{T}}
                                                                                                                                                           {30          }}]}
                                                                                                                          {1+1.6A_2^{\max(0.6,A)}                          }\Biggr)}$
                                                                             &\multicolumn{2}{c}{$\displaystyle\max{\left[{\tfrac{2}{3}-\frac{1}{6C_\phi^{(\tsn{RH})}},0}\right]}
                                                                                                               \abs{\mathrm{grad}\Biggl(\dfrac{\ell_\tsn{T}[1-\mathrm{e}^{-\frac{Re^*_\tsn{T}}
                                                                                                                                                                                {30          }}]}
                                                                                                                                              {1+1.8A_2^{\max(0.6,A)}                           }\Biggr)}$}\\
\eqref{Eq_RSMP3DDF_s_TCsFS_ss_TCs_007d}&$\varepsilon_\mathrm{v}$             &$\varepsilon^*$                                                   & $\varepsilon$ & $\varepsilon$  \\
\eqref{Eq_RSMP3DDF_s_TCsFS_ss_TCs_007d}&$C_\phi^{(\tsn{SH1})}$               &$\displaystyle3.7 A A_2^{\tfrac{1}{4}}\left[1- \mathrm{e}^{-\left(\frac{Re_{\tsn{T}}}
                                                                                                                                                     {130         }\right)^2}\right]$
                                                                             &\multicolumn{2}{c}{$\displaystyle1+2.58A A_2^{\tfrac{1}{4}}\left[1-\mathrm{e}^{-({\frac{Re_\tsn{T}}{150}})^2}\right]$}\\
\eqref{Eq_RSMP3DDF_s_TCsFS_ss_TCs_007d}&$C_\phi^{(\tsn{SI1})}$               &$\displaystyle\left[-\tfrac{4}{9}\left(C_\phi^{(\tsn{SH1})}-\tfrac{9}{4}\right)\right]\;
                                                                                            \abs{\mathrm{grad}\Biggl(\dfrac{\ell_\tsn{T}[1-\mathrm{e}^{-\frac{Re^*_\tsn{T}}
                                                                                                                                                             {30          }}]}
                                                                                                                           {1+2.9\sqrt{A_2}                                  }\Biggr)}$
                                                                             &$\displaystyle0.83\left[1-{\tfrac{2}{3}}(C^\tsn{SH1}_\phi-1)\right]
                                                                                                \abs{\mathrm{grad}\left(\dfrac{\ell_\tsn{T}[1-\mathrm{e}^{-{\tfrac{Re^*_\tsn{T}}{30}}}]}{1+2 A_2^{0.8}}\right)}$
                                                                             &$\displaystyle0.90\left[1-{\tfrac{2}{3}}(C^\tsn{SH1}_\phi-1)\right]
                                                                                                \abs{\mathrm{grad}\left(\dfrac{\ell_\tsn{T}[1-\mathrm{e}^{-{\frac{Re^*_\tsn{T}}{30}}}]}{1+1.8 A_2^{0.8}}\right)}$\\
\eqref{Eq_RSMP3DDF_s_TCsFS_ss_TCs_007d}&$C_\phi^{(\tsn{SI2})}$               &0.002                                                             & 0             & 0              \\
\eqref{Eq_RSMP3DDF_s_TCsFS_ss_TCs_007d}&$C_\phi^{(\tsn{SI3})}$               &$\displaystyle 0.14\;
                                                                               \sqrt{\dfrac{\partial\ell^*_\tsn{T}}{\partial x_\ell}
                                                                                     \dfrac{\partial\ell^*_\tsn{T}}{\partial x_\ell}}$          & 0             & 0              \\
\end{tabular}
}
\end{center}
{\footnotesize Notice that there is a typographic error in \cite[(22), p. 1836]{Gerolymos_Vallet_2001a},
the turbulent Reynolds-number in the definition of $C_\phi^{(\tsn{RI})}$ and $C_\phi^{(\tsn{SI1})}$ for the \tsn{GV} \tsn{RSM} being $Re_\tsn{T}^*$ as above.
The expression of $C_\phi^{(\tsn{RH})}$ for the \tsn{GLVY} and \tsn{GV} \tsn{RSM}s above is the if-less equivalent of
\begin{alignat}{6}\left[C_\phi^{(\tsn{RH})}\right]_\tsn{GLVY}=
                  \left[C_\phi^{(\tsn{RH})}\right]_\tsn{GV}  =\Bigg(1-\max\left(0,1-\tfrac{1}{50}Re_\tsn{T}\right)\Bigg)\times
                                                              \left\{\begin{array}{lrcl}     0.75\sqrt{A}                                    &0                    &\leq A&<0.55                 \\
                                                                                        \Big(0.75\!\!+\!\!1.3(A-0.55)\Big)A^{0.5-1.3(A-0.55)}&0.55                 &\leq A&<0.55+\frac{0.25}{1.3}\\
                                                                                             A^{\frac{1}{4}}                                 &0.55+\frac{0.25}{1.3}&\leq A&\leq1                 \\\end{array}\right.\notag\end{alignat}
Notice that there is a typographic error of the above expression of $\left[C_\phi^{(\tsn{RH})}\right]_\tsn{GLVY}=\left[C_\phi^{(\tsn{RH})}\right]_\tsn{GV}$ for $0.55\leq A<0.55+\frac{0.25}{1.3}$
in \cite[(22), p. 419, missing parentheses]{Gerolymos_Sauret_Vallet_2004a} and
in \cite[Tab. 1, p. 1370, misplaced parenthesis]{Gerolymos_Joly_Mallet_Vallet_2010a}.}
\end{sidewaystable}
\clearpage

A general tensorial representation of the pressure terms $\Pi_{ij}$, which describes, by appropriate choice of the coefficients, all 3 models \tabref{Tab_RSMP3DDF_s_TCsFS_ss_TCs_001},
reads \cite[(4--6), pp. 2851--2854]{Gerolymos_Lo_Vallet_Younis_2012a}
\begin{subequations}
                                                                                                                                    \label{Eq_RSMP3DDF_s_TCsFS_ss_TCs_007}
\begin{align}
\Pi_{ij}=&\underbrace{\overbrace{\phi_{ij}^{(\tsn{RH})}+\phi_{ij}^{(\tsn{RI})}}^{\displaystyle\phi_{ij}^{(\tsn{R})}}+
                      \overbrace{\phi_{ij}^{(\tsn{SH})}+\phi_{ij}^{(\tsn{SI})}}^{\displaystyle\phi_{ij}^{(\tsn{S})}}}_{\displaystyle\phi_{ij}}+\tfrac{2}{3}\phi_p\delta_{ij}+d_{ij}^{(p)}
                                                                                                                                    \label{Eq_RSMP3DDF_s_TCsFS_ss_TCs_007a}\\
d_{ij}^{(p)}=&C^{(\tsn{Sp1})}\bar\rho\dfrac{\mathrm{k}^3 }
                                           {\varepsilon^3}\dfrac{\partial\varepsilon^*}
                                                                {\partial x_i         }\dfrac{\partial\varepsilon^*}
                                                                                             {\partial x_j         }
            +\dfrac{\partial       }
                   {\partial x_\ell}\left[C^{(\tsn{Sp2})}(\overline{\rho u_m''u_m''u_j''}\delta_{i\ell}
                                                         +\overline{\rho u_m''u_m''u_i''}\delta_{j\ell})\right]
                                                                                                                                    \notag\\
           +&C^{(\tsn{Rp})}\bar\rho\dfrac{\mathrm{k}^2 }
                                         {\varepsilon^2}\breve{S}_{k\ell}a_{\ell k}\dfrac{\partial\mathrm{k}}
                                                                                         {\partial x_i      }\dfrac{\partial\mathrm{k}}
                                                                                                                   {\partial x_j      }
                                                                                                                                    \label{Eq_RSMP3DDF_s_TCsFS_ss_TCs_007b}\\
\phi_{ij}^{(\tsn{R})}=&\underbrace{-C_\phi^{(\tsn{RH})}\left(P_{ij}-\tfrac{1}{3}\delta_{ij} P_{mm}\right)}_{\displaystyle\phi^{(\tsn{RH})}_{ij}}
                                                                                                                                    \notag\\
                      &\underbrace{+C_\phi^{(\tsn{RI})}\left[            \phi^{(\tsn{RH})}_{nm}e_{\tsn{I}_n}e_{\tsn{I}_m}\delta_{ij}
                                                            -\tfrac{3}{2}\phi^{(\tsn{RH})}_{in}e_{\tsn{I}_n}e_{\tsn{I}_j}
                                                            -\tfrac{3}{2}\phi^{(\tsn{RH})}_{jn}e_{\tsn{I}_n}e_{\tsn{I}_i}\right]}_{\displaystyle\phi^{(\tsn{RI})}_{ij}}
                                                                                                                                    \label{Eq_RSMP3DDF_s_TCsFS_ss_TCs_007c}\\
\phi^{(\tsn{S})}_{ij}=&\underbrace{-C_\phi^{(\tsn{SH1})}\bar\rho\varepsilon_\mathrm{v} a_{ij}}_{\displaystyle\phi^{(\tsn{SH1})}_{ij}}
                       \underbrace{+C_\phi^{(\tsn{SI1})}\dfrac{\varepsilon_\mathrm{v}}{\mathrm{k}}\left[            \bar\rho r_{nm}e_{\tsn{I}_n}e_{\tsn{I}_m}\delta_{ij}
                                                                                                       -\tfrac{3}{2}\bar\rho r_{ni}e_{\tsn{I}_n}e_{\tsn{I}_j}
                                                                                                       -\tfrac{3}{2}\bar\rho r_{nj}e_{\tsn{I}_n}e_{\tsn{I}_i}\right]}_{\displaystyle\phi^{(\mathrm{\tiny SI1})}_{ij}}
                                                                                                                                    \notag\\
       &\underbrace{-C_\phi^{(\tsn{SI2})}\bar\rho\dfrac{\mathrm{k}}{\varepsilon}\dfrac{\partial\mathrm{k}}{\partial x_\ell}
                                         \left[                       a_{ik}\frac{\partial r_{kj}}{\partial x_\ell}
                                                                     +a_{jk}\frac{\partial r_{ki}}{\partial x_\ell}
                                              -\tfrac{2}{3}\delta_{ij}a_{mk}\frac{\partial r_{km}}{\partial x_\ell}\right]}_{\displaystyle\phi^{(\tsn{SI2})}_{ij}}
                                                                                                                                    \notag\\
       &\underbrace{+C_\phi^{(\tsn{SI3})}\left[            \phi^{(\tsn{SI2})}_{nm}e_{\tsn{I}_n}e_{\tsn{I}_m}\delta_{ij}
                                              -\tfrac{3}{2}\phi^{(\tsn{SI2})}_{in}e_{\tsn{I}_n}e_{\tsn{I}_j}
                                              -\tfrac{3}{2}\phi^{(\tsn{SI2})}_{jn}e_{\tsn{I}_n}e_{\tsn{I}_i}\right]}_{\displaystyle\phi^{(\tsn{SI3})}_{ij}}
                                                                                                                                    \label{Eq_RSMP3DDF_s_TCsFS_ss_TCs_007c}\\
                e_{\tsn{I}_i}:=&\dfrac{\dfrac{\partial}{\partial x_i}\Biggl(\dfrac{\ell_\tsn{T}[1-\mathrm{e}^{-{\frac{Re^*_\tsn{T}}
                                                                                                                     {          30}}}]}
                                                                                  {1+2\sqrt{A_2}+2A^{16}                              }
                                                                     \Biggr)
                                      }
                                      {\sqrt{\dfrac{\partial}{\partial x_\ell}\Biggl(\dfrac{\ell_\tsn{T}[1-\mathrm{e}^{-{\frac{Re^*_\tsn{T}}
                                                                                                                              {          30}}}]}
                                                                                           {1+2\sqrt{A_2}+2A^{16}                              }
                                                                              \Biggr)
                                             \dfrac{\partial}{\partial x_\ell}\Biggl(\dfrac{\ell_\tsn{T}[1-\mathrm{e}^{-{\frac{Re^*_\tsn{T}}
                                                                                                                              {          30}}}]}
                                                                                           {1+2\sqrt{A_2}+2A^{16}                              }
                                                                              \Biggr)
                                             }
                                      }
                                                                                                                                    \label{Eq_RSMP3DDF_s_TCsFS_ss_TCs_007d}
\end{align}
\end{subequations}
where $\phi_{ij}$ denotes the redistribution tensor, $d_{ij}^{(p)}$ denotes pressure diffusion, the superscripts \tsn{S} and \tsn{R} denote slow and rapid terms \cite{Hanjalic_1994a},
the superscripts \tsn{H} and \tsn{I} denote homogeneous and inhomogeneous terms \cite{Gerolymos_Senechal_Vallet_2013a}, and the unit-vector $\vec{e}_\tsn{I}$ was modelled \cite{Gerolymos_Vallet_2001a}
to point in the main direction of turbulence-inhomogeneity \cite{Gerolymos_Sauret_Vallet_2004a}. Notice that, although initially $\vec{e}_\tsn{I}$ was designed to mimic the wall-normal direction in wall-echo terms \cite{Gerolymos_Vallet_2001a},
it turns out that inhomogeneous terms are also active at the shear-layer edge and in regions of recirculating flow, away from or even in absence of solid walls. As a consequence, the closure \eqref{Eq_RSMP3DDF_s_TCsFS_ss_TCs_007} must be considered as a whole,
and inhomogeneous terms should be kept when computing free shear flows. Very near the walls, \smash{$\Pi_{ij}\to0\stackrel{\eqref{Eq_RSMP3DDF_s_TCsFS_ss_TCs_007a}}{\Longrightarrow}\phi_{ij}+\frac{2}{3}\phi_p\delta_{ij}\to-d_{ij}^{(p)}$},
so that all authors \cite{Gerolymos_Lo_Vallet_Younis_2012a} avoid the complexity of including terms in the model that would correctly mimic the individual behaviour of $\phi_{ij}$ and \smash{$d_{ij}^{(p)}$} as wall-distance $n\to0$,
which would cancel one another in \eqref{Eq_RSMP3DDF_s_TCsFS_ss_TCs_007a}, but rather model $\Pi_{ij}$ as a whole in that region \cite[Fig. 6, pp. 2855--2856]{Gerolymos_Lo_Vallet_Younis_2012a},
in line with the suggestion of Mansour \etal\ \cite{Mansour_Kim_Moin_1988a}. For this reason, the wall-echo-like \cite{Gibson_Launder_1978a} tensorial form of the terms containing $\vec{e}_\tsn{I}$ in \eqref{Eq_RSMP3DDF_s_TCsFS_ss_TCs_007}
is justified, because it was recently shown \cite{Gerolymos_Senechal_Vallet_2013a}, from the analysis of \tsn{DNS} data,
that it is in agreement with the near-wall behaviour of $\Pi_{ij}$ \cite[Fig. 13 p. 41]{Gerolymos_Senechal_Vallet_2013a}, unlike that of $\phi_{ij}$ \cite[Fig. 12 p. 39]{Gerolymos_Senechal_Vallet_2013a}.
The coefficients in \eqrefsatob{Eq_RSMP3DDF_s_TCsFS_ss_TCs_005} 
                               {Eq_RSMP3DDF_s_TCsFS_ss_TCs_007} 
are generally functions of the local turbulence state ($A$, $A_2$, $Re_\tsn{T}$, $\cdots$) and of its gradients, and depend on the particular \tsn {RSM} \tabref{Tab_RSMP3DDF_s_TCsFS_ss_TCs_001}.

The \tsn{WNF--LSS} \cite{Gerolymos_Sauret_Vallet_2004a} is a wall-normal-free extension of the Launder-Shima \cite{Launder_Shima_1989a} \tsn{RSM}, which, in complex flows,
performs better than the original wall-topology-dependent model, mainly because of the action of the inhomogeneous terms away from solid walls.
The main drawback of this model is that, although it quite naturally improves upon 2-equation closures, it still underestimates separation \cite{Gerolymos_Joly_Mallet_Vallet_2010a}.
The \tsn{GV} \cite{Gerolymos_Vallet_2001a} \tsn{RSM} was developed to overcome this limitation, mainly by an optimized \smash{$C_\phi^{(\tsn{RH})}$} coefficient \tabref{Tab_RSMP3DDF_s_TCsFS_ss_TCs_001}
of the isotropisation-of-production \cite{Gibson_Launder_1978a,Launder_Shima_1989a,Hanjalic_1994a}
closure of the rapid homogeneous part of redistribution \eqref{Eq_RSMP3DDF_s_TCsFS_ss_TCs_007c}.
The resulting model successfully predicted flows with large separation, but reattachment and especially relaxation were slightly slower than experimental data \cite{Sauret_Vallet_2007a,Vallet_2007a}.
The \tsn{GLVY} \cite{Gerolymos_Lo_Vallet_Younis_2012a} \tsn{RSM} improves this behaviour \cite[Fig. 9, p. 2858]{Gerolymos_Lo_Vallet_Younis_2012a}
through extended modelling of the inhomogeneous part of the slow redistribution terms \smash{$\phi_{ij}^{(\tsn{SI})}$} \eqrefsab{Eq_RSMP3DDF_s_TCsFS_ss_TCs_007a}
                                                                                                                                {Eq_RSMP3DDF_s_TCsFS_ss_TCs_007d}
and of pressure diffusion \smash{$d_{ij}^{(p)}$} \eqref{Eq_RSMP3DDF_s_TCsFS_ss_TCs_007b},
while using the same optimized closure for \smash{$\phi_{ij}^{(\tsn{RH})}$} as the \tsn{GV} \cite{Gerolymos_Vallet_2001a} \tsn{RSM}.
It was also observed that the inclusion of these modifications influences the apparent transition behaviour \cite{Rumsey_2007a} of the models,
at low external turbulence conditions \cite{icp_Gerolymos_Vallet_2013a}.

Comparing the 3 \tsn{RSM}s \tabref{Tab_RSMP3DDF_s_TCsFS_ss_TCs_001}, the \tsn{WNF--LSS} \cite{Gerolymos_Sauret_Vallet_2004a} \tsn{RSM} conceptually \cite{Hanjalic_1994a} includes pressure-diffusion in
the Daly-Harlow \cite{Daly_Harlow_1970a} closure for \smash{$d_{ij}^{(u)}$}, while the \tsn{GV} \cite{Gerolymos_Vallet_2001a} \tsn{RSM} neglects \smash{$d_{ij}^{(p)}$};
they both (\tsn{WNF--LSS} and \tsn{GV}) include the dissipation-rate anisotropy \smash{$\varepsilon_{ij}-\tfrac{2}{3}\varepsilon\delta_{ij}$} in the closure for \smash{$\phi_{ij}^{(\tsn{S})}$} \cite{Lumley_1978a,Launder_Shima_1989a}.
On the contrary, the \tsn{GLVY} \cite{Gerolymos_Lo_Vallet_Younis_2012a} \tsn{RSM} explicitly models both \smash{$d_{ij}^{(p)}$} and \smash{$\varepsilon_{ij}-\tfrac{2}{3}\varepsilon\delta_{ij}$}.

Computations were also compared with the baseline linear Launder-Sharma \cite{Launder_Sharma_1974a} $\mathrm{k}$-$\varepsilon$ closure, as implemented in \cite{Gerolymos_Vallet_1996a}.

%
%
%
%
%
\subsection{Flow solver} \label{RSMP3DDF_s_TCsFS_ss_FS}
%
%
%
%
%

Computations were performed using a structured multiblock solver \cite{Gerolymos_Vallet_2005a}, with \tsn{WENO3} \cite{Gerolymos_Senechal_Vallet_2009a}
reconstruction of the primitive variables, both mean-flow and turbulent, an \tsn{HLLC}$_\mathrm{h}$ approximate Riemann solver \cite{BenNasr_Gerolymos_Vallet_2014a},
and implicit multigrid dual-time-stepping pseudo-time-marching integration \cite{Gerolymos_Vallet_2009a}.
All of the computations presented in the paper were run using $L_\tsn{GRD}=3$ levels of multigrid with a $\tsn{V(2,0)}$ sawtooth cycle \cite{Gerolymos_Vallet_2009a}
and dual-time-stepping parameters \cite{Gerolymos_Vallet_2005a} $[\tsn{CFL},\tsn{CFL}^*;M_\mathrm{it},r_\tsn{TRG}]=[100,10;\text{---},-1]$
(where $\tsn{CFL}$ is the \tsn{CFL}-number for the pseudo-time-step, $\tsn{CFL}^*$ is the \tsn{CFL}-number for the dual pseudo-time-step, $M_\mathrm{it}$ is the number of dual subiterations,
and $r_\tsn{TRG}<0$ is the target-reduction in orders-of-magnitude of the nonlinear pseudo-time-evolution system solution).
This methodology is implemented in the open source software \texttt{aerodynamics} \cite{sft_Gerolymos_Vallet_aerodynamics_2009} with which the present results were obtained.

In all instances, a subsonic reservoir condition was applied at inflow \cite[(24), p. 1324]{Gerolymos_Vallet_1996a},
a subsonic pressure condition \cite[(26), p. 1324]{Gerolymos_Vallet_1996a} was applied at outflow (uniform static pressure at outflow),
and the no-slip walls were considered adiabatic \cite[(25), p. 1324]{Gerolymos_Vallet_1996a}.
The inflow boundary condition was implemented using the method of finite waves \cite{Atkins_Casper_1994a}.
Note that in this approach, the inflow boundary-layers are prescribed through the initial total pressure and total temperature profiles \cite{Gerolymos_Sauret_Vallet_2004c},
but the streamwise mean-flow velocity $\tilde u$ at inflow is also influenced by the outgoing pressure-wave \cite{Chakravarthy_1984a},
and may therefore evolve differently for different turbulence closures \cite[(Fig. 6), p. 209]{BenNasr_Gerolymos_Vallet_2014a}.
\begin{figure}[h!]
\begin{center}
\begin{picture}(340,420)
\put(0,-5){\includegraphics[angle=0,width=340pt,bb=62 154 561 781]{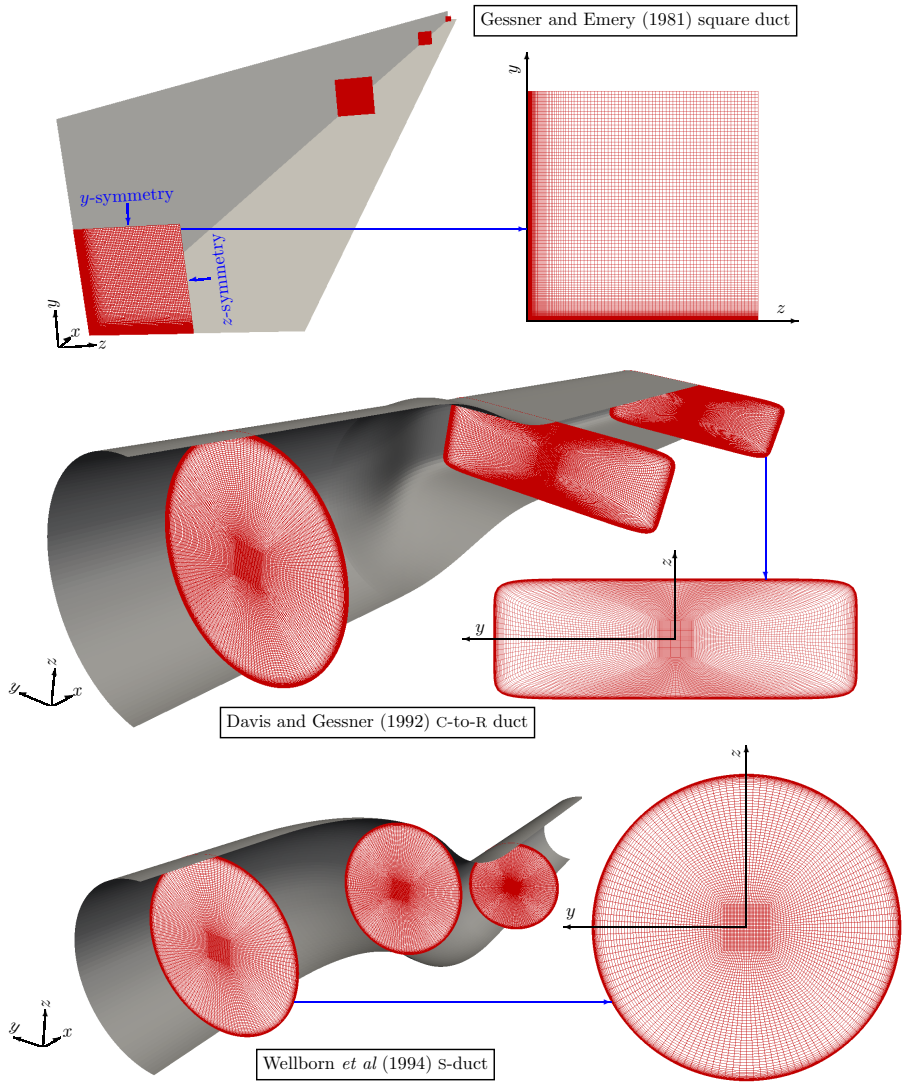}}
\end{picture}
\end{center}
\caption{Computational grid topology \tabref{Tab_RSMP3DDF_s_A_001} for the square duct \cite{Gessner_Emery_1981a}, the \tsn{C}-to-\tsn{R} transition duct \cite{Davis_Gessner_1992a} and the \tsn{S}-duct \cite{Wellborn_Reichert_Okiishi_1994a}
         test-cases (in all cases the $i=\const$ grid-surfaces are $\perp x$ planes.}
\label{Fig_RSMP3DDF_s_A_001}
\end{figure}

%
%
%
%
%
%
%
%
%
\section{Assessment}\label{RSMP3DDF_s_A}
%
%
%
%
%
%
%
%
%

The predictive capability of the 4 turbulence models \parref{RSMP3DDF_s_TCsFS_ss_TCs} was assessed by systematic comparison with experimental data for 3 duct-flow configurations \cite{Gessner_Emery_1981a,Davis_Gessner_1992a,Wellborn_Reichert_Okiishi_1994a}.
Hereafter, the abbreviations \tsn{GLVY RSM} \cite{Gerolymos_Lo_Vallet_Younis_2012a}, \tsn{GV RSM} \cite{Gerolymos_Vallet_2001a},
\tsn{WNF--LSS RSM} \cite{Gerolymos_Sauret_Vallet_2004a} and \tsn{LS} $\mathrm{k}$--$\varepsilon$ \cite{Launder_Sharma_1974a},
are used consistently to denote each model.
%
\begin{table}[ht!]
\begin{center}
\caption{Computational grids and mesh-generation parameters \cite{Gerolymos_Sauret_Vallet_2004b,Gerolymos_Tsanga_1999a}
         for the square duct \cite{Gessner_Emery_1981a}, the \tsn{C}-to-\tsn{R} transition duct \cite{Davis_Gessner_1992a} and the \tsn{S}-duct \cite{Wellborn_Reichert_Okiishi_1994a}
         test-cases.}
\label{Tab_RSMP3DDF_s_A_001}
\scalebox{.90}{
\begin{tabular}{lrclcccccccccccccccc}\hline
configuration          &\multicolumn{10}{|c} {grids}                                                                                                              \\\hline
                       &$N_\mathrm{p}$  &$n_\tsn{D}$&type                     &$N_i\times N_j\times N_k$&$N_{j_\mathrm{s}}$&$r_j$  &$N_{k_\mathrm{s}}$&$r_k$  &$r_\tsn{$\square$}$&$\Delta n_w^+$\\\hline
square duct            &$17.8\times10^6$&$1$        &\tsn{H}                  &$801\times  149\times  149$&$97$            &$1.067$&$~\,97$           &$1.067$&---                &$\sim0.5$     \\\hline
\tsn{C}-to-\tsn{R} duct& $9.4\times10^6$&$1$        &\tsn{O}$_\tsn{$\square$}$&$209\times  209\times  201$&---             &---    &$201$             &$1.060$&$0.15$             &$\sim0.2$     \\
                       &                &$2$        &\tsn{H}$_\tsn{$\square$}$&$209\times~\,53\times~\,53$&---             &---    &---               &---    &---                &---           \\\hline
\tsn{S}-duct           & $1.9\times10^6$&$1$        &\tsn{O}$_\tsn{$\square$}$&$161\times  129\times~\,81$&---             &---    &$~\,81$           &$1.220$&$0.15$             &$\sim0.4$     \\
                       &                &$2$        &\tsn{H}$_\tsn{$\square$}$&$161\times~\,33\times~\,33$&---             &---    &---               &---    &---                &---           \\\hline
\end{tabular}
}
\end{center}
 {\footnotesize $N_\mathrm{p}$: number of grid-points;
                $n_\tsn{D}$: domain index;
                type: domain grid-type;   
                \tsn{H}: \tsn{H}-type grid ($x\times y\times z$);
                \tsn{O}$_\tsn{$\square$}$: circumferentially \tsn{O}-type grid ($x\times\;-\theta\times R$) between the duct-casing and the inner square domain ($n_\tsn{D}=2$) around the centerline;
                \tsn{H}$_\tsn{$\square$}$: \tsn{H}-type grid for the inner square domain ($n_\tsn{D}=2$) around the centerline;
                $N_i\times N_j\times N_k$: grid-points;
                $N_{j_\mathrm{s}}$, $N_{k_\mathrm{s}}$: number of points geometrically stretched near the solid walls;
                $r_j$, $r_k$: geometric progression ratio;
                $r_\tsn{$\square$}$: ratio of the side of the square domain around the centerline to the size of the cross-section (defined as the average of its projections on the $y$ and $z$ axes);
                $\Delta n_w^+$ nondimensional wall-normal size of the first grid-cell in wall-units \cite{Gerolymos_Sauret_Vallet_2004b}.
 }
%
\begin{center}
\caption{Initial (\tsn{IC}s) and boundary-conditions (\tsn{BC}s) for the square duct \cite{Gessner_Emery_1981a}, the \tsn{C}-to-\tsn{R} transition duct \cite{Davis_Gessner_1992a} and the \tsn{S}-duct \cite{Wellborn_Reichert_Okiishi_1994a}
         test-cases, using the \tsn{GLVY RSM} \cite{Gerolymos_Lo_Vallet_Younis_2012a}, the \tsn{GV RSM} \cite{Gerolymos_Vallet_2001a}, the \tsn{WNF--LSS RSM} \cite{Gerolymos_Sauret_Vallet_2004a}
         and the \tsn{LS} $\mathrm{k}$-$\varepsilon$ \cite{Launder_Sharma_1974a}.}
\label{Tab_RSMP3DDF_s_A_002}
\scalebox{.72}{
\begin{tabular}{lccccccccccccccccccc}\hline
configuration          &model                               &\multicolumn{9}{|c}{\tsn{IC}s and \tsn{BC}s}                                                                                                                                                                                                                                                                         \\\hline
                       &                                    &$\delta_\mathrm{i}\;(\mathrm{mm})$&$\Pi_{\tsn{C}_\mathrm{i}}$&$M_{\tsn{CL}_\mathrm{i}}$&$T_{u_{\tsn{CL}_\mathrm{i}}}$&$\ell_{\tsn{T}_{\tsn{CL}_\mathrm{i}}}\;(\mathrm{mm})$&$p_{t_{\tsn{CL}_\mathrm{i}}}\;(\mathrm{Pa})$&$T_{t_{\tsn{CL}_\mathrm{i}}}\;(\mathrm{K})$&$q_w$ (W m$^{-2}$)&$p_\mathrm{o}\;(\mathrm{Pa})$\\\hline
square duct            &\tsn{GLVY RSM}                      &$0.875$                           &$0~~~$                    &$0.0516$                 &$1\%$                        &$50$                                                 &$101325$                                    &$288$                                      &$0$               &$0.995p_{t_{\tsn{CL}_\mathrm{i}}}$\\
                       &\tsn{GV RSM}                        &$0.300$                           &                          &                         &                             &                                                     &                                            &                                           &                  &                                  \\
                       &\tsn{WNF--LSS RSM}                  &$0.100$                           &                          &                         &                             &                                                     &                                            &                                           &                  &                                  \\
                       &\tsn{LS} $\mathrm{k}$--$\varepsilon$&$0.100$                           &                          &                         &                             &                                                     &                                            &                                           &                  &                                  \\\hline
\tsn{C}-to-\tsn{R} duct&\tsn{GLVY RSM}                      &$30.85$                           &$0.50$                    &$0.0940$                 &$0.3\%$                      &$50$                                                 &$101325$                                    &$298.3$                                    &$0$               &$100627$                          \\
                       &\tsn{GV RSM}                        &$30.85$                           &                          &                         &                             &                                                     &                                            &                                           &                  &                                  \\
                       &\tsn{WNF--LSS RSM}                  &$30.85$                           &                          &                         &                             &                                                     &                                            &                                           &                  &                                  \\
                       &\tsn{LS} $\mathrm{k}$--$\varepsilon$&$28.00$                           &                          &                         &                             &                                                     &                                            &                                           &                  &                                  \\\hline
\tsn{S}-duct           &\tsn{GLVY RSM}                      &$10.5$                            &$0.25$                    &$0.60$                   &$0.65\%$                     &$50$                                                 &$111330$                                    &$296.4$                                    &$0$               &$98600$                           \\
                       &\tsn{GV RSM}                        &$10.5$                            &$0.25$                    &                         &                             &                                                     &                                            &                                           &                  &$98600$                           \\
                       &\tsn{WNF--LSS RSM}                  &$10.3$                            &$0.35$                    &                         &                             &                                                     &                                            &                                           &                  &$98900$                           \\
                       &\tsn{LS} $\mathrm{k}$--$\varepsilon$&$10.8$                            &$0.40$                    &                         &                             &                                                     &                                            &                                           &                  &$98600$                           \\\hline
\end{tabular}
}

\end{center}
 {\footnotesize $\delta_\mathrm{i}$: boundary-layer thickness at inflow (\tsn{IC});
                $\Pi_{\tsn{C}_\mathrm{i}}$: inflow boundary-layer Coles-parameter \cite{Gerolymos_Sauret_Vallet_2004c} (\tsn{IC});
                $M_{\tsn{CL}_\mathrm{i}}$: inflow centerline Mach-number (\tsn{IC});
                $T_{u_{\tsn{CL}_\mathrm{i}}}$: turbulent intensity outside of the boundary-layers at inflow \cite{Gerolymos_Sauret_Vallet_2004c};
                $\ell_{\tsn{T}_{\tsn{CL}_\mathrm{i}}}$: turbulent lengthscale outside of the boundary-layers at inflow \cite{Gerolymos_Sauret_Vallet_2004c};
                $p_{t_{\tsn{CL}_\mathrm{i}}}$: inflow centerline total pressure (\tsn{BC});
                $T_{t_{\tsn{CL}_\mathrm{i}}}$: inflow centerline total temperature (\tsn{BC});
                $q_w$: wall heat-flux (\tsn{BC});
                $p_\mathrm{o}$: outflow static pressure (\tsn{BC}).
 }
\end{table}

%
%
%
%
%
\subsection{Developing turbulent flow in a square duct \cite{Gessner_Emery_1981a}}\label{RSMP3DDF_s_A_ss_DTFSDGE1981}
%
%
%
%
%

The experimental data described by Gessner and Emery \cite{Gessner_Emery_1981a} were obtained \cite{Po_1975a,Gessner_Po_Emery_1979a,Gessner_Emery_1981a}
in a duct of square cross-section \figrefsab{Fig_RSMP3DDF_s_A_001}
                                            {Fig_RSMP3DDF_s_A_ss_DTFSDGE1981_001}.
The duct's height, which at incompressible flow conditions is also the duct's hydraulic diameter \cite[(3.55), p. 123]{White_1991a},
is $D_\mathrm{h}=2\mathrm{a}=0.254\;\mathrm{m}$, and the length of the straight working section is $87D_\mathrm{h}$ \cite[Fig. 2, p. 121]{Gessner_Po_Emery_1979a}.
The flow \cite{Gessner_Emery_1981a} is essentially incompressible (centerline Mach number $\bar M_\tsn{CL}\approxeq0.05$) at bulk Reynolds number $Re_\tsn{B}\approxeq250000$
($Re_\tsn{B}=\bar u_\tsn{B} D_\mathrm{h} \nu^{-1}$, where $\bar u_\tsn{B}$ is the bulk velocity and $\nu$ is the practically constant kinematic viscosity).
The flow at the duct's inlet is nearly uniform, with very thin boundary-layers, whose virtual origin was estimated experimentally at $x\approxeq-0.65D_\mathrm{h}$ \cite[p. 122]{Gessner_Po_Emery_1979a}.
These very thin boundary-layers grow until they fill the entire duct at $x\approxeq32D_\mathrm{h}$ \cite[p. 123]{Gessner_Po_Emery_1979a} and interact
to reach practically fully developed flow conditions at the last measurement station located at $x=84D_\mathrm{h}$, near the exit of the duct's working section \cite[Fig. 2, p. 121]{Gessner_Po_Emery_1979a}.
Measurements \cite{Po_1975a,Gessner_Po_Emery_1979a,Gessner_Emery_1981a},
taken at 5 axial planes \figref{Fig_RSMP3DDF_s_A_ss_DTFSDGE1981_002}, include mean-flow $x$-wise velocities (Kiel probes in conjunction with a wall static pressure tap),
and secondary mean-flow velocities and Reynolds-stresses (hot-wire). They also include the detailed $x$-wise evolution of the centerline velocity (Kiel probe) and limited
skin-friction data (Preston tubes) only at the last measurement station ($x=84D_\mathrm{h}$) where the flow is considered fully developed \cite[Fig. 2, p. 448]{Gessner_Emery_1981a}.
\begin{figure}[h!]
\begin{center}
\begin{picture}(340,170)
\put(0,-5){\includegraphics[angle=0,width=340pt,bb=95 481 483 684]{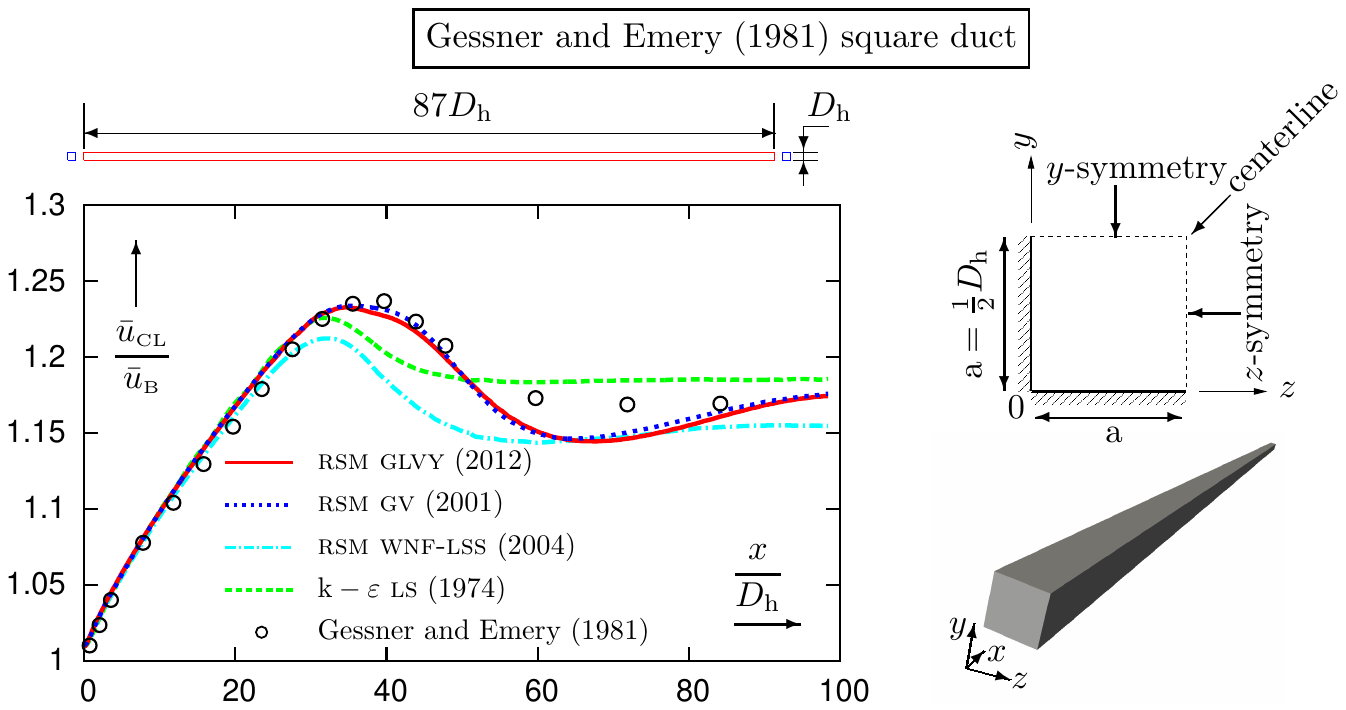}}
\end{picture}
\end{center}
\caption{Comparison of measured \cite{Gessner_Emery_1981a} streamwise evolution of $x$-wise centerline ($y=z=\mathrm{a}$) velocity $\bar u_\tsn{CL}$
with 
computations ($18\times10^6$ points grid discretizing $\tfrac{1}{4}$ of the square duct; \tabrefnp{Tab_RSMP3DDF_s_A_001})
using \parref{RSMP3DDF_s_TCsFS_ss_TCs} the \tsn{GV} \cite{Gerolymos_Vallet_2001a}, the \tsn{WNF--LSS} \cite{Gerolymos_Sauret_Vallet_2004a}
and the \tsn{GLVY} \cite{Gerolymos_Lo_Vallet_Younis_2012a} \tsn{RSM}s, and the \tsn{LS} \cite{Launder_Sharma_1974a} linear $\mathrm{k}$--$\varepsilon$ model, for developing turbulent flow in a square duct
($Re_\tsn{B}=250000$, $\bar M_\tsn{CL}\approxeq0.05$; \tabrefnp{Tab_RSMP3DDF_s_A_002}).}
\label{Fig_RSMP3DDF_s_A_ss_DTFSDGE1981_001}
\end{figure}
%
\begin{figure}[h!]
\begin{center}
\begin{picture}(340,360)
\put(0,-5){\includegraphics[angle=0,width=340pt,bb=86 260 524 724]{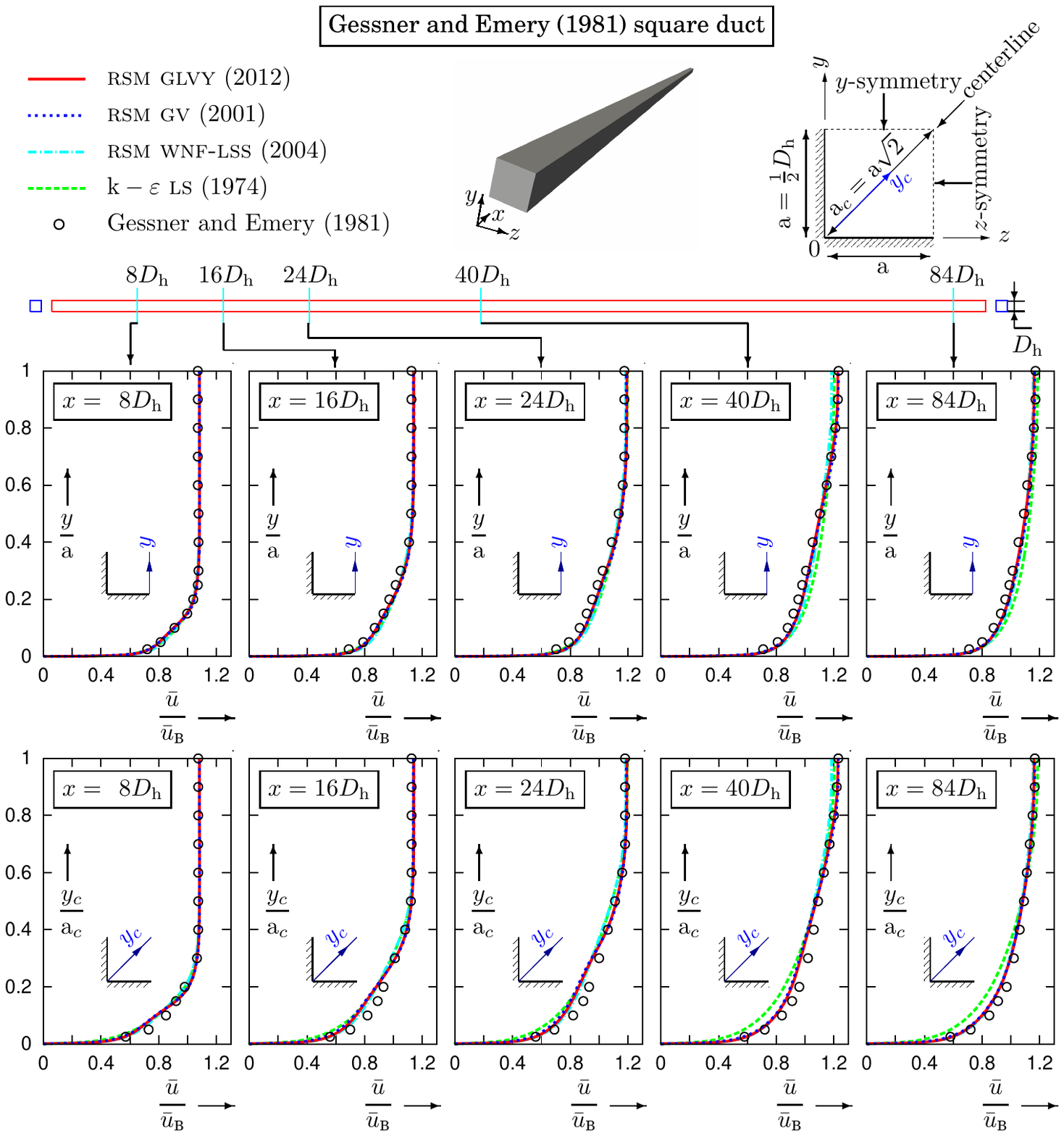}}
\end{picture}
\end{center}
\caption{Comparison of measured \cite{Gessner_Emery_1981a} streamwise ($x$-wise) velocity $\bar u$, along the wall-bisector ($z=\mathrm{a}$) and along the corner-bisector ($z=y$),
at the 5 experimental measurement stations, with 
computations ($18\times10^6$ points grid discretizing $\tfrac{1}{4}$ of the square duct; \tabrefnp{Tab_RSMP3DDF_s_A_001})
using \parref{RSMP3DDF_s_TCsFS_ss_TCs} the \tsn{GV} \cite{Gerolymos_Vallet_2001a}, the \tsn{WNF--LSS} \cite{Gerolymos_Sauret_Vallet_2004a}
and the \tsn{GLVY} \cite{Gerolymos_Lo_Vallet_Younis_2012a} \tsn{RSM}s, and the \tsn{LS} \cite{Launder_Sharma_1974a} linear $\mathrm{k}$--$\varepsilon$ model, for developing turbulent flow in a square duct
($Re_\tsn{B}=250000$, $\bar M_\tsn{CL}\approxeq0.05$; \tabrefnp{Tab_RSMP3DDF_s_A_002}).}
\label{Fig_RSMP3DDF_s_A_ss_DTFSDGE1981_002}
\end{figure}
%
\begin{figure}[h!]
\begin{center}
\begin{picture}(340,360)
\put(0,-5){\includegraphics[angle=0,width=340pt,bb=86 260 527 724]{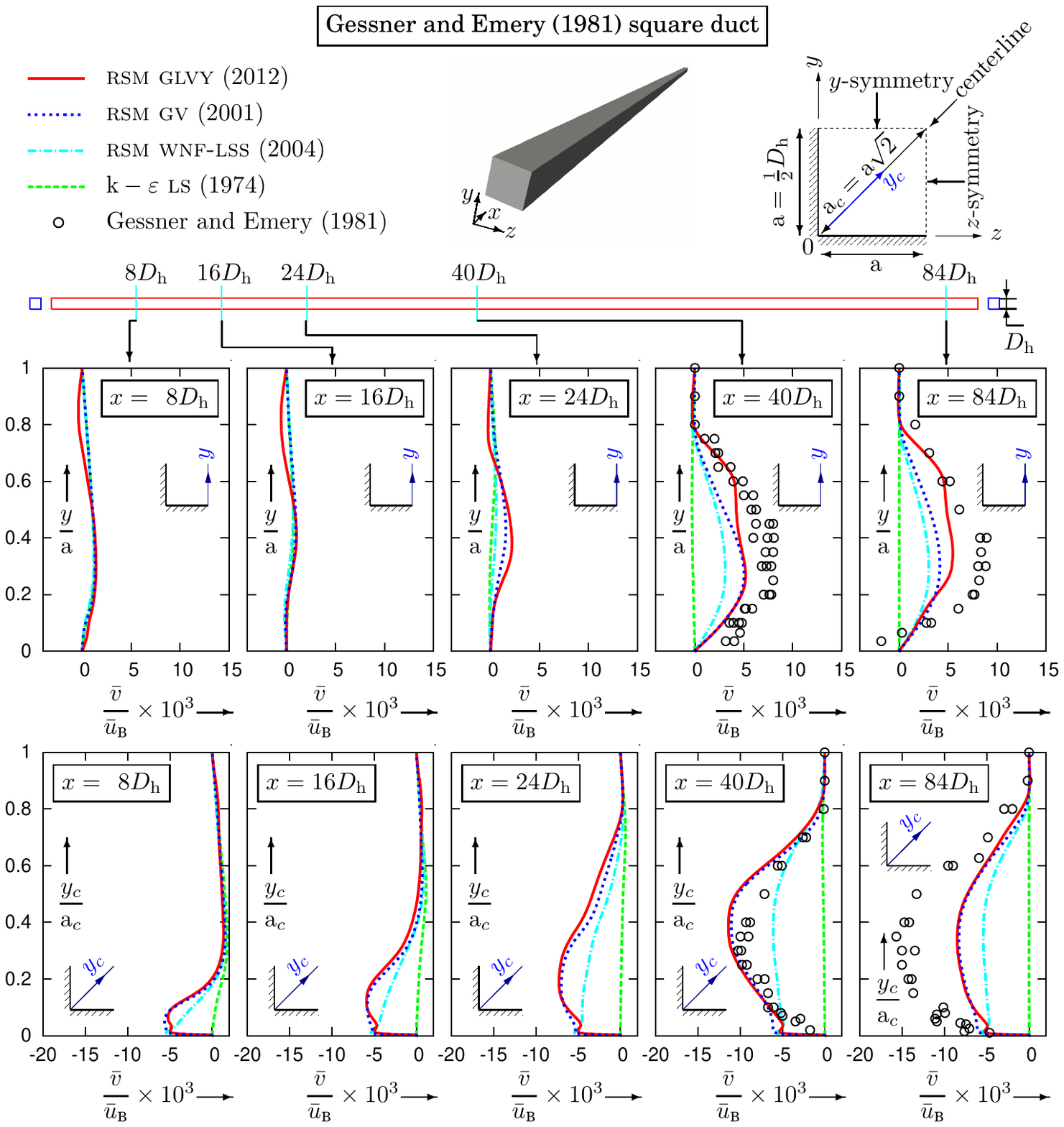}}
\end{picture}
\end{center}
\caption{Comparison of measured \cite{Gessner_Emery_1981a} wall-normal ($y$-wise) velocity $\bar v$, along the wall-bisector ($z=\mathrm{a}$) and along the corner-bisector ($z=y$, where by symmetry $\bar w=\bar v$),
at the 5 experimental measurement stations, with 
computations ($18\times10^6$ points grid discretizing $\tfrac{1}{4}$ of the square duct; \tabrefnp{Tab_RSMP3DDF_s_A_001})
using \parref{RSMP3DDF_s_TCsFS_ss_TCs} the \tsn{GV} \cite{Gerolymos_Vallet_2001a}, the \tsn{WNF--LSS} \cite{Gerolymos_Sauret_Vallet_2004a}
and the \tsn{GLVY} \cite{Gerolymos_Lo_Vallet_Younis_2012a} \tsn{RSM}s, and the \tsn{LS} \cite{Launder_Sharma_1974a} linear $\mathrm{k}$--$\varepsilon$ model, for developing turbulent flow in a square duct
($Re_\tsn{B}=250000$, $\bar M_\tsn{CL}\approxeq0.05$; \tabrefnp{Tab_RSMP3DDF_s_A_002}).}
\label{Fig_RSMP3DDF_s_A_ss_DTFSDGE1981_003}
\end{figure}
%
\begin{figure}[h!]
\begin{center}
\begin{picture}(340,360)
\put(0,-5){\includegraphics[angle=0,width=340pt,bb= 85 258 525 724]{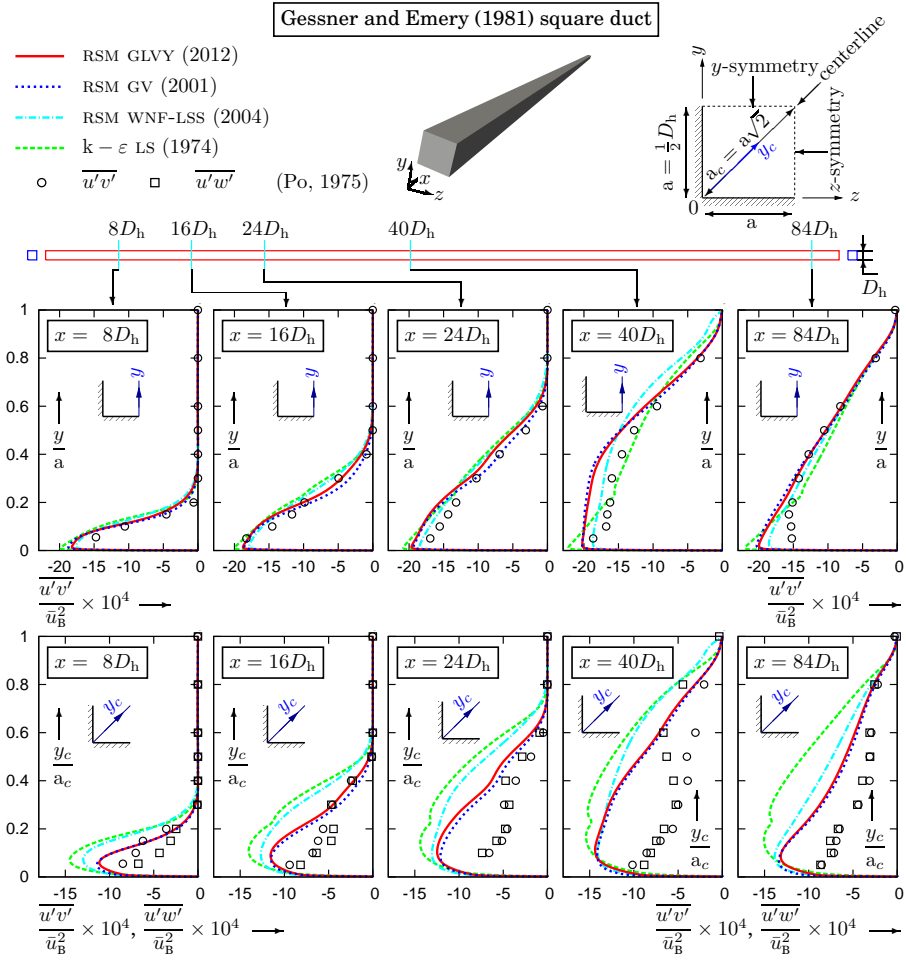}}
\end{picture}
\end{center}
\caption{Comparison of measured \cite{Gessner_Emery_1981a} Reynolds shear-stresses, $\overline{u'v'}$ along the wall-bisector ($z=\mathrm{a}$) and along the corner-bisector ($z=y$),
and $\overline{u'w'}$ along the corner-bisector ($z=y$, where by symmetry $\overline{u'w'}=\overline{u'v'}$),
at the 5 experimental measurement stations, with 
computations ($18\times10^6$ points grid discretizing $\tfrac{1}{4}$ of the square duct; \tabrefnp{Tab_RSMP3DDF_s_A_001})
using \parref{RSMP3DDF_s_TCsFS_ss_TCs} the \tsn{GV} \cite{Gerolymos_Vallet_2001a}, the \tsn{WNF--LSS} \cite{Gerolymos_Sauret_Vallet_2004a}
and the \tsn{GLVY} \cite{Gerolymos_Lo_Vallet_Younis_2012a} \tsn{RSM}s, and the \tsn{LS} \cite{Launder_Sharma_1974a} linear $\mathrm{k}$--$\varepsilon$ model, for developing turbulent flow in a square duct
($Re_\tsn{B}=250000$, $\bar M_\tsn{CL}\approxeq0.05$; \tabrefnp{Tab_RSMP3DDF_s_A_002}).}
\label{Fig_RSMP3DDF_s_A_ss_DTFSDGE1981_004}
\end{figure}
%
\begin{figure}[h!]
\begin{center}
\begin{picture}(340,360)
\put(0,-5){\includegraphics[angle=0,width=340pt,bb= 86 258 524 724]{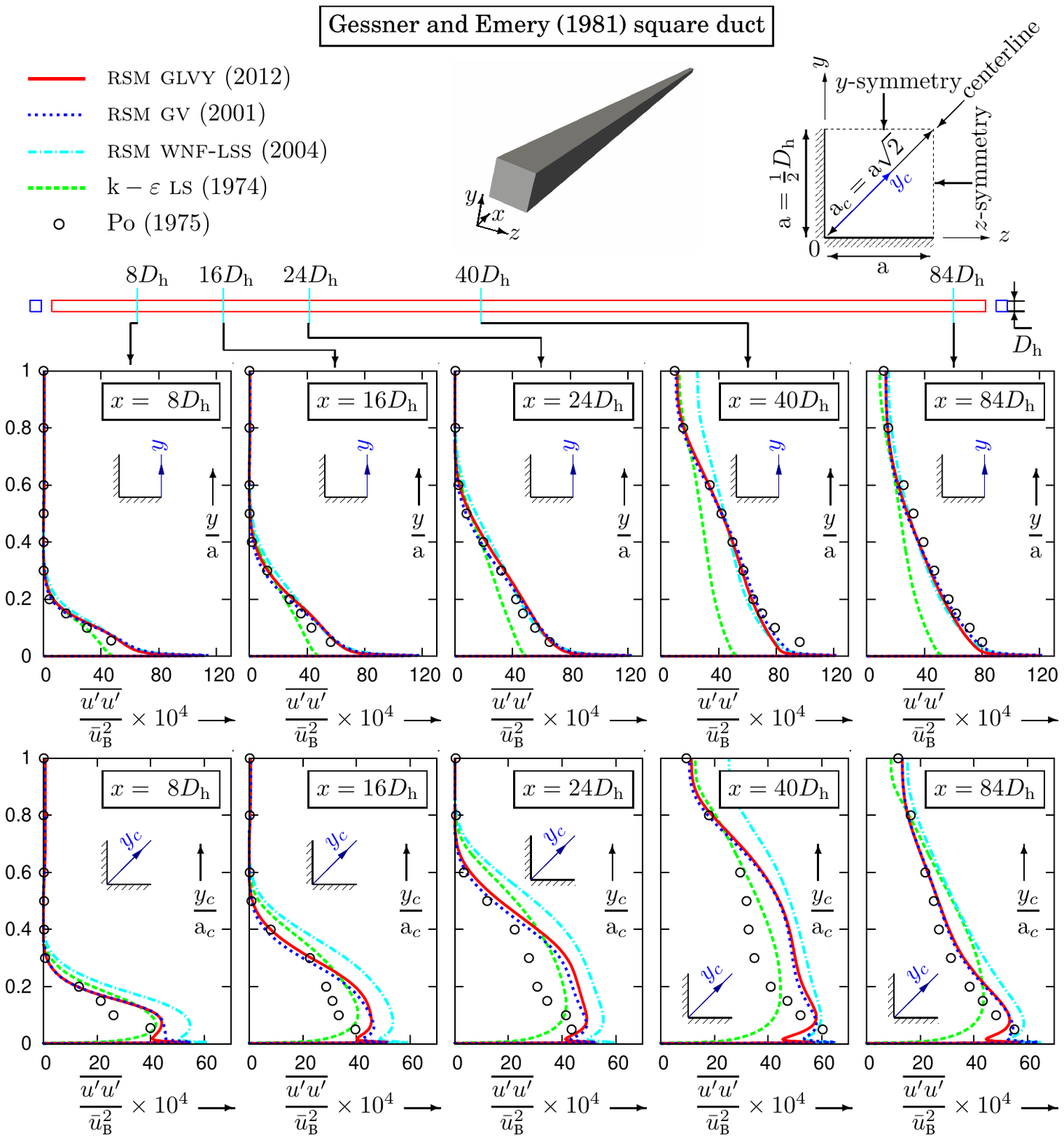}}
\end{picture}
\end{center}
\caption{Comparison of measured \cite{Gessner_Emery_1981a} streamwise diagonal Reynolds-stress $\overline{u'u'}$, along the wall-bisector ($z=\mathrm{a}$) and along the corner-bisector ($z=y$),
at the 5 experimental measurement stations, with 
computations ($18\times10^6$ points grid discretizing $\tfrac{1}{4}$ of the square duct; \tabrefnp{Tab_RSMP3DDF_s_A_001})
using \parref{RSMP3DDF_s_TCsFS_ss_TCs} the \tsn{GV} \cite{Gerolymos_Vallet_2001a}, the \tsn{WNF--LSS} \cite{Gerolymos_Sauret_Vallet_2004a}
and the \tsn{GLVY} \cite{Gerolymos_Lo_Vallet_Younis_2012a} \tsn{RSM}s, and the \tsn{LS} \cite{Launder_Sharma_1974a} linear $\mathrm{k}$--$\varepsilon$ model, for developing turbulent flow in a square duct
($Re_\tsn{B}=250000$, $\bar M_\tsn{CL}\approxeq0.05$; \tabrefnp{Tab_RSMP3DDF_s_A_002}).}
\label{Fig_RSMP3DDF_s_A_ss_DTFSDGE1981_005}
\end{figure}
 
In the Gessner and Emery \cite{Gessner_Emery_1981a} square duct, the main mechanisms are the interaction of stress-induced secondary flows, typical of the square cross-section \cite{Gessner_Jones_1965a},
with boundary-layer entrainment \cite{Kovasznay_Kibens_Blackwelder_1970a}. The streamwise thickening of the wall-layers induces blockage, resulting in flow acceleration,
which overshoots before stabilizing at the streamwise-inva\-riant fully developed level \figref{Fig_RSMP3DDF_s_A_ss_DTFSDGE1981_001}.
Sufficient grid resolution is therefore required, both near the walls and at the centerline, to correctly reproduce the development and interaction of the boundary-layers,
and as a consequence to obtain grid-convergence of the streamwise  evolution of centerline velocity \figref{Fig_RSMP3DDF_s_A_ss_DTFSDGE1981_001}.
Results are presented for an $18\times10^6$ points grid \tabref{Tab_RSMP3DDF_s_A_001} discretizing one quadrant of the duct, with symmetry-conditions at the $y$- and $z$-wise symmetry-planes.
The computational domain $L_x\times L_y\times L_z=98.43D_\mathrm{h}\times\tfrac{1}{2}D_\mathrm{h}\times\tfrac{1}{2}D_\mathrm{h}$ was slightly longer ($0\leq x\leq25\;\mathrm{m}\approxeq98D_\mathrm{h}>87D_\mathrm{h}$)
than the actual duct \cite{Gessner_Po_Emery_1979a} to avoid interaction between the uniform outflow pressure boundary-condition and computed results at the last
measurement station ($x=84D_\mathrm{h}$). The grid \figref{Fig_RSMP3DDF_s_A_001} is uniform in the streamwise ($x$) direction,
while in the $y$ and $z$ directions, $65\%$ of the $N_j=N_k=149$ points are stretched geometrically near the walls \cite{Gerolymos_Sauret_Vallet_2004b}
with ratio $r_j=r_k=1.067$ \tabref{Tab_RSMP3DDF_s_A_001}, the remaining $35\%$ being uniformly distributed in the centerline region. For the investigated flow conditions, the
first node at the walls is located at $\Delta y_w^+=\Delta z_w^+\lessapprox\tfrac{1}{2}$ \tabref{Tab_RSMP3DDF_s_A_001}.

At inflow \tabref{Tab_RSMP3DDF_s_A_002}, standard atmospheric total conditions ($p_{t_{\tsn{CL}_\mathrm{i}}}=101325\;\mathrm{Pa}$, $T_{t_{\tsn{CL}_\mathrm{i}}}=288\;\mathrm{K}$),
with a turbulent intensity $T_{u_{\tsn{CL}_\mathrm{i}}}=1\%$ and turbulent lengthscale $\ell_{\tsn{T}_{\tsn{CL}_\mathrm{i}}}=50\;\mathrm{mm}$, were assumed at the centerline.
The outflow pressure was adjusted to obtain the correct $Re_\tsn{B}=250000$ ($p_\mathrm{o}=0.995p_{t_{\tsn{CL}_\mathrm{i}}}$) corresponding to an inlet Mach number
at centerline $M_{\tsn{CL}_\mathrm{i}}\approxeq0.0516$ \tabref{Tab_RSMP3DDF_s_A_002}.
The initial inflow boundary-layer was adjusted to a different value for each turbulence model ($0.1\;\mathrm{mm}\leq\delta_\mathrm{i}\leq0.875\;\mathrm{mm}$)
to obtain a close fit to the experimental centerline velocity $\bar u_\tsn{CL}$ in the entry region of the duct ($x\in [0,10 D_\mathrm{h}]$; \figrefnp{Fig_RSMP3DDF_s_A_ss_DTFSDGE1981_001}).
Another approach would have been to start the computations with $0$ initial inflow boundary-layer thickness and to adjust in the post-processing phase the virtual origin of the developing boundary-layers ($x$-shift the results)
to best fit the experimental centerline velocity $\bar u_\tsn{CL}$ in the entry region.

The computations using the 3 \tsn{RSM}s and the $\mathrm{k}$--$\varepsilon$ model \parref{RSMP3DDF_s_TCsFS_ss_TCs} highlight \figrefsatob{Fig_RSMP3DDF_s_A_ss_DTFSDGE1981_001}
                                                                                                                                         {Fig_RSMP3DDF_s_A_ss_DTFSDGE1981_006}
the great sensitivity of the predictions to the turbulence model.
It should be stated from the outset that the underlying Boussinesq hypothesis \cite[pp. 273--278]{Wilcox_1998a} renders the linear \tsn{LS} $\mathrm{k}$--$\varepsilon$ model
ill-adapted for the present Reynolds-stress-anisotropy-driven flow \cite{Gessner_Emery_1981a}; results with the baseline \tsn{LS} $\mathrm{k}$--$\varepsilon$ model are only included as a reference to the limitations of standard Boussinesq models.

In the initial part of the duct ($0\lessapprox x\lessapprox30D_\mathrm{h}$; \figrefnp{Fig_RSMP3DDF_s_A_ss_DTFSDGE1981_001}), all of the 4 models \parref{RSMP3DDF_s_TCsFS_ss_TCs} correctly predict the
thickening of and associated blockage by the developing wall-layers, which determine, because of massflow conservation, the centerline velocity $\bar u_\tsn{CL}$.
Recall that the initial conditions for the boundary-layers at inflow ($x=0$) were independently adjusted for each turbulence model \tabref{Tab_RSMP3DDF_s_A_002},
precisely to obtain the best fit of $\bar u_\tsn{CL}$ in this region ($0\lessapprox x\lessapprox30D_\mathrm{h}$; \figrefnp{Fig_RSMP3DDF_s_A_ss_DTFSDGE1981_001}).
The best prediction is obtained by the \tsn{GLVY} and \tsn{GV} \tsn{RSM}s (whose results are quite similar; \figrefnp{Fig_RSMP3DDF_s_A_ss_DTFSDGE1981_001}),
both of which correctly simulate the $\bar u_\tsn{CL}$-peak ($30D_\mathrm{h}\lessapprox x\lessapprox50D_\mathrm{h}$; \figrefnp{Fig_RSMP3DDF_s_A_ss_DTFSDGE1981_001})
and the final fully developed level at $x=84D_\mathrm{h}$ \figref{Fig_RSMP3DDF_s_A_ss_DTFSDGE1981_001}. However, the results of the \tsn{GLVY} and \tsn{GV} \tsn{RSM}s do not
tend to this final level monotonically, as the experimental data in the region $50D_\mathrm{h}\lessapprox x\lessapprox80D_\mathrm{h}$ seem to indicate, but exhibit a $\sim\!\!2.5\%$ undershoot before reaching the correct
fully developed level at $x=84D_\mathrm{h}$ \figref{Fig_RSMP3DDF_s_A_ss_DTFSDGE1981_001}.
In contrast with the \tsn{GLVY} and \tsn{GV} \tsn{RSM}s, the \tsn{WNF--LSS RSM} severely underpredicts the experimentally observed $\bar u_\tsn{CL}$-peak ($30D_\mathrm{h}\lessapprox x\lessapprox60D_\mathrm{h}$; \figrefnp{Fig_RSMP3DDF_s_A_ss_DTFSDGE1981_001})
and also underpredicts by $\sim\!\!2.5\%$ the final fully developed value ($x=84D_\mathrm{h}$; \figrefnp{Fig_RSMP3DDF_s_A_ss_DTFSDGE1981_001}).
On the other hand, the \tsn{WNF--LSS RSM} tends to this final value in a less oscillatory fashion ($40D_\mathrm{h}\lessapprox x\lessapprox80D_\mathrm{h}$; \figrefnp{Fig_RSMP3DDF_s_A_ss_DTFSDGE1981_001}).
Finally, the \tsn{LS} $\mathrm{k}$--$\varepsilon$ model also underestimates the  $\bar u_\tsn{CL}$-peak ($30D_\mathrm{h}\lessapprox x\lessapprox60D_\mathrm{h}$; \figrefnp{Fig_RSMP3DDF_s_A_ss_DTFSDGE1981_001})
and tends monotonically to an $\sim\!\!1.5\%$ overestimated value of the final fully developed level ($x=84D_\mathrm{h}$; \figrefnp{Fig_RSMP3DDF_s_A_ss_DTFSDGE1981_001}).

The detailed evolution of the streamwise mean-flow velocity $\bar u$ profiles \figref{Fig_RSMP3DDF_s_A_ss_DTFSDGE1981_002} provides insight into the predictions of centerline velocity $\bar u_\tsn{CL}$ \figref{Fig_RSMP3DDF_s_A_ss_DTFSDGE1981_001}
by the different models. The term wall-bisector was used by Gessner and Emery \cite{Gessner_Emery_1981a} to denote the symmetry-plane at $z=\mathrm{a}=\tfrac{1}{2}D_\mathrm{h}$
and the term corner-bisector to denote the diagonal with distance \smash{$y_c:=\tfrac{\sqrt{2}}{2}(y-y_w)+\tfrac{\sqrt{2}}{2}(z-z_w)$} from the corner (notice that $\mathrm{a}_c^{-1}y_c=\mathrm{a}^{-1}y=\mathrm{a}^{-1}z$
along the diagonal whose length between the corner and the centerline is $\mathrm{a}_c:=\mathrm{a}\sqrt{2}$).
The flowfield along the corner-bisector $y_c$ is strongly influenced by the secondary flows. The \tsn{GLVY} and \tsn{GV} \tsn{RSM}s yield quite accurate results for the $\bar u$ profiles \figref{Fig_RSMP3DDF_s_A_ss_DTFSDGE1981_002},
both along the wall-bisector $y$ ($z=\mathrm{a}$; \figrefnp{Fig_RSMP3DDF_s_A_ss_DTFSDGE1981_002}) and along the corner-bisector $y_c$ ($z=y$; \figrefnp{Fig_RSMP3DDF_s_A_ss_DTFSDGE1981_002}).
Notice, nonetheless, that no experimental data are available in the region $50D_\mathrm{h}\lessapprox x\lessapprox80D_\mathrm{h}$ where the slight undershoot in centerline velocity $\bar u_\tsn{CL}$
was observed \figref{Fig_RSMP3DDF_s_A_ss_DTFSDGE1981_001}. The predictions of the \tsn{WNF--LSS RSM} for the streamwise velocity $\bar u$ \figref{Fig_RSMP3DDF_s_A_ss_DTFSDGE1981_001}
are similar to those of the \tsn{GLVY} and \tsn{GV} \tsn{RSM}s, except for the outer part (wake region) of the boundary-layer, especially at $x=40D_\mathrm{h}$ along the wall-bisector $y$ ($z=\mathrm{a}$; \figrefnp{Fig_RSMP3DDF_s_A_ss_DTFSDGE1981_002}) and at
$x\in\{8D_\mathrm{h},16D_\mathrm{h},24D_\mathrm{h},40D_\mathrm{h}\}$ along the corner-bisector $y_c$ ($z=y$; \figrefnp{Fig_RSMP3DDF_s_A_ss_DTFSDGE1981_002}).
Finally, expectedly, the linear \tsn{LS} $\mathrm{k}$--$\varepsilon$ model makes the worst prediction, especially at the last 2 stations
$x\in\{40D_\mathrm{h},84D_\mathrm{h}\}$, where it overpredicts $\bar u$ in the lower part of the boundary-layer along the wall-bisector $y$ ($z=\mathrm{a}$; \figrefnp{Fig_RSMP3DDF_s_A_ss_DTFSDGE1981_002})
and rather severely underpredicts it in the lower part of the boundary-layer along the corner-bisector $y_c$ ($z=y$; \figrefnp{Fig_RSMP3DDF_s_A_ss_DTFSDGE1981_002}).

By the continuity equation the $x$-wise development of the streamwise velocity $\bar u$ is related to the profiles of the in-plane velocity components, $\bar v$ and $\bar w$. Measurements of the $y$-wise component $\bar v$
are only available at the last 2 measurement planes ($x\in\{40D_\mathrm{h},84D_\mathrm{h}\}$; \figrefnp{Fig_RSMP3DDF_s_A_ss_DTFSDGE1981_003}). Notice first that, along the wall-bisector $y$ ($z=\mathrm{a}$) we have $\bar w=0$ by symmetry,
while along the corner-bisector $y_c$ ($z=y$) we have $\bar w=\bar v$ again by symmetry. Contrary to the results for the profiles of streamwise velocity $\bar u$ \figref{Fig_RSMP3DDF_s_A_ss_DTFSDGE1981_002},
the predictions of the $y$-wise component $\bar v$ have noticeable differences between the various models \figref{Fig_RSMP3DDF_s_A_ss_DTFSDGE1981_003}. The \tsn{GLVY RSM} gives the best prediction of secondary velocities,
both along the wall-bisector $y$ ($z=\mathrm{a}$; \figrefnp{Fig_RSMP3DDF_s_A_ss_DTFSDGE1981_003}) and along the corner-bisector $y_c$ ($z=y$; \figrefnp{Fig_RSMP3DDF_s_A_ss_DTFSDGE1981_003}).
Although the agreement of the \tsn{GLVY RSM} results with measurements is quite satisfactory at $x=40D_\mathrm{h}$, the secondary velocities are underestimated at the last measurement station $x=84D_\mathrm{h}$ \figref{Fig_RSMP3DDF_s_A_ss_DTFSDGE1981_003}.
The \tsn{GV RSM} gives results very close to those of the \tsn{GLVY RSM} along the corner-bisector $y_c$ ($z=y$; \figrefnp{Fig_RSMP3DDF_s_A_ss_DTFSDGE1981_003}),
some discrepancies very near the corner ($y_c\lessapprox0.1\mathrm{a}_c$; \figrefnp{Fig_RSMP3DDF_s_A_ss_DTFSDGE1981_003}) notwithstanding, but underestimates $\bar v$ along the wall-bisector $y$ ($z=\mathrm{a}$; \figrefnp{Fig_RSMP3DDF_s_A_ss_DTFSDGE1981_003})
at the outer part of the boundary-layer. The \tsn{WNF--LSS RSM}, does predict secondary flows, less intense than the \tsn{GLVY} and \tsn{GV} \tsn{RSM}s \figref{Fig_RSMP3DDF_s_A_ss_DTFSDGE1981_003},
while the linear \tsn{LS} $\mathrm{k}$--$\varepsilon$ model completely fails \figref{Fig_RSMP3DDF_s_A_ss_DTFSDGE1981_003},
implying that the strong values of $\bar v$ at the last measurement stations ($x\in\{40D_\mathrm{h},84D_\mathrm{h}\}$; \figrefnp{Fig_RSMP3DDF_s_A_ss_DTFSDGE1981_003}) are the consequence of secondary turbulence-driven flows,
in a region where the flow approaches the fully developed state \cite{Gessner_Jones_1965a,Gessner_Emery_1981a,Bradshaw_1987a}.
\vspace{-.05in}

\begin{figure}[h!]
\begin{center}
\begin{picture}(340,480)
\put(0,-5){\includegraphics[angle=0,width=340pt,bb=85 99 524 724]{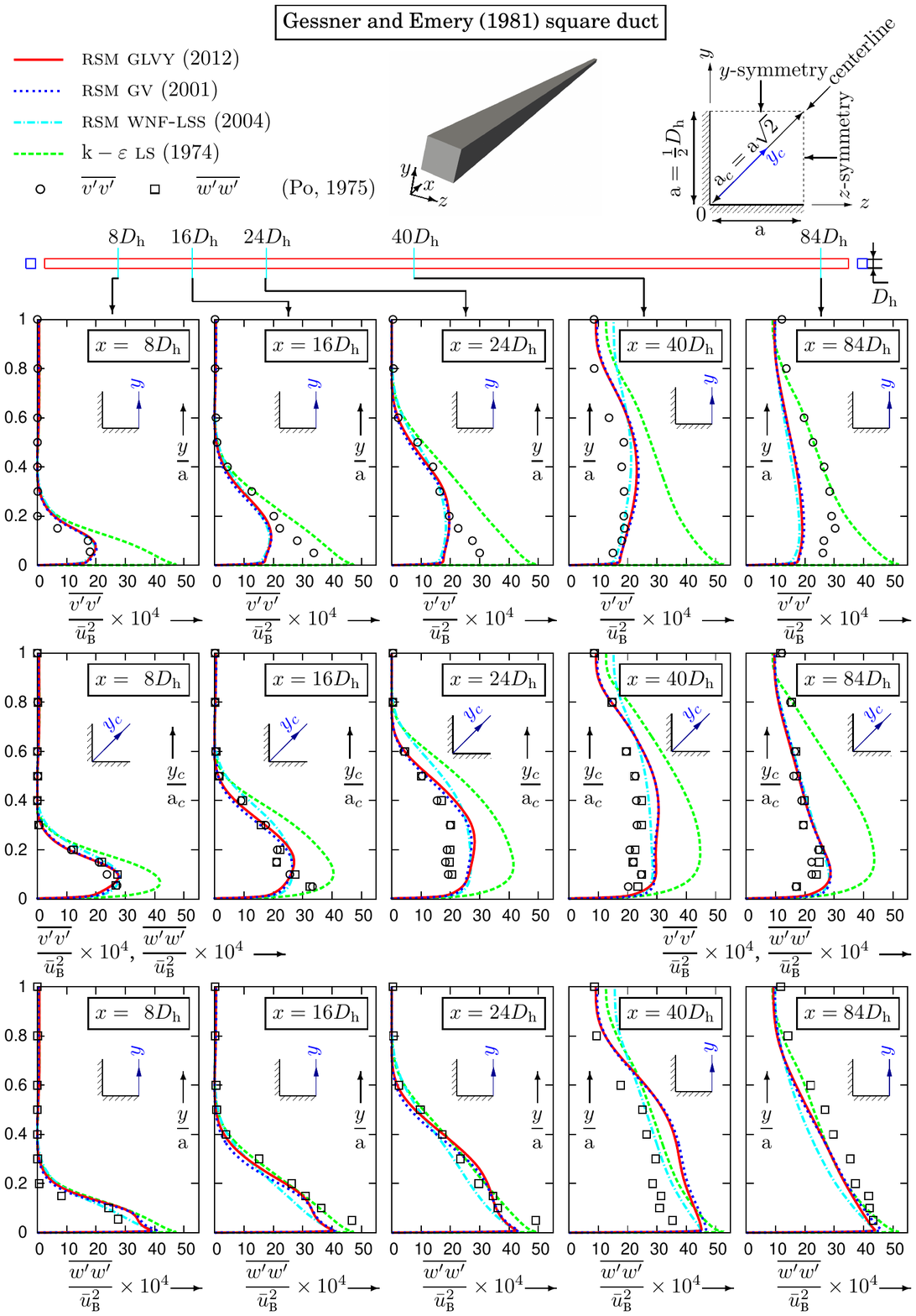}}
\end{picture}
\end{center}
\caption{Comparison of measured \cite{Gessner_Emery_1981a} normal Reynolds-stresses $\overline{v'v'}$ and $\overline{w'w'}$, along the wall-bisector ($z=\mathrm{a}$) and along the corner-bisector ($z=y$, where by symmetry $\overline{w'w'}=\overline{v'v'}$),
at the 5 experimental measurement stations, with 
computations ($18\times10^6$ points grid discretizing $\tfrac{1}{4}$ of the square duct; \tabrefnp{Tab_RSMP3DDF_s_A_001})
using \parref{RSMP3DDF_s_TCsFS_ss_TCs} the \tsn{GV} \cite{Gerolymos_Vallet_2001a}, the \tsn{WNF--LSS} \cite{Gerolymos_Sauret_Vallet_2004a}
and the \tsn{GLVY} \cite{Gerolymos_Lo_Vallet_Younis_2012a} \tsn{RSM}s, and the \tsn{LS} \cite{Launder_Sharma_1974a} linear $\mathrm{k}$--$\varepsilon$ model, for developing turbulent flow in a square duct
($Re_\tsn{B}=250000$, $\bar M_\tsn{CL}\approxeq0.05$; \tabrefnp{Tab_RSMP3DDF_s_A_002}).}
\label{Fig_RSMP3DDF_s_A_ss_DTFSDGE1981_006}
\end{figure}

The comparison of computational results with measured Reynolds-stresses \figrefsatob{Fig_RSMP3DDF_s_A_ss_DTFSDGE1981_004}
                                                                                    {Fig_RSMP3DDF_s_A_ss_DTFSDGE1981_006}
is consistent with the comparisons of the mean-flow velocity field \figrefsab{Fig_RSMP3DDF_s_A_ss_DTFSDGE1981_002}
                                                                             {Fig_RSMP3DDF_s_A_ss_DTFSDGE1981_003}.
The \tsn{GLVY} and \tsn{GV} \tsn{RSM}s give the best overall prediction of the shear Reynolds-stresses, $\overline{u'v'}$ along the wall-bisector $y$ ($z=\mathrm{a}$; \figrefnp{Fig_RSMP3DDF_s_A_ss_DTFSDGE1981_004})
and $\overline{u'v'}=\overline{u'w'}$ (by symmetry) along the corner-bisector $y_c$ ($z=y$; \figrefnp{Fig_RSMP3DDF_s_A_ss_DTFSDGE1981_004}),
but overestimate their magnitude, especially along the corner-bisector $y_c$ ($z=y$; \figrefnp{Fig_RSMP3DDF_s_A_ss_DTFSDGE1981_004}).
The experimental data are generally consistent with the symmetry condition $\overline{u'v'}=\overline{u'w'}$ along the corner-bisector $y_c$ ($z=y$; \figrefnp{Fig_RSMP3DDF_s_A_ss_DTFSDGE1981_004}),
except at $x=40D_\mathrm{h}$ in the outer part of the boundary-layer ($y_c\gtrapprox0.4\mathrm{a}_c$; $z=y$; \figrefnp{Fig_RSMP3DDF_s_A_ss_DTFSDGE1981_004}).
The overprediction of the shear Reynolds-stress $\overline{u'v'}=\overline{u'w'}$ along the corner-bisector $y_c$ at $x=40D_\mathrm{h}$ ($z=y$; \figrefnp{Fig_RSMP3DDF_s_A_ss_DTFSDGE1981_004}) is not consistent
with the satisfactory prediction of the mean-flow velocity field at this station ($x=40D_\mathrm{h}$; $z=y$; \figrefsnpab{Fig_RSMP3DDF_s_A_ss_DTFSDGE1981_002}
                                                                                                                         {Fig_RSMP3DDF_s_A_ss_DTFSDGE1981_003}),
especially as the $x$-wise gradients predicted by the \tsn{GLVY} and \tsn{GV} \tsn{RSM}s are in good agreement with experimental data at $x=40D_\mathrm{h}$ \figref{Fig_RSMP3DDF_s_A_ss_DTFSDGE1981_001}.
Regarding the last measurement station at $x=84D_\mathrm{h}$, the shear Reynolds-stress $\overline{u'v'}=\overline{u'w'}$ predicted by the \tsn{GLVY} and \tsn{GV} \tsn{RSM}s
along the corner-bisector $y_c$ ($x=84D_\mathrm{h}$; $z=y$; \figrefnp{Fig_RSMP3DDF_s_A_ss_DTFSDGE1981_004}) is closer to the experimental data than at $x=40D_\mathrm{h}$,
but computed values are still larger in magnitude by $\sim\!\!30\%$. The \tsn{WNF--LSS RSM} and \tsn{LS} $\mathrm{k}$--$\varepsilon$ models predictions of the shear Reynolds-stress $\overline{u'v'}$
along the wall-bisector $y$ ($z=\mathrm{a}$; \figrefnp{Fig_RSMP3DDF_s_A_ss_DTFSDGE1981_004}) are generally similar with those of the \tsn{GLVY} and \tsn{GV} \tsn{RSM}s, in satisfactory agreement with measurements.
On the other hand, the \tsn{WNF--LSS RSM} and the \tsn{LS} $\mathrm{k}$--$\varepsilon$ model perform less satisfactorily than the \tsn{GLVY} and \tsn{GV} \tsn{RSM}s regarding the prediction
of the shear Reynolds-stresses $\overline{u'v'}=\overline{u'w'}$ (by symmetry) along the corner-bisector $y_c$ ($z=y$; \figrefnp{Fig_RSMP3DDF_s_A_ss_DTFSDGE1981_004}),
the \tsn{WNF--LSS RSM}, expectedly, performing better than the linear \tsn{LS} $\mathrm{k}$--$\varepsilon$ model.
The \tsn{GLVY} and \tsn{GV} \tsn{RSM}s predict quite accurately the streamwise normal Reynolds-stress $\overline{u'^2}$ \figref{Fig_RSMP3DDF_s_A_ss_DTFSDGE1981_005}
both along the wall-bisector $y$ ($z=\mathrm{a}$) and the corner-bisector $y_c$ ($z=y$), some slight discrepancies
along the corner-bisector $y_c$ ($x\in\{16D_\mathrm{h},24D_\mathrm{h},40D_\mathrm{h}\}$; $z=y$; \figrefnp{Fig_RSMP3DDF_s_A_ss_DTFSDGE1981_005}) notwithstanding.
The predictions of the streamwise normal Reynolds-stress $\overline{u'^2}$ by the \tsn{WNF--LSS RSM} and the \tsn{LS} $\mathrm{k}$--$\varepsilon$ model are, again,
less satisfactory \figref{Fig_RSMP3DDF_s_A_ss_DTFSDGE1981_005}, especially along the wall-bisector $y$ ($z=\mathrm{a}$; \figrefnp{Fig_RSMP3DDF_s_A_ss_DTFSDGE1981_005}).
Regarding the prediction of the other normal Reynolds-stresses, wall-normal $\overline{v'^2}$ along the wall-bisector $y$ ($z=\mathrm{a}$), transverse $\overline{w'^2}$ along the wall-bisector $y$ ($z=\mathrm{a}$),
and secondary $\overline{v'^2}=\overline{w'^2}$ along the corner-bisector $y_c$ ($z=y$), all 3 \tsn{RSM}s (\tsn{GLVY}, \tsn{GV} and \tsn{WNF--LSS}) are in good agreement with experimental data \figref{Fig_RSMP3DDF_s_A_ss_DTFSDGE1981_006},
in contrast with the linear \tsn{LS} $\mathrm{k}$--$\varepsilon$ model, which completely fails in predicting the Reynolds-stress tensor anisotropy \figref{Fig_RSMP3DDF_s_A_ss_DTFSDGE1981_006},
because of the pathological shortcomings of the Boussinesq hypothesis \cite[pp. 273--278]{Wilcox_1998a}.
\clearpage

To explain the better agreement with experimental data of the \tsn{GLVY} and \tsn{GV} \tsn{RSM}s, compared to the \tsn{WNF--LSS RSM} \figrefsatob{Fig_RSMP3DDF_s_A_ss_DTFSDGE1981_001}
                                                                                                                                                 {Fig_RSMP3DDF_s_A_ss_DTFSDGE1981_006},
it is interesting to examine the differences between the closures \tabref{Tab_RSMP3DDF_s_TCsFS_ss_TCs_001}.
The differences \tabref{Tab_RSMP3DDF_s_TCsFS_ss_TCs_001} between the \tsn{GLVY} and \tsn{GV} \tsn{RSM}s (pressure diffusion $d_{ij}^{(p)}$, explicit algebraic modelling for $\varepsilon_{ij}$,
extra inhomogeneous terms in $\Pi_{ij}$) do not have any substantial influence on the prediction of the \cite{Gessner_Emery_1981a} square duct flow, the only noticeable difference
being the better prediction by the \tsn{GLVY RSM} of the wall-normal velocity $\bar v$ along the wall-bisector $y$ ($x\in\{40D_\mathrm{h},84D_\mathrm{h}\}$; $z=\mathrm{a}$; \figrefnp{Fig_RSMP3DDF_s_A_ss_DTFSDGE1981_003}),
especially in the outer part of the boundary-layer ($y\gtrapprox0.6\mathrm{a}$). There are 2 main differences between the \tsn{WNF--LSS RSM} and the \tsn{GV RSM} \tabref{Tab_RSMP3DDF_s_TCsFS_ss_TCs_001},
the coefficient-function $C_\phi^{(\tsn{RH})}$ of the isotropisation-of-production model for the homogeneous rapid part of redistribution \eqref{Eq_RSMP3DDF_s_TCsFS_ss_TCs_007c},
and the closure for the triple velocity correlations \eqref{Eq_RSMP3DDF_s_TCsFS_ss_TCs_005}. The coefficient $C_\phi^{(\tsn{RH})}$ \tabref{Tab_RSMP3DDF_s_TCsFS_ss_TCs_001} in the \tsn{GLVY} and \tsn{GV} \tsn{RSM}s was designed
\cite[Fig. 4, p. 1838]{Gerolymos_Vallet_2001a} to increase rapidly towards a value of $1$ when Lumley's \cite{Lumley_1978a} flatness parameter $A$ \eqref{Eq_RSMP3DDF_s_TCsFS_ss_TCs_001b} increases
beyond $0.55$ (which is approximately the value of $A$ just before the beginning of the logarithmic zone of the flat-plate boundary-layer velocity profile). Regarding turbulent diffusion \tabref{Tab_RSMP3DDF_s_TCsFS_ss_TCs_001},
the \tsn{GLVY} and \tsn{GV} \tsn{RSM}s use the Hanjali\'c-Launder \cite{Hanjalic_Launder_1976a}
closure, whereas the \tsn{WNF--LSS RSM} uses the Daly-Harlow \cite{Daly_Harlow_1970a} closure \parref{RSMP3DDF_s_TCsFS_ss_TCs}.
Obviously \figrefsatob{Fig_RSMP3DDF_s_A_ss_DTFSDGE1981_001}
                      {Fig_RSMP3DDF_s_A_ss_DTFSDGE1981_006},
the combined use of these 2 modelling choices in the \tsn{GLVY} and \tsn{GV} \tsn{RSM}s substantially improves the prediction of the \cite{Gessner_Emery_1981a} square duct flow
compared to the \tsn{WNF--LSS RSM}. To put into perspective the specific influence of each of the 2 mechanisms, a test-model (not recommended for practical use), hereafter \tsn{GV--DH RSM},
which combines the $C_\phi^{(\tsn{RH})}$ coefficient-function of the \tsn{GLVY} and \tsn{GV} \tsn{RSM}s \tabref{Tab_RSMP3DDF_s_TCsFS_ss_TCs_001} with the Daly-Harlow \cite{Daly_Harlow_1970a} closure
for turbulent diffusion (with appropriate recalibration of various coefficients to get the correct log-law in flat-plate boundary-layer flow; \cite[Tab. 2, p. 418]{Gerolymos_Sauret_Vallet_2004a}),
has been developed \cite{Gerolymos_Sauret_Vallet_2004a,Vallet_2007a}.
Calculations of the \cite{Gessner_Emery_1981a} square duct flow with the \tsn{GV}, \tsn{GV--DH}  and \tsn{WNF--LSS} \tsn{RSM}s, using the same inflow boundary-layer-thickness
\cite[$\delta_{y_i}=\delta_{z_i}=0.5\;\mathrm{mm}$, Fig. 7, p. 422]{Gerolymos_Sauret_Vallet_2004a} for all of the models, indicate that $C_\phi^{(\tsn{RH})}$ influences the initial part of the
region where the boundary-layers on the 4 walls first merge ($30D_\mathrm{h}\lessapprox x\lessapprox40D_\mathrm{h}$; \figrefnp{Fig_RSMP3DDF_s_A_ss_DTFSDGE1981_001}) whereas turbulent diffusion is active especially in the region
after the centerline velocity peak ($40D_\mathrm{h}\lessapprox x\lessapprox60D_\mathrm{h}$; \figrefnp{Fig_RSMP3DDF_s_A_ss_DTFSDGE1981_001}).

%
%
%
%
%
\subsection{Circular-to-rectangular transition duct \cite{Davis_Gessner_1992a}}\label{RSMP3DDF_s_A_ss_CtoRTDDG1992}
%
%
%
%
%

This configuration \cite{Davis_1991a,Davis_Gessner_1992a}
\begin{figure}[h!]
\begin{center}
\begin{picture}(340,300)
\put(0,-5){\includegraphics[angle=0,width=340pt,bb= 99 204 522 586]{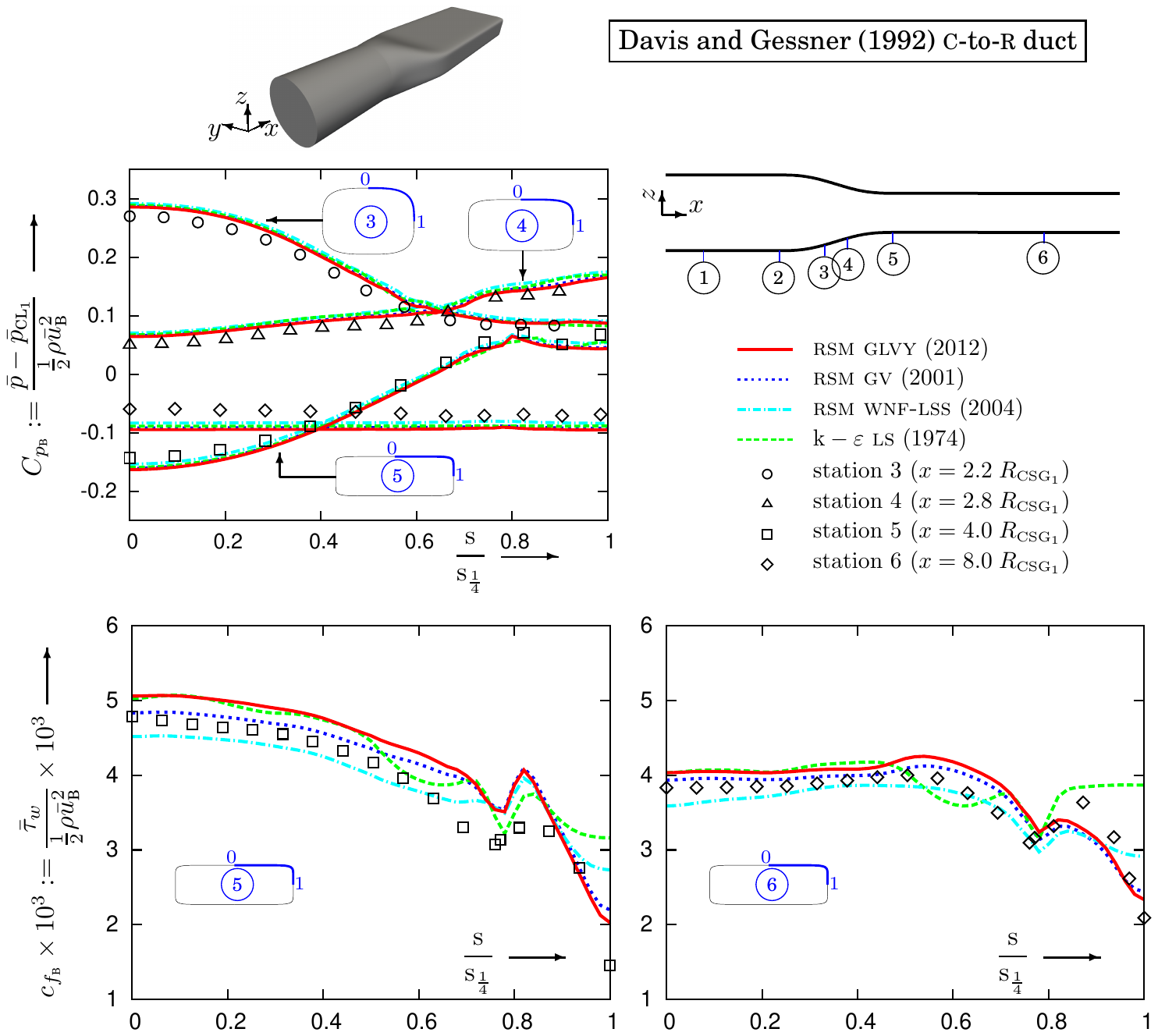}}
\end{picture}
\end{center}
\caption{Comparison of measured \cite{Davis_Gessner_1992a} wall-pressure coefficient $C_{p_\tsn{B}}$ and skin-friction coefficient $c_{f_\tsn{B}}$,
at 4 experimental measurement stations, with 
computations ($10\times10^6$ points grid discretizing the entire duct; \tabrefnp{Tab_RSMP3DDF_s_A_001})
using \parref{RSMP3DDF_s_TCsFS_ss_TCs} the \tsn{GV} \cite{Gerolymos_Vallet_2001a}, the \tsn{WNF--LSS} \cite{Gerolymos_Sauret_Vallet_2004a}
and the \tsn{GLVY} \cite{Gerolymos_Lo_Vallet_Younis_2012a} \tsn{RSM}s, and the \tsn{LS} \cite{Launder_Sharma_1974a} linear $\mathrm{k}$--$\varepsilon$ model, for turbulent flow in a \tsn{C}-to-\tsn{R} transition duct
($Re_\tsn{B}=390000$, $\bar M_\tsn{CL}\approxeq0.09$; \tabrefnp{Tab_RSMP3DDF_s_A_002}; $\mathrm{s}$ is the curvilinear coordinate of the duct contour in the $yz$-plane normalized by $\tfrac{1}{4}$ of the circumference $\mathrm{s}_{\frac{1}{4}}$).}
\label{Fig_RSMP3DDF_s_A_ss_CtoRTDDG1992_001}
\end{figure}
%
\begin{table}[h!]
\begin{center}
\caption{Comparison of experimental inflow boundary-layer parameters, for the \tsn{C}-to-\tsn{R} transition duct \cite[Tab. 1, p. 370]{Davis_Gessner_1992a} at station 1 located at $x=-R_{\tsn{CSG}_1}$
         (cross-sectional transition starts at $x=+R_{\tsn{CSG}_1}$, one diameter downstream; \figrefnp{Fig_RSMP3DDF_s_A_ss_CtoRTDDG1992_001}) and for
         the \tsn{S}-duct \cite[Tab. 2, p. 671]{Wellborn_Reichert_Okiishi_1994a} at station \tsn{A} located at $x=-R_{\tsn{CSG}_\tsn{A}}$
         (centerline and cross-section evolution starts at $x=0$, half a diameter downstream; \figrefsnpab{Fig_RSMP3DDF_s_A_ss_D3DSDWRO1994_001}{Fig_RSMP3DDF_s_A_ss_D3DSDWRO1994_002}),
         with computational results (grids \tabrefnp{Tab_RSMP3DDF_s_A_001}) using the \tsn{GLVY RSM} \cite{Gerolymos_Lo_Vallet_Younis_2012a}, the \tsn{GV RSM} \cite{Gerolymos_Vallet_2001a}, the \tsn{WNF--LSS RSM} \cite{Gerolymos_Sauret_Vallet_2004a}
         and the \tsn{LS} $\mathrm{k}$-$\varepsilon$ \cite{Launder_Sharma_1974a}.}
\label{Tab_RSMP3DDF_s_A_003}
\scalebox{0.92}{
\begin{tabular}{llrrrrr}\hline\addlinespace[0.5em]
configuration                      &                                                          & experiment & \tsn{GLVY RSM} & \tsn{GV RSM} & \tsn{WNF--LSS RSM}&\tsn{LS} $\mathrm{k}$-$\varepsilon$\\\hline\addlinespace[0.5em]
\tsn{C}-to-\tsn{R} duct (station 1)& $\bar u_\tsn{B}\;(\mathrm{m}\;\mathrm{s}^{-1})$          &$29.95$     &$30.20$         &$30.25$       &$30.41$            &$30.51$                            \\
                                   & $Re_\tsn{B}\times10^{-6}$                                &$0.390$     &$0.397$         &$0.398$       &$0.400$            &$0.401$                            \\
                                   & $\delta_{995}               \;R_\tsn{CSG}^{-1}\times100$ &$28.55$     &$30.03$         &$30.05$       &$29.95$            &$28.98$                            \\
                                   & $\delta_{{1_k}_\mathrm{axi}}\;R_\tsn{CSG}^{-1}\times100$ &$3.83$      &$3.94$          &$3.92$        &$3.87$             &$3.59$                             \\
                                   & $\delta_{{2_k}_\mathrm{axi}}\;R_\tsn{CSG}^{-1}\times100$ &$2.81$      &$2.93$          &$2.92$        &$2.90$             &$2.69$                             \\
                                   & $\delta_{{3_k}_\mathrm{axi}}\;R_\tsn{CSG}^{-1}\times100$ &$4.97$      &$5.21$          &$5.20$        &$5.16$             &$4.80$                             \\
                                   & $H_{{12_k}_\mathrm{axi}}$                                &$1.36$      &$1.34$          &$1.34$        &$1.34$             &$1.33$                             \\
                                   & $H_{{32_k}_\mathrm{axi}}$                                &$1.77$      &$1.78$          &$1.78$        &$1.78$             &$1.78$                             \\\hline\addlinespace[0.5em]
\tsn{S}-duct (station \tsn{A})     & $\breve M_\tsn{CL}$                                      & $0.6~\,$   & $0.59$         & $0.60$       &$0.60$             &$0.62$                             \\
                                   & $Re_\tsn{CL}\times10^{-6}$                               & $2.6~\,$   & $2.58$         & $2.59$       &$2.62$             &$2.67$                             \\
                                   & $\delta_{95}                \;R_\tsn{CSG}^{-1}\times100$ & $6.95$     & $6.97$         & $6.95$       &$6.94$             &$7.26$                             \\
                                   & $\delta_{{1_k}_\mathrm{axi}}\;R_\tsn{CSG}^{-1}\times100$ & $1.46$     & $1.11$         & $1.10$       &$1.11$             &$1.13$                             \\
                                   & $\delta_{{2_k}_\mathrm{axi}}\;R_\tsn{CSG}^{-1}\times100$ & $1.06$     & $0.81$         & $0.80$       &$0.81$             &$0.84$                             \\
                                   & $\delta_{{3_k}_\mathrm{axi}}\;R_\tsn{CSG}^{-1}\times100$ &            & $1.42$         & $1.41$       &$1.42$             &$1.47$                             \\
                                   & $H_{{12_k}_\mathrm{axi}}$                                & $1.38$     & $1.37$         & $1.36$       &$1.37$             &$1.36$                             \\
                                   & $H_{{32_k}_\mathrm{axi}}$                                &            & $1.75$         & $1.76$       &$1.76$             &$1.76$                             \\\hline
\end{tabular}
}

\end{center}
 {\footnotesize $\bar u_\tsn{B}$: bulk velocity;
                $R_\tsn{CSG}$ ($D_\tsn{CSG}$): casing radius (diameter);
                $Re_\tsn{B}=\bar u_\tsn{B} D_\tsn{CSG} \nu^{-1}$: bulk Reynolds number ($\nu$ is the practically constant kinematic viscosity);
                $\delta_{995}$ ($\delta_{95}$): boundary-layer thickness measured from the wall to the location where the velocity is $99.5\%$ ($95\%$) of $\bar u_\tsn{CL}$;
                $\delta_{{1_k}_\mathrm{axi}}=\int^{\delta}_0(1-\bar{u}/\bar{u}_\tsn{CL})(R/R_\tsn{CSG})d(R_\tsn{CSG}-R)$: axisymmetric kinematic boundary-layer displacement-thickness;
                $\delta_{{2_k}_\mathrm{axi}}=\int^{\delta}_0(1-\bar{u}/\bar{u}_\tsn{CL})(\bar{u}/\bar{u}_\tsn{CL})(R/R_\tsn{CSG})d(R_\tsn{CSG}-R)$: axisymmetric kinematic boundary-layer momentum-thickness;
                $\delta_{{3_k}_\mathrm{axi}}=\int^{\delta}_0(1-\bar{u}^2/\bar{u}_\tsn{CL}^2)(\bar{u}/\bar{u}_\tsn{CL})(R/R_\tsn{CSG})d(R_\tsn{CSG}-R)$: axisymmetric kinematic boundary-layer energy-thickness;
                $H_{{12_k}_\mathrm{axi}}=\delta_{{1_k}_\mathrm{axi}}/\delta_{{2_k}_\mathrm{axi}}$ ($H_{{32_k}_\mathrm{axi}}=\delta_{{3_k}_\mathrm{axi}}/\delta_{{2_k}_\mathrm{axi}}$): axisymmetric kinematic boundary-layer shape-factors;
                $\breve M_\tsn{CL}=\tilde u_\tsn{CL} \breve a_\tsn{CL}$: centerline Mach number ($\breve a_\tsn{CL}$ is the centerline sound-speed);
                $Re_\tsn{CL}=\tilde u_\tsn{CL} D_\tsn{CSG} \nu_\tsn{CL}^{-1}$: Reynolds number based on centerline flow quantities;
                axisymmetric integral boundary-layer thicknesses defined following Reichert \cite[p. 67]{Reichert_1991a}

               }
\end{table}
is a transition duct where the cross-section changes \figref{Fig_RSMP3DDF_s_A_001} from circular at the inlet to quasi-rectangular at the exit (rectangle aspect-ratio of $3$ at the exit section).
Such geometries are typical of the transition section necessary to connect an aircraft engine exit to a rectangular nozzle \cite{Reichert_1991a}.
The precise geometrical specification of the duct's cross-section is superelliptical \cite[(A.1), p. 136]{Davis_1991a} so that
the exit section has slightly rounded corners with a "{\em variable radius fillet}" \cite[p. 2]{Davis_1991a}. The diameter of the circular inlet section is $D_{\tsn{CSG}_1}=2R_{\tsn{CSG}_1}=0.2043\;\mathrm{m}$ \cite[p. 137]{Davis_1991a},
and the length of the transition section (from inlet station 2 to exit station 5; \figrefnp{Fig_RSMP3DDF_s_A_ss_CtoRTDDG1992_001}) is $\tfrac{3}{2}D_{\tsn{CSG}_1}$. Although there is no net cross-sectional area change,
between inflow and outflow, locally \cite[Fig. 4, p. 242]{Lien_Leschziner_1996a}, the transition section is divergent \cite[p. 2]{Davis_1991a} from inlet to midpoint (cross-sectional area increase of $15\%$)
and then convergent from midpoint to exit (cross-sectional area decreases back to the inlet area).
The duct is cylindrical upstream (circular cross-section of diameter $D_{\tsn{CSG}_1}$ for several diameters upstream of station 2) and
downstream (quasi-rectangular superelliptical constant cross-section for several inlet-diameters $D_{\tsn{CSG}_1}$ downstream of station 5) of the transition section \cite[Fig. 3.1, p. 22]{Davis_1991a}.

The flow \cite{Davis_Gessner_1992a} is essentially incompressible (centerline Mach number $\bar M_\tsn{CL}\approxeq0.10$) at bulk Reynolds number $Re_\tsn{B}\approxeq390000$
($Re_\tsn{B}=\bar u_\tsn{B} D_{\tsn{CSG}_1} \nu^{-1}$, where $\bar u_\tsn{B}$ is the bulk velocity and $\nu$ is the practically constant kinematic viscosity).
Measurements, taken at 6 axial stations \figref{Fig_RSMP3DDF_s_A_ss_CtoRTDDG1992_001}, include total pressure (circular and flattened Pitot tubes and Kiel probes),
static pressures (static pressure probes and wall pressure taps), mean-velocities and Reynolds-stresses (hot wires) and skin-friction (Preston tubes resting on the duct walls).
These data are available \cite{Davis_Gessner_1992a} in digital form \cite{ERCOFTAC_1999a}.

Because of the combined streamwise evolution of both cross-sectional form and area \figrefsab{Fig_RSMP3DDF_s_A_001}
                                                                                             {Fig_RSMP3DDF_s_A_ss_CtoRTDDG1992_001},
the curvature of the duct's walls changes sign $x$-wise \cite[Fig. 3, p. 241]{Lien_Leschziner_1996a}.
The upper and lower ($z$-wise) walls are concave from the inlet (station 2) to approximately midpoint (located between stations 3 and 4; \figrefnp{Fig_RSMP3DDF_s_A_ss_CtoRTDDG1992_001})
and then convex from approximately midpoint to exit (station 5; \figrefnp{Fig_RSMP3DDF_s_A_ss_CtoRTDDG1992_001}). The opposite applies to the sidewalls ($y$-wise) which are convex in the first part (from station 2 to approximately midpoint)
then switching \cite[Fig. 3, p. 241]{Lien_Leschziner_1996a} to concave (from approximately midpoint to exit station 5). This streamwise evolution of the duct's geometry directly affects the mean pressure field,
inducing strong pressure gradients, both streamwise (area change) and crossflow (wall curvature), generating intense ($\gtrapprox10\%\;\bar u_\tsn{B}$; \cite[Fig. 7, p. 371]{Davis_Gessner_1992a})
pressure-driven secondary flows \cite[Prandtl's first kind]{Bradshaw_1987a}
which rapidly form 2 contrarotating ($y$-wise symmetry) pairs of contrarotating ($z$-wise symmetry) vortices \cite[Fig. 7, p. 371]{Davis_Gessner_1992a}, one pair near the $z=0$ midplane of each sidewall.
Downstream of station 5 (exit of the transition section; \figrefnp{Fig_RSMP3DDF_s_A_ss_CtoRTDDG1992_001}) the vortex system persists, evolving streamwise, and is clearly visible at the last measurement station 6, 2 inlet diameters ($2D_{\tsn{CSG}_1}$)
downstream of station 5 \figref{Fig_RSMP3DDF_s_A_ss_CtoRTDDG1992_001}.

The in-depth analysis of the experimental data by Davis \cite{Davis_1991a} has largely contributed to our understanding of the dynamics of the mean-flow and associated Reynolds-stresses.
Careful computations of the Davis and Gessner \cite{Davis_Gessner_1992a} \tsn{C}-to-\tsn{R} transition duct, in quite satisfactory agreement with experimental measurements, have been reported
by Sotiropoulos and Patel \cite{Sotiropoulos_Patel_1994a}, using a 7-equation \tsn{RSM}, which is a variant of the Launder-Shima \cite{Launder_Shima_1989a} \tsn{RSM} in 2 respects, (a) the use of the
Hanjali\'c-Launder \cite{Hanjalic_Launder_1976a} closure for the triple velocity correlations in lieu of the Daly-Harlow \cite{Daly_Harlow_1970a} closure adopted for turbulent diffusion in the original model \cite{Launder_Shima_1989a},
and (b) the use of the modified coefficient-functions in the $\varepsilon$-equation introduced by Shima \cite{Shima_1993a} to improve the prediction of skin-friction.
Notice the the closure used for turbulent diffusion has a strong influence on the predicted secondary flows \cite{Vallet_2007a}.
Sotiropoulos and Patel \cite{Sotiropoulos_Patel_1995b} have further exploited their computational results to analyse the streamwise ($x$-wise component of) mean-flow vorticity equation \cite[(1), p. 504]{Sotiropoulos_Patel_1995b},
and have shown that all of the vorticity-production mechanisms (vortex stretching and skewing, turbulence-induced production) are important in different regions of the flow.
\begin{figure}[h!]
\begin{center}
\begin{picture}(340,330)
\put(0,-5){\includegraphics[angle=0,width=340pt,bb=59 140 513 579]{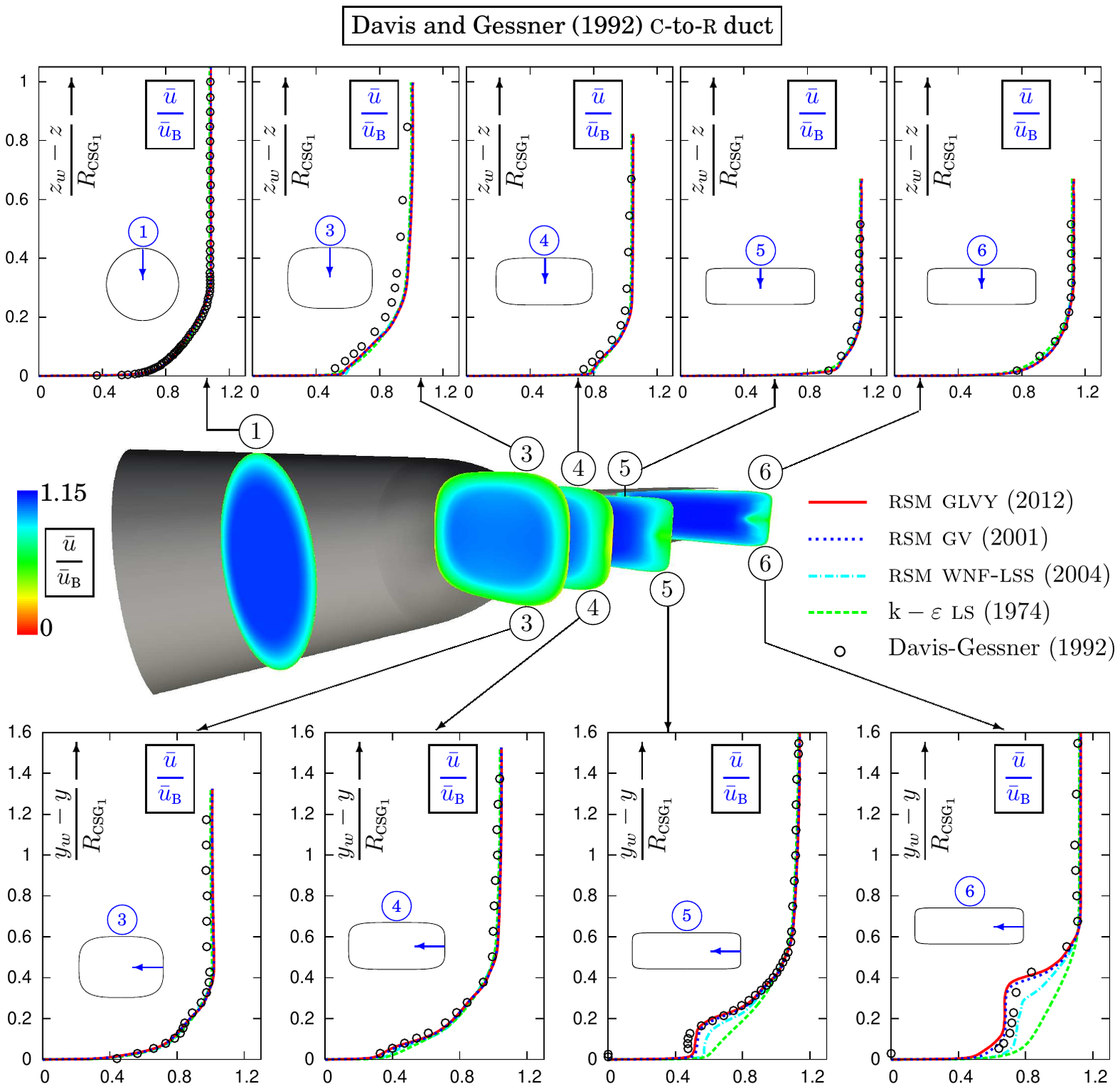}}
\end{picture}
\end{center}
\caption{Comparison of measured \cite{Davis_Gessner_1992a} streamwise ($x$-wise) velocity $\bar u$, along the $y$-wise ($z=0$ symmetry plane) and the $z$-wise ($y=0$ symmetry plane) directions,
at 5 experimental measurement stations, with 
computations ($10\times10^6$ points grid discretizing the entire duct; \tabrefnp{Tab_RSMP3DDF_s_A_001})
using \parref{RSMP3DDF_s_TCsFS_ss_TCs} the \tsn{GV} \cite{Gerolymos_Vallet_2001a}, the \tsn{WNF--LSS} \cite{Gerolymos_Sauret_Vallet_2004a}
and the \tsn{GLVY} \cite{Gerolymos_Lo_Vallet_Younis_2012a} \tsn{RSM}s, and the \tsn{LS} \cite{Launder_Sharma_1974a} linear $\mathrm{k}$--$\varepsilon$ model, for turbulent flow in a \tsn{C}-to-\tsn{R} transition duct
($Re_\tsn{B}=390000$, $\bar M_\tsn{CL}\approxeq0.09$; \tabrefnp{Tab_RSMP3DDF_s_A_002}; contour plots \tsn{GLVY RSM}).}
\label{Fig_RSMP3DDF_s_A_ss_CtoRTDDG1992_002}
\end{figure}

The computations were run on a $10\times10^6$ grid \tabref{Tab_RSMP3DDF_s_A_001} discretizing the entire duct without symmetry conditions \figref{Fig_RSMP3DDF_s_A_001}. Based on the grid-convergence studies of Sotiropoulos and Patel \cite{Sotiropoulos_Patel_1994a},
who used an $O(\Delta\ell^2)$ upwind numerical scheme for the incompressible \tsn{RSM--RANS} equations, this grid \tabref{Tab_RSMP3DDF_s_A_001} is sufficiently fine. As defined in the experimental study \cite{Davis_1991a,Davis_Gessner_1992a},
the origin of the coordinates system, $x=0$, is located at mid-distance between station 1 and station 2, located at the beginning of the transition section ($x_1=-R_{\tsn{CSG}_1}$, $x_2=+R_{\tsn{CSG}_1}$; \figrefnp{Fig_RSMP3DDF_s_A_ss_CtoRTDDG1992_001}).
The computational domain ($-D_{\tsn{CSG}_1}\leq x\leq5D_{\tsn{CSG}_1}$) starts $\tfrac{3}{2}$ inlet-diameters ($\tfrac{3}{2}D_{\tsn{CSG}_1}$) upstream of the transition section inlet (station 2; \figrefnp{Fig_RSMP3DDF_s_A_ss_CtoRTDDG1992_001})
and extends 3 inlet-diameters ($3D_{\tsn{CSG}_1}$) downstream of the transition section exit (station 5; \figrefnp{Fig_RSMP3DDF_s_A_ss_CtoRTDDG1992_001}), to avoid interaction between the uniform outflow pressure boundary-condition and computed results at the last
measurement station 6 located 2 inlet-diameters ($2D_{\tsn{CSG}_1}$) downstream of the transition section exit \figref{Fig_RSMP3DDF_s_A_ss_CtoRTDDG1992_001}.
The grid is uniform in the streamwise ($x$) direction and consists of 2 blocks (\figrefnp{Fig_RSMP3DDF_s_A_001}; \tabrefnp{Tab_RSMP3DDF_s_A_001}).
The inner block (\tsn{H}$_\tsn{$\square$}$; \tabrefnp{Tab_RSMP3DDF_s_A_001}) is an \tsn{H}-grid of $x$-wise varying square cross-section with uniform $yz$-spacing, introduced to avoid the axis-singularity of an axisymmetric-type grid.
The outer block (\tsn{O}$_\tsn{$\square$}$; \tabrefnp{Tab_RSMP3DDF_s_A_001}) is stretched geometrically near the wall with ratio $r_k$ \tabref{Tab_RSMP3DDF_s_A_001}.
For the investigated flow conditions, the first node at the walls is located at $\Delta n_w^+\lessapprox\tfrac{2}{10}$ \tabref{Tab_RSMP3DDF_s_A_001}, $n$ being the wall-normal direction.
\begin{figure}[h!]
\begin{center}
\begin{picture}(340,210)
\put(0,-5){\includegraphics[angle=0,width=340pt,bb= 56 130 515 410]{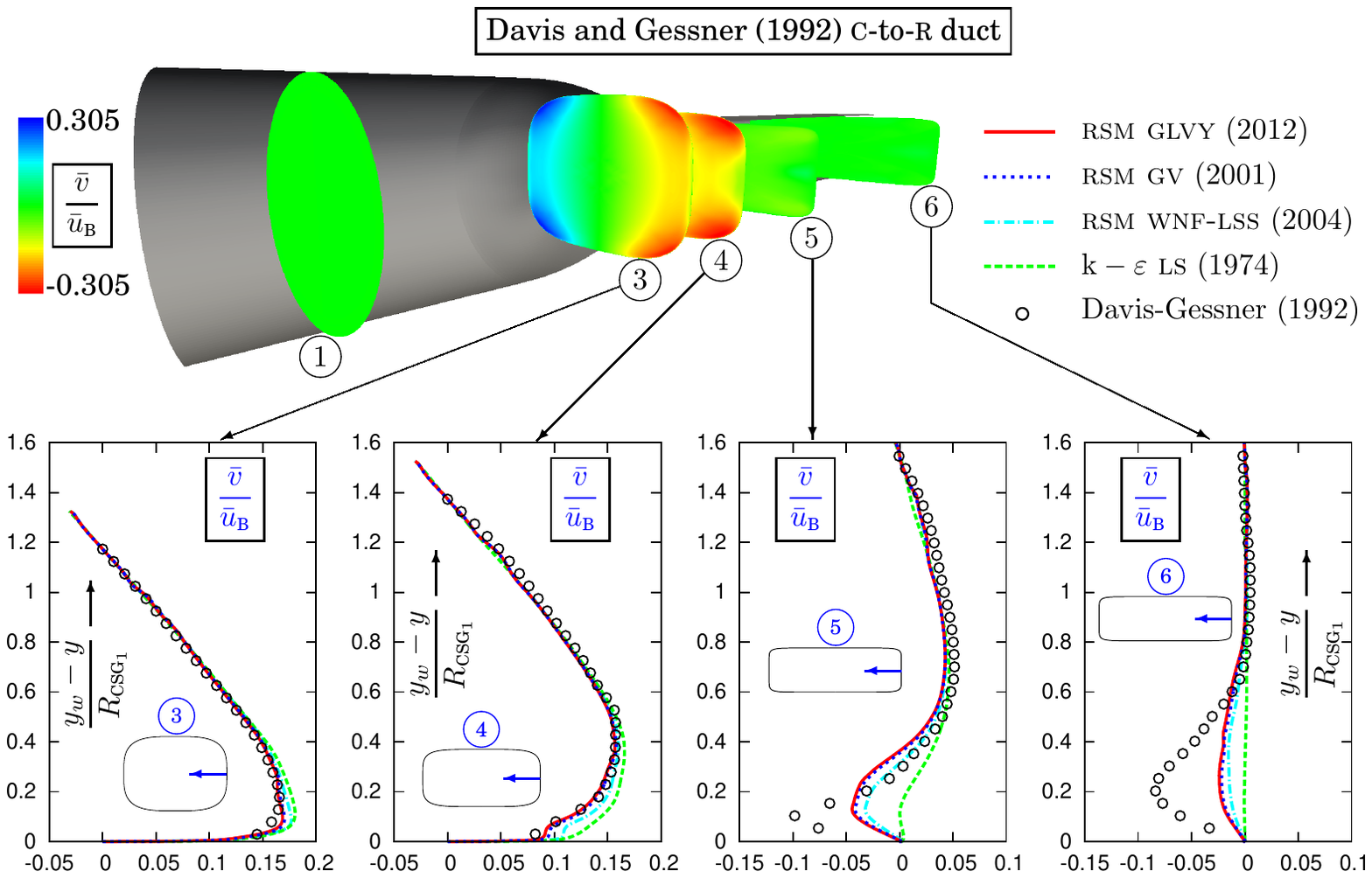}}
\end{picture}
\end{center}
\caption{Comparison of measured \cite{Davis_Gessner_1992a} wall-normal velocity $\bar v$, along the $y$-wise ($z=0$ symmetry plane) direction,
at 4 experimental measurement stations, with 
computations ($10\times10^6$ points grid discretizing the entire duct; \tabrefnp{Tab_RSMP3DDF_s_A_001})
using \parref{RSMP3DDF_s_TCsFS_ss_TCs} the \tsn{GV} \cite{Gerolymos_Vallet_2001a}, the \tsn{WNF--LSS} \cite{Gerolymos_Sauret_Vallet_2004a}
and the \tsn{GLVY} \cite{Gerolymos_Lo_Vallet_Younis_2012a} \tsn{RSM}s, and the \tsn{LS} \cite{Launder_Sharma_1974a} linear $\mathrm{k}$--$\varepsilon$ model, for turbulent flow in a \tsn{C}-to-\tsn{R} transition duct
($Re_\tsn{B}=390000$, $\bar M_\tsn{CL}\approxeq0.09$; \tabrefnp{Tab_RSMP3DDF_s_A_002}; contour plots \tsn{GLVY RSM}).}
\label{Fig_RSMP3DDF_s_A_ss_CtoRTDDG1992_003}
\end{figure}
%
\begin{figure}[h!]
\begin{center}
\begin{picture}(340,210)
\put(0,-5){\includegraphics[angle=0,width=340pt,bb= 60 293 520 564]{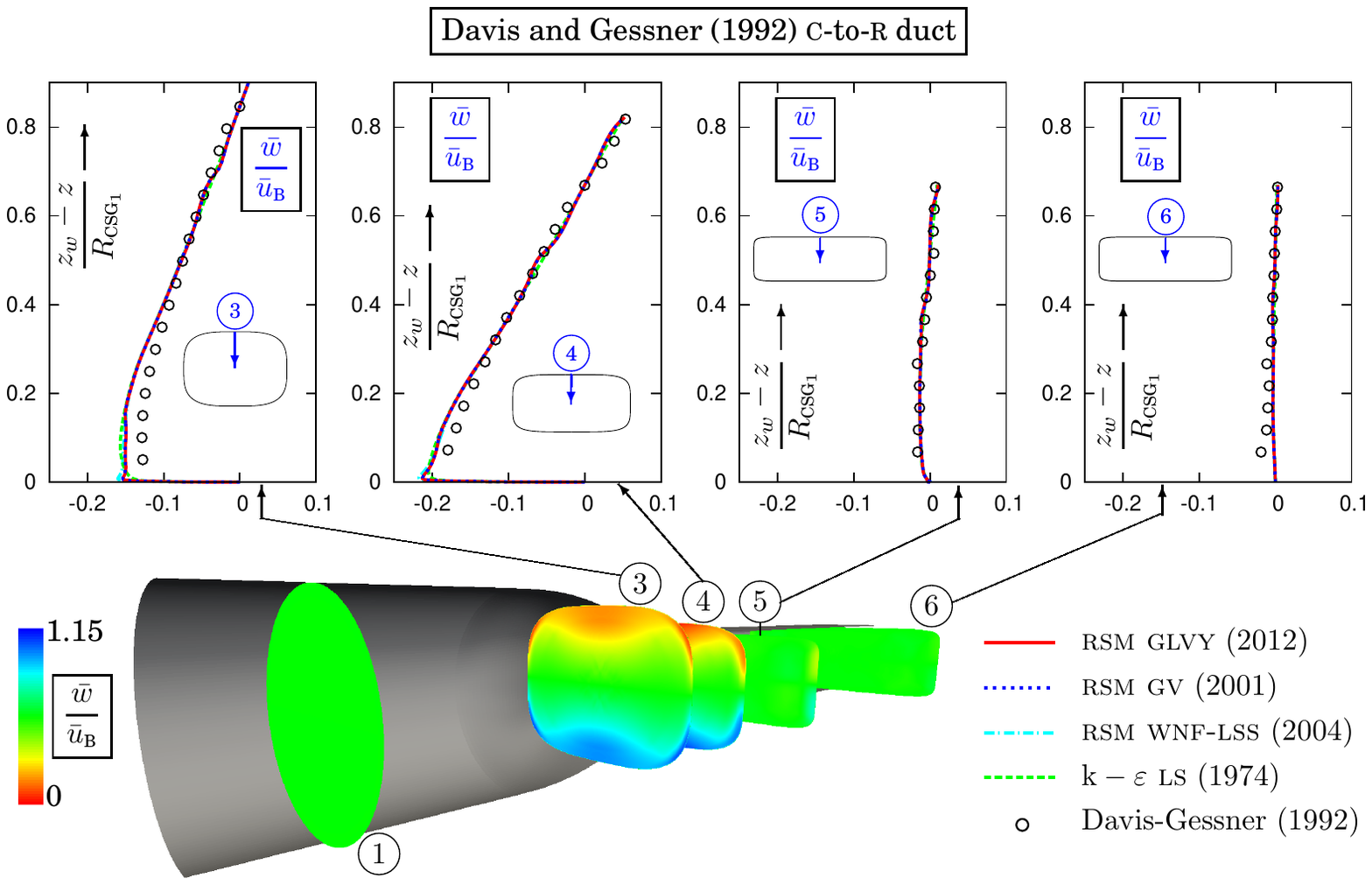}}
\end{picture}
\end{center}
\caption{Comparison of measured \cite{Davis_Gessner_1992a} wall-normal velocity $\bar w$, along the $z$-wise ($y=0$ symmetry plane) direction,
at 4 experimental measurement stations, with 
computations ($10\times10^6$ points grid discretizing the entire duct; \tabrefnp{Tab_RSMP3DDF_s_A_001})
using \parref{RSMP3DDF_s_TCsFS_ss_TCs} the \tsn{GV} \cite{Gerolymos_Vallet_2001a}, the \tsn{WNF--LSS} \cite{Gerolymos_Sauret_Vallet_2004a}
and the \tsn{GLVY} \cite{Gerolymos_Lo_Vallet_Younis_2012a} \tsn{RSM}s, and the \tsn{LS} \cite{Launder_Sharma_1974a} linear $\mathrm{k}$--$\varepsilon$ model, for turbulent flow in a \tsn{C}-to-\tsn{R} transition duct
($Re_\tsn{B}=390000$, $\bar M_\tsn{CL}\approxeq0.09$; \tabrefnp{Tab_RSMP3DDF_s_A_002}; contour plots \tsn{GLVY RSM}).}
\label{Fig_RSMP3DDF_s_A_ss_CtoRTDDG1992_004}
\end{figure}
%
\begin{figure}[h!]
\begin{center}
\begin{picture}(340,360)
\put(0,-5){\includegraphics[angle=0,width=340pt,bb=106 150 511 557]{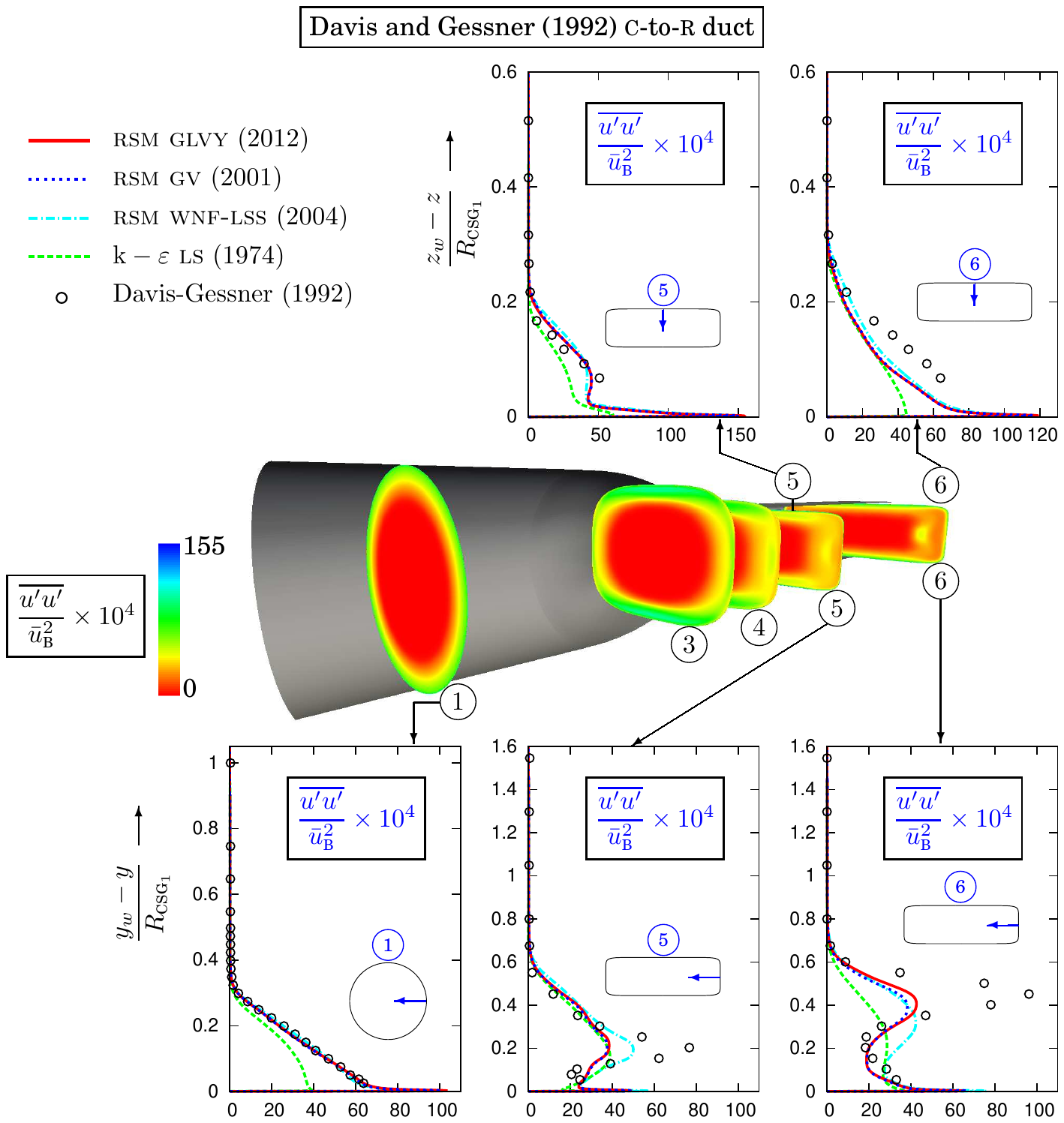}}
\end{picture}
\end{center}
\caption{Comparison of measured \cite{Davis_Gessner_1992a} streamwise ($x$-wise) velocity-variance $\overline{u'u'}$, along the $y$-wise ($z=0$ symmetry plane) and the $z$-wise ($y=0$ symmetry plane) directions,
at 3 experimental measurement stations, with 
computations ($10\times10^6$ points grid discretizing the entire duct; \tabrefnp{Tab_RSMP3DDF_s_A_001})
using \parref{RSMP3DDF_s_TCsFS_ss_TCs} the \tsn{GV} \cite{Gerolymos_Vallet_2001a}, the \tsn{WNF--LSS} \cite{Gerolymos_Sauret_Vallet_2004a}
and the \tsn{GLVY} \cite{Gerolymos_Lo_Vallet_Younis_2012a} \tsn{RSM}s, and the \tsn{LS} \cite{Launder_Sharma_1974a} linear $\mathrm{k}$--$\varepsilon$ model, for turbulent flow in a \tsn{C}-to-\tsn{R} transition duct
($Re_\tsn{B}=390000$, $\bar M_\tsn{CL}\approxeq0.09$; \tabrefnp{Tab_RSMP3DDF_s_A_002}; contour plots \tsn{GLVY RSM}).}
\label{Fig_RSMP3DDF_s_A_ss_CtoRTDDG1992_005}
\end{figure}
%
\begin{figure}[h!]
\begin{center}
\begin{picture}(340,360)
\put(0,-5){\includegraphics[angle=0,width=340pt,bb=106 150 511 557]{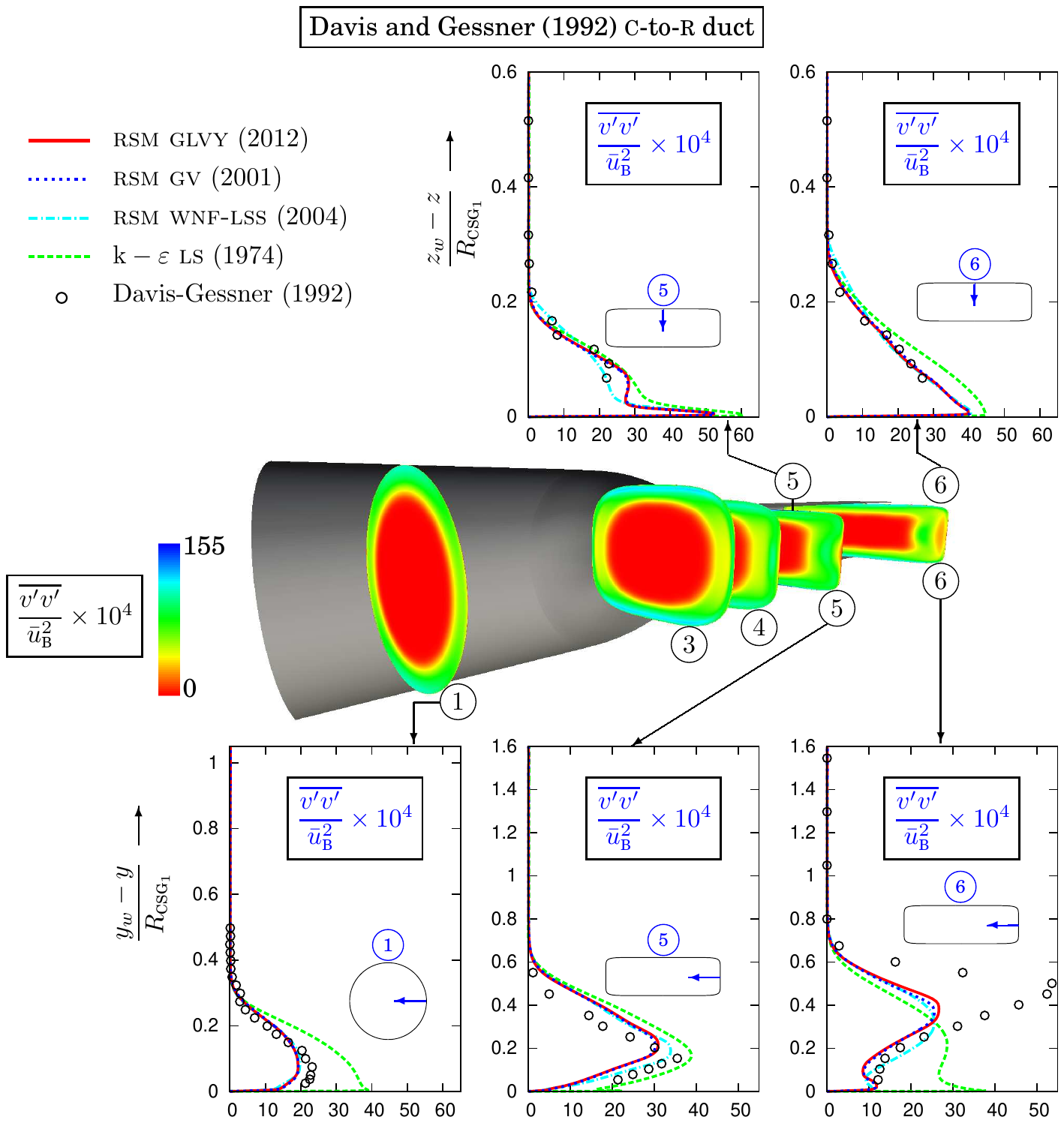}}
\end{picture}
\end{center}
\caption{Comparison of measured \cite{Davis_Gessner_1992a} $y$-wise velocity-variance $\overline{v'v'}$, wall-normal along the $y$-wise ($z=0$ symmetry plane) direction and transverse along the $z$-wise ($y=0$ symmetry plane) direction,
at 3 experimental measurement stations, with 
computations ($10\times10^6$ points grid discretizing the entire duct; \tabrefnp{Tab_RSMP3DDF_s_A_001})
using \parref{RSMP3DDF_s_TCsFS_ss_TCs} the \tsn{GV} \cite{Gerolymos_Vallet_2001a}, the \tsn{WNF--LSS} \cite{Gerolymos_Sauret_Vallet_2004a}
and the \tsn{GLVY} \cite{Gerolymos_Lo_Vallet_Younis_2012a} \tsn{RSM}s, and the \tsn{LS} \cite{Launder_Sharma_1974a} linear $\mathrm{k}$--$\varepsilon$ model, for turbulent flow in a \tsn{C}-to-\tsn{R} transition duct
($Re_\tsn{B}=390000$, $\bar M_\tsn{CL}\approxeq0.09$; \tabrefnp{Tab_RSMP3DDF_s_A_002}; contour plots \tsn{GLVY RSM}).}
\label{Fig_RSMP3DDF_s_A_ss_CtoRTDDG1992_006}
\end{figure}
%
\begin{figure}[h!]
\begin{center}
\begin{picture}(340,360)
\put(0,-5){\includegraphics[angle=0,width=340pt,bb=106 150 511 557]{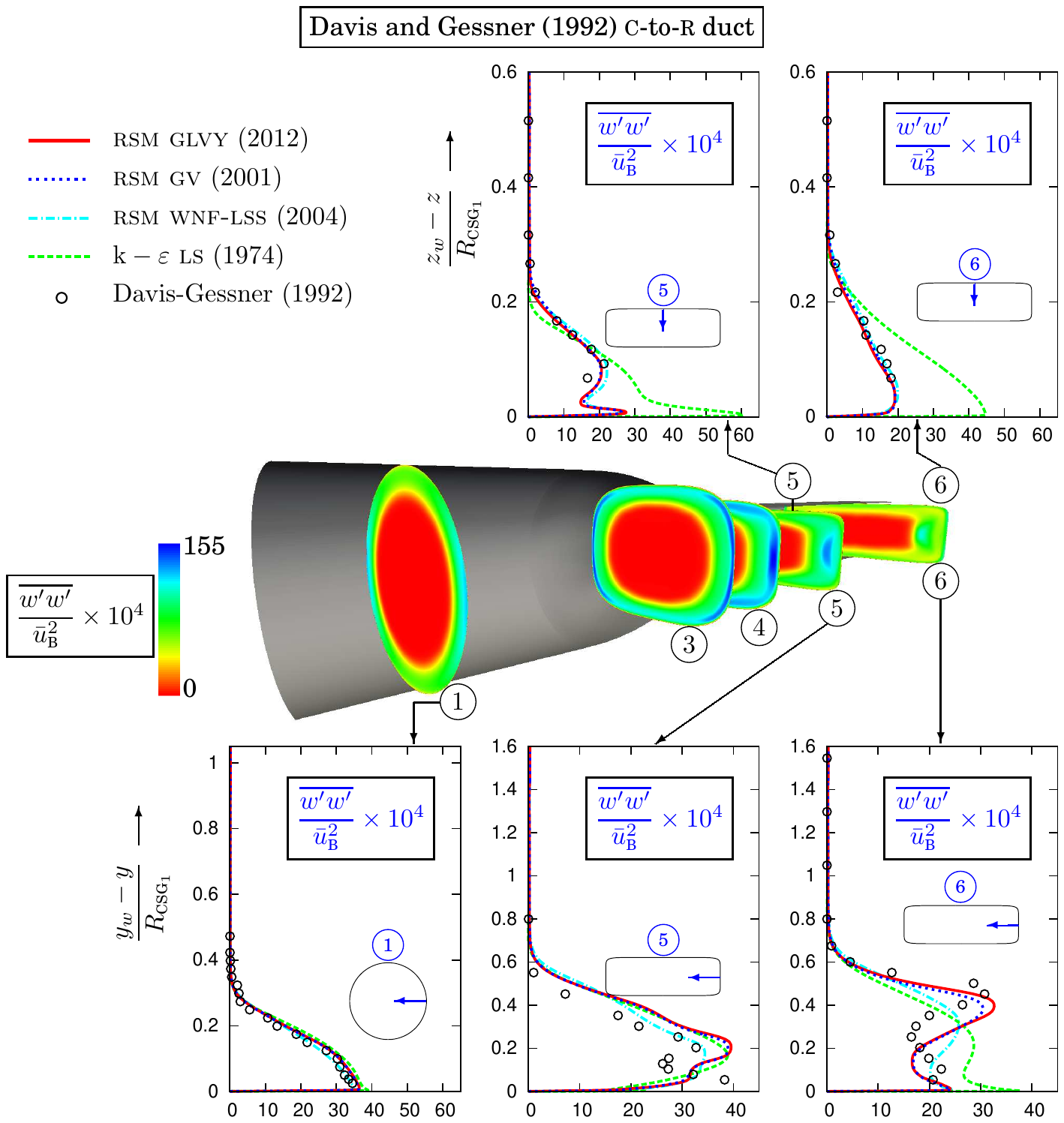}}
\end{picture}
\end{center}
\caption{Comparison of measured \cite{Davis_Gessner_1992a} $z$-wise velocity-variance $\overline{w'w'}$, transverse along the $y$-wise ($z=0$ symmetry plane) direction and wall-normal along the $z$-wise ($y=0$ symmetry plane) direction,
at 3 experimental measurement stations, with 
computations ($10\times10^6$ points grid discretizing the entire duct; \tabrefnp{Tab_RSMP3DDF_s_A_001})
using \parref{RSMP3DDF_s_TCsFS_ss_TCs} the \tsn{GV} \cite{Gerolymos_Vallet_2001a}, the \tsn{WNF--LSS} \cite{Gerolymos_Sauret_Vallet_2004a}
and the \tsn{GLVY} \cite{Gerolymos_Lo_Vallet_Younis_2012a} \tsn{RSM}s, and the \tsn{LS} \cite{Launder_Sharma_1974a} linear $\mathrm{k}$--$\varepsilon$ model, for turbulent flow in a \tsn{C}-to-\tsn{R} transition duct
($Re_\tsn{B}=390000$, $\bar M_\tsn{CL}\approxeq0.09$; \tabrefnp{Tab_RSMP3DDF_s_A_002}; contour plots \tsn{GLVY RSM}).}
\label{Fig_RSMP3DDF_s_A_ss_CtoRTDDG1992_007}
\end{figure}
%
\begin{figure}[h!]
\begin{center}
\begin{picture}(340,440)
\put(0,-5){\includegraphics[angle=0,width=340pt,bb=102 127 513 669]{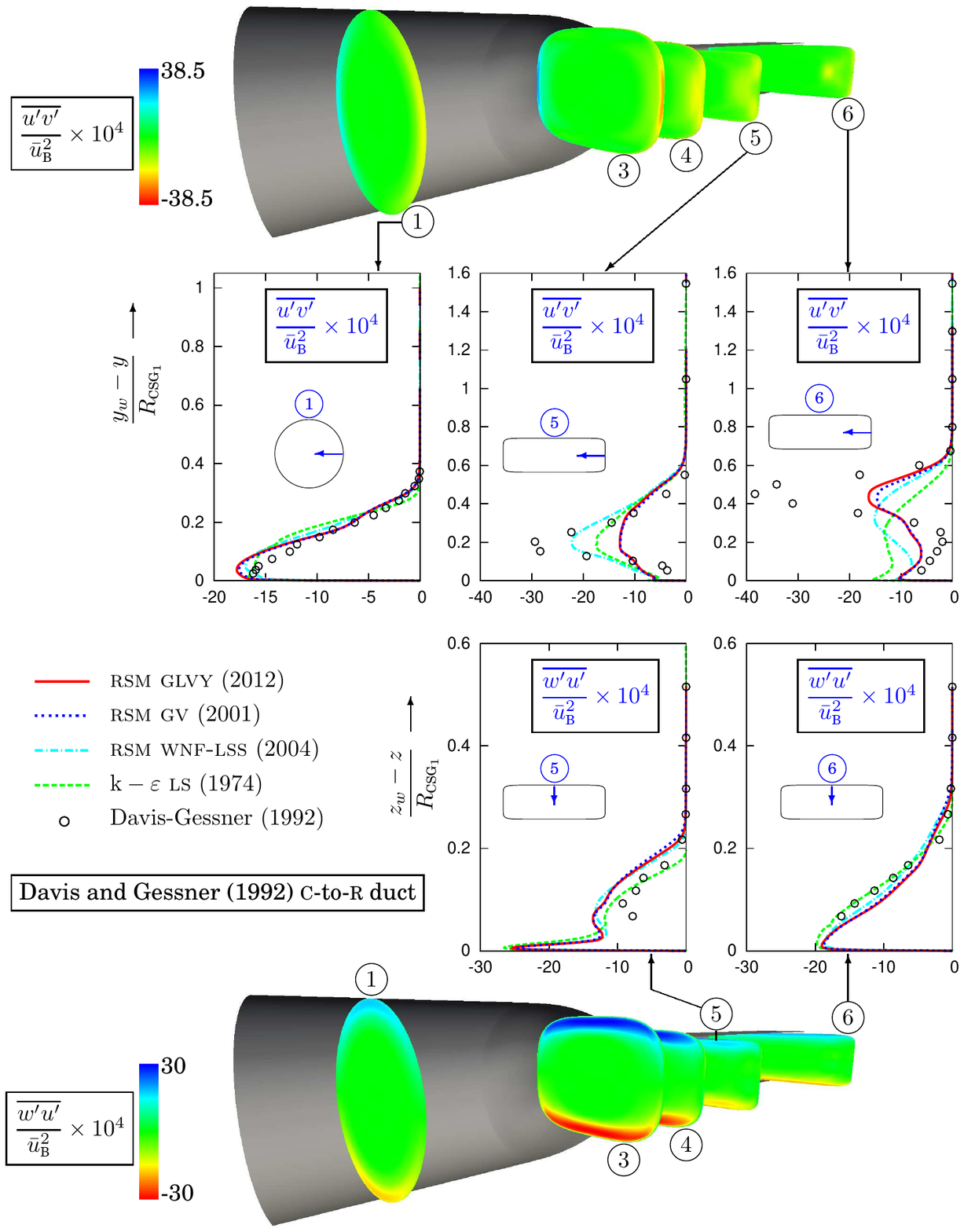}}
\end{picture}
\end{center}
\caption{Comparison of measured \cite{Davis_Gessner_1992a} shear Reynolds-stress, $\overline{u'v'}$ along the $y$-wise ($z=0$ symmetry plane) and $\overline{u'w'}$ the $z$-wise ($y=0$ symmetry plane) directions,
at 3 experimental measurement stations, with 
computations ($10\times10^6$ points grid discretizing the entire duct; \tabrefnp{Tab_RSMP3DDF_s_A_001})
using \parref{RSMP3DDF_s_TCsFS_ss_TCs} the \tsn{GV} \cite{Gerolymos_Vallet_2001a}, the \tsn{WNF--LSS} \cite{Gerolymos_Sauret_Vallet_2004a}
and the \tsn{GLVY} \cite{Gerolymos_Lo_Vallet_Younis_2012a} \tsn{RSM}s, and the \tsn{LS} \cite{Launder_Sharma_1974a} linear $\mathrm{k}$--$\varepsilon$ model, for turbulent flow in a \tsn{C}-to-\tsn{R} transition duct
($Re_\tsn{B}=390000$, $\bar M_\tsn{CL}\approxeq0.09$; \tabrefnp{Tab_RSMP3DDF_s_A_002}; contour plots \tsn{GLVY RSM}).}
\label{Fig_RSMP3DDF_s_A_ss_CtoRTDDG1992_008}
\end{figure}

At inflow \tabref{Tab_RSMP3DDF_s_A_002}, measured \cite{ERCOFTAC_1999a,Davis_1991a,Davis_Gessner_1992a}
total conditions ($p_{t_{\tsn{CL}_\mathrm{i}}}=101325\;\mathrm{Pa}$, $T_{t_{\tsn{CL}_\mathrm{i}}}=298.3\;\mathrm{K}$), with a turbulent intensity $T_{u_{\tsn{CL}_\mathrm{i}}}=0.3\%$, were applied at the centerline.   
In the absence of experimental data, a turbulent lengthscale $\ell_{\tsn{T}_{\tsn{CL}_\mathrm{i}}}=50\;\mathrm{mm}$ was assumed at the centerline, 
with reference to the duct radius ($R_{\tsn{CSG}_1}=0.10215\;\mathrm{m}$) and the measured boundary-layer thickness $\delta_{995}=30.85\;\mathrm{mm}$ at station 1 (\figrefnp{Fig_RSMP3DDF_s_A_ss_CtoRTDDG1992_001}; \tabrefnp{Tab_RSMP3DDF_s_A_003}).
Detailed measurements of the boundary-layer profiles of mean-flow and Reynolds-stresses are available \cite{ERCOFTAC_1999a,Davis_1991a,Davis_Gessner_1992a},
and were interpolated onto the computational grid to define the inflow conditions. These data were extended to the wall, in the region where experimental data were not available, using semi-analytical profiles \cite{Gerolymos_Sauret_Vallet_2004c},
and used to define, by assuming local equilibrium in the boundary-layer and matching to the prescribed centerline $\ell_{\tsn{T}_{\tsn{CL}_\mathrm{i}}}$ \cite{Gerolymos_Sauret_Vallet_2004c}, the $\varepsilon$ profiles.
The outflow pressure was adjusted to obtain the correct $Re_\tsn{B}=390000$ ($p_\mathrm{o}=100627\;\mathrm{Pa}$) corresponding to an inlet Mach number at centerline $M_{\tsn{CL}_\mathrm{i}}\approxeq0.0940$ \tabref{Tab_RSMP3DDF_s_A_002}.

Computational results for the integral axisymmetric \cite{Fujii_Okiishi_1972a} boundary-layer thicknesses and associated shape-factors at the first measurement station 1
\figref{Fig_RSMP3DDF_s_A_ss_CtoRTDDG1992_001}, where the flow is still practically axisymmetric, are in good agreement \tabref{Tab_RSMP3DDF_s_A_003} with those determined from the experimental data
\cite[Tab. 1, p. 370]{Davis_Gessner_1992a}. Following Davis and Gessner \cite{Davis_Gessner_1992a} the approximate (linearized; $\delta\ll R_\tsn{CSG}$) definitions of the axisymmetric integral boundary-layer thicknesses \cite[(3.5--3.7), p. 20]{Davis_1991a},
as defined by Fujii and Okiishi \cite{Fujii_Okiishi_1972a}, were applied.

Predicted wall-pressures are quite similar for all 4 turbulence models and are in quite satisfactory agreement with available measurements \figref{Fig_RSMP3DDF_s_A_ss_CtoRTDDG1992_001}. Skin-friction was measured by Preston tubes aligned with the $x$-wise
direction and "{\em presumes that the 2-D form of the law-of-the-wall is valid and that streamwise pressure-gradients are small}" \cite[p. 19]{Davis_1991a}.
Computed skin-friction was determined by the wall-normal gradient of streamwise velocity $\bar{u}$, at each measurement plane. At stations 5 (exit of the transition section) and 6 (2 inlet diameters further downstream), the \tsn{GLVY} and
\tsn{GV} \tsn{RSM}s predict quite well the evolution of skin-friction along the peripheral wall-coordinate $\mathrm{s}$ \figref{Fig_RSMP3DDF_s_A_ss_CtoRTDDG1992_001}, yielding the correct $\mathrm{s}$-gradient of $c_{f_\tsn{B}}$ everywhere.
The small differences in absolute level at station 5 \figref{Fig_RSMP3DDF_s_A_ss_CtoRTDDG1992_001}, where the streamwise pressure-gradient is not negligible, are of the same order as the differences between measurements with Preston tubes of various diameters
\cite[Fig. 15, p. 374]{Davis_Gessner_1992a}, and can also be attributed to the error introduced by the log-law assumption in the measurements \cite{Sotiropoulos_Patel_1995a}.
On the other hand, the linear \tsn{LS} $\mathrm{k}$--$\varepsilon$ model is unsatisfactory predicting a peculiar inverted $\mathrm{s}$-curvature of $c_{f_\tsn{B}}$
around $\mathrm{s}\approxeq 0.6\mathrm{s}_{\frac{1}{4}}$ at both stations \figref{Fig_RSMP3DDF_s_A_ss_CtoRTDDG1992_001}
and a nearly constant level of $c_{f_\tsn{B}}$ on the sidewall ($0.9\mathrm{s}_{\frac{1}{4}}\lessapprox s \lessapprox\mathrm{s}_{\frac{1}{4}}$; \figrefnp{Fig_RSMP3DDF_s_A_ss_CtoRTDDG1992_001}).
The negative $\mathrm{s}$-gradient of $c_{f_\tsn{B}}$ on the sidewall ($\mathrm{s}=\mathrm{s}_{\frac{1}{4}}$ correspond to the middle of the sidewall at $z=0$; \figrefnp{Fig_RSMP3DDF_s_A_ss_CtoRTDDG1992_001})
is an important feature of the flow, as it is directly related
\cite[pp. 50--51]{Davis_1991a} to the presence of the secondary flow streamwise vortices \cite[Fig. 7, p. 371]{Davis_Gessner_1992a}. The overall prediction of $c_{f_\tsn{B}}$ by the \tsn{WNF--LSS RSM} is satisfactory, except for the  too weak negative
$\mathrm{s}$-gradient of $c_{f_\tsn{B}}$ for $\mathrm{s}\gtrapprox 0.9 \mathrm{s}_{\frac{1}{4}}$ \figref{Fig_RSMP3DDF_s_A_ss_CtoRTDDG1992_001} which is indicative of an underestimation of the strength of the streamwise vortices.

All 4 turbulence models predict quite accurately the streamwise mean-flow velocity $\bar {u}$ along the $z$-traverse on the $y$-symmetry plane at all measurement stations \figref{Fig_RSMP3DDF_s_A_ss_CtoRTDDG1992_002}. Along the $y$-traverse on the
$z$-symmetry plane \figref{Fig_RSMP3DDF_s_A_ss_CtoRTDDG1992_002}, differences between turbulence models start appearing at station 4, where the linear \tsn{LS} $\mathrm{k}$--$\varepsilon$ model does not reproduce the experimentally observed inflection 
of the velocity profile at $y_w-y\approxeq 0.1 R_{\tsn{CSG}_1}$ \figref{Fig_RSMP3DDF_s_A_ss_CtoRTDDG1992_002}.
Further downstream, at stations 5 and 6, the linear \tsn{LS} $\mathrm{k}$--$\varepsilon$ model fails to predict the experimentally observed double inflection of the velocity profile along the  $y$-traverse \figref{Fig_RSMP3DDF_s_A_ss_CtoRTDDG1992_002},
returning instead a more filled 2-D-boundary-layer-like profile. Davis \cite[pp. 50--51]{Davis_1991a} has identified this feature of the $\bar {u}$ velocity profile as the result of a "{\em transfer of low-momentum fluid from the boundary-layer toward the
centerline creating a flat spot in the velocity field}", which "{\em is seen to be much larger at station 6 than at station 5}" (contour plots of $\bar {u}$; \figrefnp{Fig_RSMP3DDF_s_A_ss_CtoRTDDG1992_002}).
This transfer, along the sidewall, is directly related to the presence of the secondary flow vortex-pair near the $z=0$ symmetry plane \cite[Fig. 7, p. 371]{Davis_Gessner_1992a}. The 3 \tsn{RSM}s successfully predict the double inflection
of the $\bar {u}$ profile along the $y$-traverses at planes 5 and 6 \figref{Fig_RSMP3DDF_s_A_ss_CtoRTDDG1992_002}. The \tsn{GLVY} and \tsn{GV} \tsn{RSM}s agree quite well with measurements along the $y$-traverses at planes 5 and 6,
indeed everywhere \figref{Fig_RSMP3DDF_s_A_ss_CtoRTDDG1992_002}. Although the \tsn{WNF--LSS RSM} predicts the double inflection shape of the $\bar {u}$ profile along the $y$-traverses at stations 5 and 6, it overpredicts $\bar {u}$, implying
a slight underprediction of secondary flows.

Differences between turbulence closures in predicting the wall-normal velocity $\bar{v}$ along the $y$-traverses at the $z=0$ symmetry plane (where $\bar{w}=0$ by symmetry) appear already at station 3 \figref{Fig_RSMP3DDF_s_A_ss_CtoRTDDG1992_003}.
The \tsn{GLVY} and \tsn{GV} \tsn{RSM}s predict $\bar{v}$ quite accurately at stations 3 and 4, where the linear \tsn{LS} $\mathrm{k}$--$\varepsilon$ model and to a lesser extent the \tsn{WNF--LSS RSM}, slightly overestimate it near the sidewall
($y_w-y\lessapprox 0.4 R_{\tsn{CSG}_1}$; \figrefnp{Fig_RSMP3DDF_s_A_ss_CtoRTDDG1992_003}).
At station 5, the 3 \tsn{RSM}s perform quite well in the outer part of the boundary-layer ($y_w-y\gtrapprox 0.2 R_{\tsn{CSG}_1}$; \figrefnp{Fig_RSMP3DDF_s_A_ss_CtoRTDDG1992_003}) but overestimate $\bar{v}$ near the sidewall
($y_w-y\lessapprox 0.2 R_{\tsn{CSG}_1}$; \figrefnp{Fig_RSMP3DDF_s_A_ss_CtoRTDDG1992_003}) by $\sim\!\!50$\% at the peak. They are nonetheless in much better agreement with experimental data than the linear \tsn{LS} $\mathrm{k}$--$\varepsilon$ model which predicts
levels that are 5-fold lower ($y\lessapprox 0.4 R_{\tsn{CSG}_1}$; station 5; \figrefnp{Fig_RSMP3DDF_s_A_ss_CtoRTDDG1992_003}). At station 6, 2 inlet diameters further downstream, the $\bar{v}$ velocity along the $y$-traverse is severely underestimated by 
the 3 \tsn{RSM}s \figref{Fig_RSMP3DDF_s_A_ss_CtoRTDDG1992_003} which underpredict the strength of the secondary flows at this station. Nonetheless, the 3 \tsn{RSM}s largely outperform the linear \tsn{LS} $\mathrm{k}$--$\varepsilon$,
which completely fails, returning negligible small levels of $\bar v$ at station 6 \figref{Fig_RSMP3DDF_s_A_ss_CtoRTDDG1992_003}.
The pair of contrarotating vortices observed at stations 5 and 6 near the intersection between the $z$-symmetry plane and the sidewall \cite[Fig. 7, p. 371]{Davis_Gessner_1992a} induces velocities away from the sidewall ($\bar{v}<0$ on the near-sidewall
along the $y$-traverse; \figrefnp{Fig_RSMP3DDF_s_A_ss_CtoRTDDG1992_003}), whose measured peak value remains approximately constant ($\sim\!\!-0.1$) between stations 5 and 6 \figref{Fig_RSMP3DDF_s_A_ss_CtoRTDDG1992_003}.
The failure of the \tsn{RSM} computations to correctly predict the relaxation of the flow in the straight constant cross-section duct between stations 5 and 6, possibly reveals an inadequacy of the models. Nonetheless, grid-resolution on the cross-section 
at these stations is rather poor \figref{Fig_RSMP3DDF_s_A_001}, containing only a few cells across the vortices \cite[Fig. 7, p. 371]{Davis_Gessner_1992a}.
For this reason, computations using finer ($j$-wise and $k$-wise; \tabrefnp{Tab_RSMP3DDF_s_A_001}) grids are required to determine computational grid-convergence of the flow in the contrarotating vortex pair region, and this will be the subject of future work.
The wall-normal velocity $\bar{w}$ along the $z$-traverses at the $y=0$ symmetry plane (where $\bar{v}=0$ by symmetry) is very well predicted at all stations by all 4 turbulence closures \figref{Fig_RSMP3DDF_s_A_ss_CtoRTDDG1992_004}.

All 3 \tsn{RSM}s predict quite accurately the streamwise Reynolds-stress $\overline{u'u'}$ along the $z$-traverse on the $y=0$ symmetry plane at station 5 \figref{Fig_RSMP3DDF_s_A_ss_CtoRTDDG1992_005}, and also, despite a slight underestimation, at station 6
\figref{Fig_RSMP3DDF_s_A_ss_CtoRTDDG1992_005} further downstream.  
Along the $y$-traverse on the $z=0$ symmetry plane, except for station 1 \figref{Fig_RSMP3DDF_s_A_ss_CtoRTDDG1992_005} near the computational inflow where the measured Reynolds-stresses were interpolated onto the grid and applied as boundary conditions,
the 3 \tsn{RSM}s predict correctly the profile shape but underestimate by $\sim\!\!50$\% the peak value at stations 5 and 6 \figref{Fig_RSMP3DDF_s_A_ss_CtoRTDDG1992_005}. All 3 \tsn{RSM}s predict quite accurately the in-plane diagonal Reynolds-stresses
$\overline{v'v'}$ \figref{Fig_RSMP3DDF_s_A_ss_CtoRTDDG1992_006} and $\overline{w'w'}$ \figref{Fig_RSMP3DDF_s_A_ss_CtoRTDDG1992_007}, with the exception of $\overline{v'v'}$ at station 6 along the $y$-traverse on the $z$-symmetry 
plane \figref{Fig_RSMP3DDF_s_A_ss_CtoRTDDG1992_007} where the peak value is underestimated by $\sim\!\!50$\%.
The predictions of the diagonal Reynolds-stresses ($\overline{u'u'}$, $\overline{v'v'}$, $\overline{w'w'}$) by the \tsn{GLVY} and \tsn{GV} \tsn{RSM}s are in very close agreement 
\figrefsatob{Fig_RSMP3DDF_s_A_ss_CtoRTDDG1992_005}
            {Fig_RSMP3DDF_s_A_ss_CtoRTDDG1992_007},
and also with those predicted by \tsn{WNF--LSS RSM} \figrefsatob{Fig_RSMP3DDF_s_A_ss_CtoRTDDG1992_005}
                                                                {Fig_RSMP3DDF_s_A_ss_CtoRTDDG1992_007}
except at station 6 along the $y$-traverse
on the $z$-symmetry plane where the \tsn{GLVY} and \tsn{GV} \tsn{RSM}s are in closer agreement with measurements.
Expectedly, the linear $\mathrm{k}$--$\varepsilon$ model completely fails in predicting the Reynolds-stress tensor anisotropy, yielding unsatisfactory results for the diagonal Reynolds-stresses 
\figrefsatob{Fig_RSMP3DDF_s_A_ss_CtoRTDDG1992_005}
            {Fig_RSMP3DDF_s_A_ss_CtoRTDDG1992_007},
because of the pathological shortcomings of the Boussinesq hypothesis \cite[pp. 273--278]{Wilcox_1998a}.

The prediction of the shear Reynolds-stress $\overline{u'v'}$ along the $z$-traverse on the $y$-symmetry plane (where  $\overline{u'v'}=0$ by symmetry) at stations 5 and 6 by the 3 \tsn{RSM}s is quite satisfactory
\figref{Fig_RSMP3DDF_s_A_ss_CtoRTDDG1992_008}. On the contrary, the \tsn{LS} \cite{Launder_Sharma_1974a} linear $\mathrm{k}$--$\varepsilon$ model does not reproduce as well the shape of the $\overline{u'v'}$ profile at station 5
\figref{Fig_RSMP3DDF_s_A_ss_CtoRTDDG1992_008}, a deficiency which does not appear to have a substantial influence on the prediction of the streamwise mean-velocity profile $\bar{u}$ ($z$-traverse, station 5, \figrefnp{Fig_RSMP3DDF_s_A_ss_CtoRTDDG1992_002}).
The prediction of the shear Reynolds-stress $\overline{u'v'}$ \figref{Fig_RSMP3DDF_s_A_ss_CtoRTDDG1992_008} along the $y$-traverse on the $z$-symmetry plane (where  $\overline{u'w'}=0$ by symmetry) should be analyzed in relation to the prediction of the streamwise
mean-velocity $\bar{u}$ \figref{Fig_RSMP3DDF_s_A_ss_CtoRTDDG1992_002}. At station 6, along the $y$-traverse on the $z$-symmetry plane, all turbulence models underestimate by $\sim\!\!50$\% the outer peak of $\overline{u'v'}$
at $y_w-y\approxeq 0.45 R_{\tsn{CSG}_1}$ \figref{Fig_RSMP3DDF_s_A_ss_CtoRTDDG1992_008}. The grid-resolution issues mentioned above not withstanding, notice that the \tsn{GLVY} and \tsn{GV} \tsn{RSM}s predict quite well $\overline{u'v'}$
at station 6 for $0\lessapprox y_w-y\lessapprox 0.3 R_{\tsn{CSG}_1}$ \figref{Fig_RSMP3DDF_s_A_ss_CtoRTDDG1992_008}, and this is obviously related to the satisfactory prediction of $\bar{u}$ by these models 
($y$-traverse, station 6;  \figrefnp{Fig_RSMP3DDF_s_A_ss_CtoRTDDG1992_002}). On the contrary, the linear \tsn{LS} $\mathrm{k}$--$\varepsilon$ model which strongly overpredicts $\overline{u'v'}$ in this range 
($y$-traverse, $0\lessapprox y_w-y\lessapprox 0.3 R_{\tsn{CSG}_1}$, station 5; \figrefnp{Fig_RSMP3DDF_s_A_ss_CtoRTDDG1992_008}) fails to correctly predict the streamwise mean-velocity $\bar{u}$ at this location \figref{Fig_RSMP3DDF_s_A_ss_CtoRTDDG1992_002}.
Notice that the \tsn{WNF--LSS RSM} which performs much better than the \tsn{LS} $\mathrm{k}$--$\varepsilon$ model in predicting the shear Reynolds-stress $\overline{u'v'}$ 
($y$-traverse, $0\lessapprox y_w-y\lessapprox 0.3 R_{\tsn{CSG}_1}$, station 6; \figrefnp{Fig_RSMP3DDF_s_A_ss_CtoRTDDG1992_008}) also predicts the correct double inflection shape of the $\bar{u}$-profile 
($y$-traverse, station 6;  \figrefnp{Fig_RSMP3DDF_s_A_ss_CtoRTDDG1992_002}), albeit less accurately than the \tsn{GLVY} and \tsn{GV} \tsn{RSM}s.

Despite the grid-convergence issues raised above (which can only be resolved by additional calculations on much finer grids), the systematic comparison of the computations of the Davis and Gessner \cite{Davis_Gessner_1992a} \tsn{C}-to-\tsn{R} duct configuration
with the experimental data \figrefsatob{Fig_RSMP3DDF_s_A_ss_CtoRTDDG1992_001}
                                       {Fig_RSMP3DDF_s_A_ss_CtoRTDDG1992_008}
yields useful conclusions. The linear \tsn{LS} $\mathrm{k}$--$\varepsilon$ model, handicapped by Boussinesq's hypothesis
\cite[pp. 273--279]{Wilcox_1998a} fails to predict with sufficient accuracy the regions of the flowfield that are dominated by secondary flows \figrefsatob{Fig_RSMP3DDF_s_A_ss_CtoRTDDG1992_001}
                                                                                                                                                            {Fig_RSMP3DDF_s_A_ss_CtoRTDDG1992_008}. 
The 3 \tsn{RSM}s perform much better, capturing several complex features of the flow \figrefsatob{Fig_RSMP3DDF_s_A_ss_CtoRTDDG1992_001}{Fig_RSMP3DDF_s_A_ss_CtoRTDDG1992_008}, although they are not  sufficiently accurate on the
$10\times10^6$ points grid used \tabref{Tab_RSMP3DDF_s_A_001} in predicting all the details of the flow near the intersection of the sidewall with the $z=0$ symmetry plane \figrefsatob{Fig_RSMP3DDF_s_A_ss_CtoRTDDG1992_001}
                                                                                                                                                                                        {Fig_RSMP3DDF_s_A_ss_CtoRTDDG1992_008}.
As for the Gessner and Emery \cite{Gessner_Emery_1981a} square duct case \parref{RSMP3DDF_s_A_ss_DTFSDGE1981}, the \tsn{GLVY} and \tsn{GV} \tsn{RSM}s (which yield very similar results) perform sometimes better than the \tsn{WNF--LSS RSM},
especially near the sidewall in the region of strong secondary flows.

%
%
%
%
%
\subsection{Diffusing 3-D S-Duct \cite{Wellborn_Reichert_Okiishi_1994a}}\label{RSMP3DDF_s_A_ss_D3DSDWRO1994}
%
%
%
%
%

\begin{figure}[h!]
\begin{center}
\begin{picture}(340,220)
\put(0,-5){\includegraphics[angle=0,width=340pt,bb= 12 323 590 705]{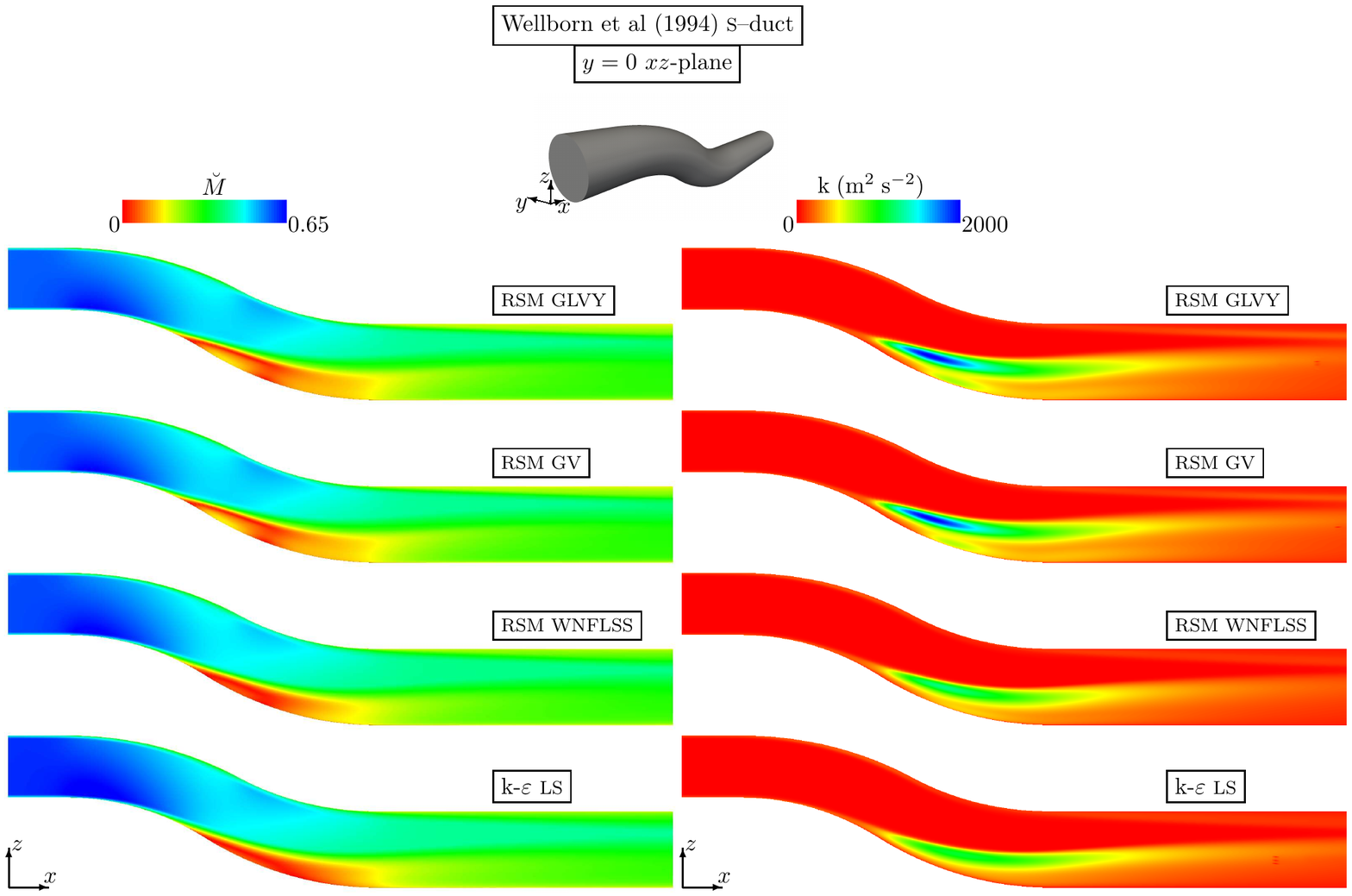}}
\end{picture}
\end{center}
\caption{Level plots of Mach number $\breve M$ and of turbulent kinetic energy $\mathrm{k}$ on the $y=0$ symmetry plane of the Wellborn \etal\ \cite{Wellborn_Reichert_Okiishi_1994a} diffusing \tsn{S}-duct
($Re_{\tsn{CL}_\tsn{A}}=2.6\times10^6$, $\bar M_{\tsn{CL}_\tsn{A}}\approxeq0.6$; \tabrefnp{Tab_RSMP3DDF_s_A_002}) obtained from computations
($2\times10^6$ points grid discretizing the entire duct; \tabrefnp{Tab_RSMP3DDF_s_A_001})
using \parref{RSMP3DDF_s_TCsFS_ss_TCs} the \tsn{GV} \cite{Gerolymos_Vallet_2001a}, the \tsn{WNF--LSS} \cite{Gerolymos_Sauret_Vallet_2004a}
and the \tsn{GLVY} \cite{Gerolymos_Lo_Vallet_Younis_2012a} \tsn{RSM}s, and the \tsn{LS} \cite{Launder_Sharma_1974a} linear $\mathrm{k}$--$\varepsilon$ model.}
\label{Fig_RSMP3DDF_s_A_ss_D3DSDWRO1994_001}
\end{figure}

\begin{figure}[h!]
\begin{center}
\begin{picture}(340,350)
\put(0,-5){\includegraphics[angle=0,width=340pt,bb= 86 105 571 598]{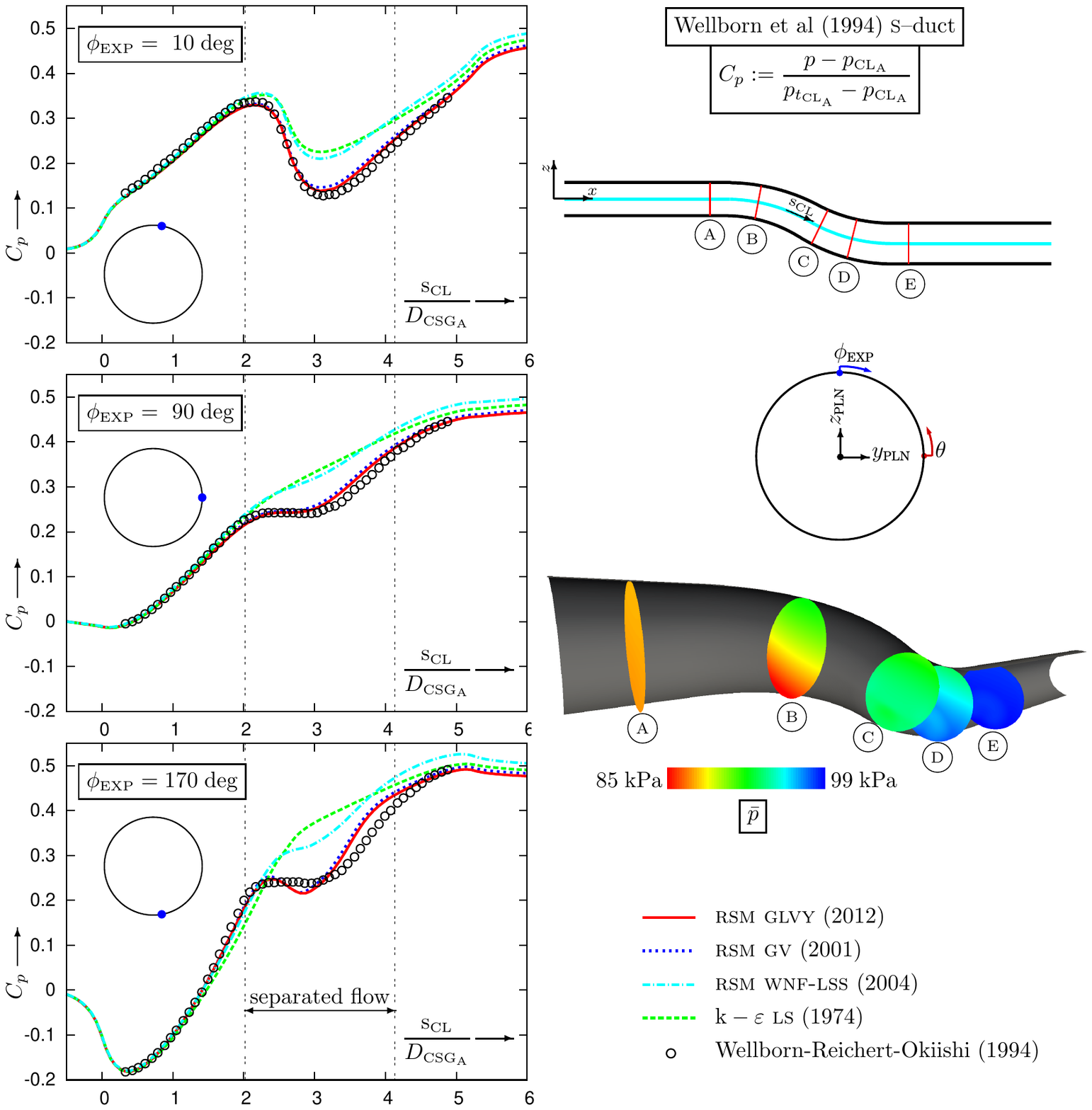}}
\end{picture}
\end{center}
\caption{Comparison of measured \cite{Wellborn_Reichert_Okiishi_1994a} wall-pressure coefficient $C_p$ (based on centerline quantities at plane \tsn{A}),
plotted against the curvilinear coordinate $\mathrm{s}_\tsn{CL}$ along the duct centerline (planes $\perp$ to the centerline define stations of circular cross-section),
at 3 azimuthal locations, with computations ($2\times10^6$ points grid discretizing the entire duct; \tabrefnp{Tab_RSMP3DDF_s_A_001})
using \parref{RSMP3DDF_s_TCsFS_ss_TCs} the \tsn{GV} \cite{Gerolymos_Vallet_2001a}, the \tsn{WNF--LSS} \cite{Gerolymos_Sauret_Vallet_2004a}
and the \tsn{GLVY} \cite{Gerolymos_Lo_Vallet_Younis_2012a} \tsn{RSM}s, and the \tsn{LS} \cite{Launder_Sharma_1974a} linear $\mathrm{k}$--$\varepsilon$ model, for turbulent flow in a diffusing \tsn{S}-duct
($Re_{\tsn{CL}_\tsn{A}}=2.6\times10^6$, $\bar M_{\tsn{CL}_\tsn{A}}\approxeq0.6$; \tabrefnp{Tab_RSMP3DDF_s_A_002}; contour plots \tsn{GLVY RSM}).}
\label{Fig_RSMP3DDF_s_A_ss_D3DSDWRO1994_002}
\end{figure}

The previously studied square duct \parref{RSMP3DDF_s_A_ss_DTFSDGE1981} and \tsn{C}-to-\tsn{R} transition duct \parref{RSMP3DDF_s_A_ss_CtoRTDDG1992} test-cases have a straight centerline ($x$-axis of the coordinates system).
Furthermore, the diverging part of the \tsn{C}-to-\tsn{R} transition duct (from station 2 to midpoint of the transition section; \figrefnp{Fig_RSMP3DDF_s_A_ss_CtoRTDDG1992_001}) was sufficiently long to avoid separation.
The \tsn{S}-duct test-case \cite{Wellborn_Okiishi_Reichert_1993a,Wellborn_Reichert_Okiishi_1994a}
includes these 2 features, \viz it has a serpentine centerline (\tsn{S}-duct) combined with substantial ($52\%$) area increase, from inflow to outflow \cite{Wellborn_Reichert_Okiishi_1994a},
inducing a large region of separated flow near the duct floor, immediately after the beginning of the \tsn{S}-bend \figref{Fig_RSMP3DDF_s_A_ss_D3DSDWRO1994_001}.
The serpentine centerline of the \tsn{S}-duct lies on the $xz$-plane (no off-plane skewing; \figrefnp{Fig_RSMP3DDF_s_A_ss_D3DSDWRO1994_002}) and consists of 2 circular arcs of opposite curvature smoothly joined at a common tangency point
\cite[Fig. 2, p. 670]{Wellborn_Reichert_Okiishi_1994a}. Planes $\perp$ to the centerline define stations of circular cross-section, with varying radius, whose dependence on the angular coordinate (equivalently the curvilinear length) along the centerline
\cite[(2), p. 670]{Wellborn_Reichert_Okiishi_1994a} defines the geometry of the duct. The origin of the curvilinear coordinates along the centerline $\mathrm{s}_\tsn{CL}$ is at the beginning of the \tsn{S}-bend, which also corresponds to $x=0$.
The inlet-diameter is $D_{\tsn{CSG}_\tsn{A}}=2R_{\tsn{CSG}_\tsn{A}}=0.2042\;\mathrm{m}$ (this is also the diameter at the first measurement plane \tsn{A},
located at $x=-\tfrac{1}{2}D_{\tsn{CSG}_\tsn{A}}$, 1 inlet-radius upstream of the beginning of the \tsn{S}-bend; \figrefnp{Fig_RSMP3DDF_s_A_ss_D3DSDWRO1994_002}) while the exit diameter is $D_{\tsn{CSG}_\tsn{E}}=0.2514\;\mathrm{m}$
(this is also the diameter at the last measurement plane \tsn{E},
located at $x\approxeq5.61D_{\tsn{CSG}_\tsn{A}}$, $0.61$ inlet-diameters downstream of the exit of the \tsn{S}-bend located at $x=5D_{\tsn{CSG}_\tsn{A}}$; \figrefnp{Fig_RSMP3DDF_s_A_ss_D3DSDWRO1994_002}).

\begin{figure}[h!]
\begin{center}
\begin{picture}(340,280)
\put(0,-5){\includegraphics[angle=0,width=340pt,bb= 52 270 574 699]{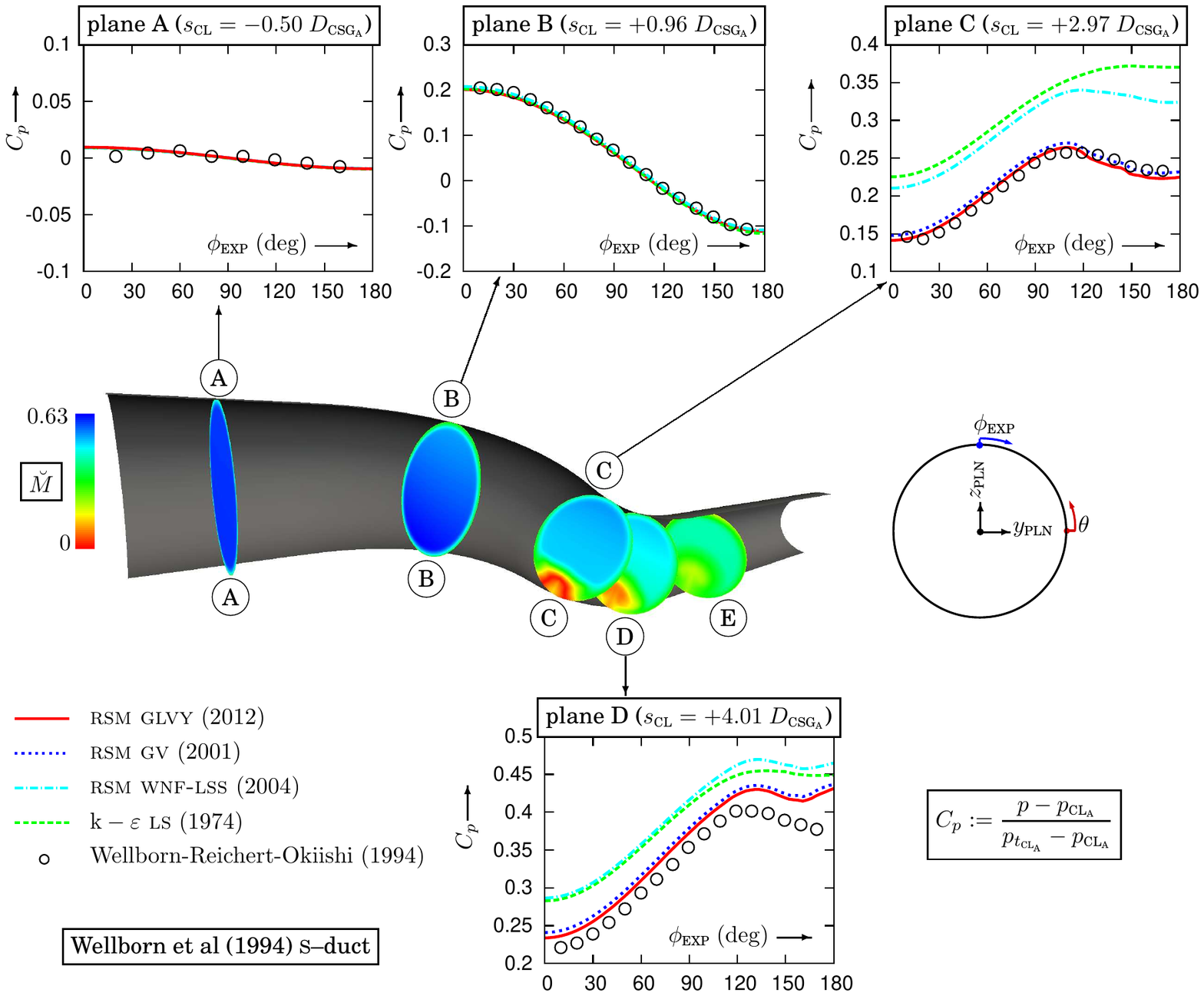}}
\end{picture}
\end{center}
\caption{Comparison of measured \cite{Wellborn_Reichert_Okiishi_1994a} wall-pressure coefficient $C_p$ (based on centerline quantities at plane \tsn{A}),
at 4 experimental measurement stations (planes $\perp$ to the centerline whose intersection with the duct defines circular cross-sections),
plotted against the azimuthal location along the circumference, with computations ($2\times10^6$ points grid discretizing the entire duct; \tabrefnp{Tab_RSMP3DDF_s_A_001})
using \parref{RSMP3DDF_s_TCsFS_ss_TCs} the \tsn{GV} \cite{Gerolymos_Vallet_2001a}, the \tsn{WNF--LSS} \cite{Gerolymos_Sauret_Vallet_2004a}
and the \tsn{GLVY} \cite{Gerolymos_Lo_Vallet_Younis_2012a} \tsn{RSM}s, and the \tsn{LS} \cite{Launder_Sharma_1974a} linear $\mathrm{k}$--$\varepsilon$ model, for turbulent flow in a diffusing \tsn{S}-duct
($Re_{\tsn{CL}_\tsn{A}}=2.6\times10^6$, $\bar M_{\tsn{CL}_\tsn{A}}\approxeq0.6$; \tabrefnp{Tab_RSMP3DDF_s_A_002}; $\mathrm{s}_\tsn{CL}$ is the curvilinear coordinate along the duct centerline; contour plots \tsn{GLVY RSM}).}
\label{Fig_RSMP3DDF_s_A_ss_D3DSDWRO1994_003}
\end{figure}

The flow \cite{Wellborn_Reichert_Okiishi_1994a} is subsonic (centerline Mach number at measurement plane \tsn{A} $\bar M_{\tsn{CL}_\tsn{A}}\approxeq0.60$) and the centerline Reynolds number is $Re_{\tsn{CL}_\tsn{A}}\approxeq2.6\times10^6$
($Re_{\tsn{CL}_\tsn{A}}=\bar u_{\tsn{CL}_\tsn{A}} D_{\tsn{CSG}_\tsn{A}} \bar\nu^{-1}_{\tsn{CL}_\tsn{A}}$, where $\bar u_{\tsn{CL}_\tsn{A}}$ is the centerline velocity and $\bar\nu_{\tsn{CL}_\tsn{A}}$ is the kinematic viscosity at centerline).
Available field measurements, taken at 5 axial planes $\perp$ to the centerline (circular cross-section; \figrefnp{Fig_RSMP3DDF_s_A_ss_D3DSDWRO1994_002}), using calibrated 3-hole
and 5-hole pneumatic probes \cite{Wellborn_Okiishi_Reichert_1993a,Wellborn_Reichert_Okiishi_1994a},
provide pressures (total and static) and the mean-flow velocity vectors. Wall-pressure measurements are also available \cite{Wellborn_Okiishi_Reichert_1993a,Wellborn_Reichert_Okiishi_1994a},
both around the circumference of 4 of the measurement planes \figref{Fig_RSMP3DDF_s_A_ss_D3DSDWRO1994_003},
and streamwise, at 3 angular locations \figref{Fig_RSMP3DDF_s_A_ss_D3DSDWRO1994_002}.
%
%

The computations were run on a $2\times10^6$ grid \tabref{Tab_RSMP3DDF_s_A_001} discretizing the entire duct without symmetry conditions \figref{Fig_RSMP3DDF_s_A_001}. Based on previous grid-convergence studies \cite{Gerolymos_Joly_Mallet_Vallet_2010a},
on a similar \tsn{2S}-duct configuration, this grid \tabref{Tab_RSMP3DDF_s_A_001} is sufficient to obtain accurate results for comparison between the different models \parref{RSMP3DDF_s_TCsFS_ss_TCs}.
The computational domain ($-0.98D_{\tsn{CSG}_\tsn{A}}\lessapprox x\lessapprox9.8D_{\tsn{CSG}_\tsn{A}}$) starts approximately 1 inlet-diameter ($D_{\tsn{CSG}_\tsn{A}}$) upstream of the start of the \tsn{S}-bend and
extends approximately 5 inlet-diameters ($5D_{\tsn{CSG}_\tsn{A}}$) downstream of the \tsn{S}-bend exit, thus avoiding any
interaction between the uniform outflow pressure boundary-condition and computed results at the last
measurement station \tsn{E} \figref{Fig_RSMP3DDF_s_A_ss_CtoRTDDG1992_001}.
The grid is uniform in the streamwise ($x$) direction (the $i=\const$ grid-surfaces are $\perp x$ planes) and consists of 2 blocks (\figrefnp{Fig_RSMP3DDF_s_A_001}; \tabrefnp{Tab_RSMP3DDF_s_A_001}).
The inner block (\tsn{H}$_\tsn{$\square$}$; \tabrefnp{Tab_RSMP3DDF_s_A_001}) is an \tsn{H}-grid of $x$-wise varying square cross-section with uniform $yz$-spacing, introduced to avoid the axis-singularity of an axisymmetric-type grid.
The outer block (\tsn{O}$_\tsn{$\square$}$; \tabrefnp{Tab_RSMP3DDF_s_A_001}) is stretched geometrically near the wall with ratio $r_k$ \tabref{Tab_RSMP3DDF_s_A_001}.
For the investigated flow conditions, the first node at the walls is located at $\Delta n_w^+\lessapprox\tfrac{4}{10}$ \tabref{Tab_RSMP3DDF_s_A_001}, $n$ being the wall-normal direction.

At inflow \tabref{Tab_RSMP3DDF_s_A_002} total conditions ($p_{t_{\tsn{CL}_\mathrm{i}}}=111330\;\mathrm{Pa}$, $T_{t_{\tsn{CL}_\mathrm{i}}}=296.4\;\mathrm{K}$) were assumed at the centerline,
corresponding to the Mach ($M_{\tsn{CL}_\mathrm{i}}=0.60$) and Reynolds ($Re_{\tsn{CL}_\mathrm{i}}=2.6\times10^6$) number values reported in the measurements \cite{Wellborn_Okiishi_Reichert_1993a,Wellborn_Reichert_Okiishi_1994a}.
A turbulent intensity $T_{u_{\tsn{CL}_\mathrm{i}}}=0.63\%$ was applied at the centerline; Wellborn \etal\ \cite[p. 29]{Wellborn_Okiishi_Reichert_1993a} report this value from measurements of Reichert \cite{Reichert_1991a} on the same facility.
In the absence of experimental data, a turbulent lengthscale $\ell_{\tsn{T}_{\tsn{CL}_\mathrm{i}}}=50\;\mathrm{mm}$ was assumed at the centerline,
with reference to the duct radius ($R_{\tsn{CSG}_\tsn{A}}=0.1021\;\mathrm{m}$). The initial inflow boundary-layer thickness and Coles-parameter \cite{Gerolymos_Sauret_Vallet_2004c} were adjusted, independently for each model \tabref{Tab_RSMP3DDF_s_A_002},
to match the experimental boundary-layer data at the first measurement plane \tsn{A}. Finally the outflow pressure was also adjusted, independently for each model \tabref{Tab_RSMP3DDF_s_A_002},
to obtain the correct $\breve M_{\tsn{CL}_\tsn{A}}\approxeq0.60$ \tabref{Tab_RSMP3DDF_s_A_003}.

Computational results for the integral axisymmetric \cite{Fujii_Okiishi_1972a} boundary-layer thicknesses and associated shape-factors at the first measurement plane \tsn{A}
\figref{Fig_RSMP3DDF_s_A_ss_CtoRTDDG1992_001}, where the flow is still practically axisymmetric, are in good agreement \tabref{Tab_RSMP3DDF_s_A_003} with those determined from the experimental data
\cite[Tab. 2, p. 671]{Wellborn_Reichert_Okiishi_1994a}. Following Wellborn \etal\ \cite{Wellborn_Okiishi_Reichert_1993a} the approximate (linearized; $\delta\ll R_\tsn{CSG}$) definitions of the axisymmetric integral boundary-layer thicknesses
\cite[(V.1--V.2), p.29]{Wellborn_Okiishi_Reichert_1993a}, as defined by Fujii and Okiishi \cite{Fujii_Okiishi_1972a}, were applied.
The definitions given by Wellborn \etal\ \cite[(V.1--V.2), p.29]{Wellborn_Okiishi_Reichert_1993a} concern compressible integral thicknesses, but the actual shape-factor values ($\sim\!\!1.38$) imply that
the thicknesses provided in the experimental database \cite{Wellborn_Okiishi_Reichert_1993a,Wellborn_Reichert_Okiishi_1994a}
are kinematic (the corresponding compressible value would be $\sim\!\!1.65$), as defined in the associated study (from which the inlet turbulent intensity was determined, on the same experimental facility, by Reichert \cite[(V.7--V.8), p. 67]{Reichert_1991a}.
This is implied by the statement that "{\em comparisons indicate little deviation from a conventional turbulent boundary-layer}" \cite[p. 29]{Wellborn_Okiishi_Reichert_1993a}.
 
All 4 turbulence closures predict separation near the duct floor \figref{Fig_RSMP3DDF_s_A_ss_D3DSDWRO1994_001} in agreement with experiment \cite{Wellborn_Reichert_Okiishi_1994a}, but differ in the location of separation and reattachment, in the extent
($x$-wise) and thickness ($z$-wise) of the separated flow region, and in the predicted structure of the recirculating flow \figref{Fig_RSMP3DDF_s_A_ss_D3DSDWRO1994_001}. The \tsn{GLVY} and \tsn{GV} \tsn{RSM}s yield very similar results 
\figref{Fig_RSMP3DDF_s_A_ss_D3DSDWRO1994_001}, and are in quite satisfactory agreement with available measurements  \figrefsatob{Fig_RSMP3DDF_s_A_ss_D3DSDWRO1994_002}
                                                                                                                                {Fig_RSMP3DDF_s_A_ss_D3DSDWRO1994_008}.
The \tsn{WNF--LSS RSM} predicts separation further downstream (with respect to the \tsn{GLVY} and \tsn{GV} \tsn{RSM}s; \figrefnp{Fig_RSMP3DDF_s_A_ss_D3DSDWRO1994_001}) and the linear \tsn{LS} $\mathrm{k}$--$\varepsilon$ model, which is known to 
underestimate flow detachment \cite{Gerolymos_1990c}, separates a little further downstream still. Even more important, there are noticeable differences in the separated flow structure \figref{Fig_RSMP3DDF_s_A_ss_D3DSDWRO1994_001}
between the \tsn{GLVY} and \tsn{GV} \tsn{RSM}s on the one hand, and the \tsn{WNF--LSS RSM} and the linear \tsn{LS} $\mathrm{k}$--$\varepsilon$ model on the other.
The \tsn{GLVY} and \tsn{GV} \tsn{RSM}s predict a much thicker ($z$-wise) low-speed region with a stronger recirculation zone near the wall just downstream of separation \figref{Fig_RSMP3DDF_s_A_ss_D3DSDWRO1994_001}.
This flow structure contains strong mean-velocity gradients producing high levels of turbulent kinetic energy $\mathrm{k}$, which presents 2 local maxima, one in the post-separation wake-region (dark blue levels, \tsn{GLVY} and \tsn{GV RSM}s;
\figrefnp{Fig_RSMP3DDF_s_A_ss_D3DSDWRO1994_001}) and another near the wall in the pre-reattachment region (light green levels of $\mathrm{k}$, \tsn{GLVY} and \tsn{GV RSM}s; \figrefnp{Fig_RSMP3DDF_s_A_ss_D3DSDWRO1994_001}).
On the other hand, the \tsn{WNF--LSS RSM} and the linear \tsn{LS} $\mathrm{k}$--$\varepsilon$ model predict a thinner ($z$-wise) low-speed region, with weak recirculation near the wall, and lower levels of $\mathrm{k}$ \figref{Fig_RSMP3DDF_s_A_ss_D3DSDWRO1994_001}.

The \tsn{GLVY} and \tsn{GV} \tsn{RSM}s' predictions compare quite well with experimental wall-pressure data \figrefsab{Fig_RSMP3DDF_s_A_ss_D3DSDWRO1994_002}
                                                                                                                      {Fig_RSMP3DDF_s_A_ss_D3DSDWRO1994_003},
correctly predicting the pressure-plateau on the duct floor ($\phi_\tsn{EXP}=170\;\mathrm{deg}$; \figrefnp{Fig_RSMP3DDF_s_A_ss_D3DSDWRO1994_002})
and the significant $z$-wise extent of the low-speed region indicated by the presence of the pressure-plateau at duct midplane ($\phi_\tsn{EXP}=90\;\mathrm{deg}$; \figrefnp{Fig_RSMP3DDF_s_A_ss_D3DSDWRO1994_002}).
This large separated flow region induces substantial flow blockage \cite[pp. 310--311]{Cumpsty_1989a}, accelerating the flow in the duct's ceiling area ($\phi_\tsn{EXP}=10\;\mathrm{deg}$; \figrefnp{Fig_RSMP3DDF_s_A_ss_D3DSDWRO1994_002}).
The satisfactory agreement of the  \tsn{GLVY} and \tsn{GV RSM}s' predictions with measurements near the duct ceiling ($\phi_\tsn{EXP}=10\;\mathrm{deg}$; \figrefnp{Fig_RSMP3DDF_s_A_ss_D3DSDWRO1994_002}) indicates that
the \tsn{GLVY} and \tsn{GV} \tsn{RSM}s yield a satisfactory prediction of the blockage induced by the large separation on the duct floor \figref{Fig_RSMP3DDF_s_A_ss_D3DSDWRO1994_001}.
Near the beginning of the \tsn{S}-bend, at planes \tsn{A} (one inlet radius $R_{\tsn{CSG}_1}$ upstream) and \tsn{B} (approximately one inlet diameter $D_{\tsn{CSG}_1}$ downstream), all 4 turbulence models are in excellent agreement with measurements
\figref{Fig_RSMP3DDF_s_A_ss_D3DSDWRO1994_003}, correctly predicting in plane \tsn{B} the circumferential pressure-gradient that drives the boundary-layer fluid along the duct's circumference \figref{Fig_RSMP3DDF_s_A_ss_D3DSDWRO1994_007} 
from ceiling (higher pressure due to the streamwise-concave wall; \figrefnp{Fig_RSMP3DDF_s_A_ss_D3DSDWRO1994_003}) to floor (lower pressure due to the streamwise-convex wall; \figrefnp{Fig_RSMP3DDF_s_A_ss_D3DSDWRO1994_003}).
At plane \tsn{C}, in the separated flow region \figrefsab{Fig_RSMP3DDF_s_A_ss_D3DSDWRO1994_001}{Fig_RSMP3DDF_s_A_ss_D3DSDWRO1994_002}, the \tsn{GLVY} and \tsn{GV} \tsn{RSM}s are again in excellent agreement with measurements, correctly predicting the 
circumferential evolution of $C_p$ \figref{Fig_RSMP3DDF_s_A_ss_D3DSDWRO1994_003} both in level and shape. The \tsn{WNF--LSS RSM} predicts the correct shape of the circumferential evolution of $C_p$ at plane \tsn{C}, but largely overestimates its value
by $\sim\!\!50$\%, whereas the linear \tsn{LS} $\mathrm{k}$--$\varepsilon$ model which overestimates $C_p$ even more fails to predict the inversion the circumferential pressure-gradient \figref{Fig_RSMP3DDF_s_A_ss_D3DSDWRO1994_003} from channel
mid-height ($\phi_{\tsn{EXP}}\approxeq 110\;\mathrm{deg}$) to floor ($\phi_{\tsn{EXP}}\approxeq 180\;\mathrm{deg}$). 
At plane \tsn{D}, where the flow reattaches in the experiment \figref{Fig_RSMP3DDF_s_A_ss_D3DSDWRO1994_002}, the \tsn{GLVY} and \tsn{GV} \tsn{RSM}s again provide the best prediction, compared to the \tsn{WNF--LSS RSM} 
and the linear \tsn{LS} $\mathrm{k}$--$\varepsilon$ model, but they slightly overestimate $C_p$, especially near the floor ($130\;\mathrm{deg}\lessapprox \phi_{\tsn{EXP}}\lessapprox 180\;\mathrm{deg}$; \figrefnp{Fig_RSMP3DDF_s_A_ss_D3DSDWRO1994_003}).

\begin{figure}[h!]
\begin{center}
\begin{picture}(340,490)
\put(50,-15){\includegraphics[angle=0,height=510pt,bb=131 51 468 752]{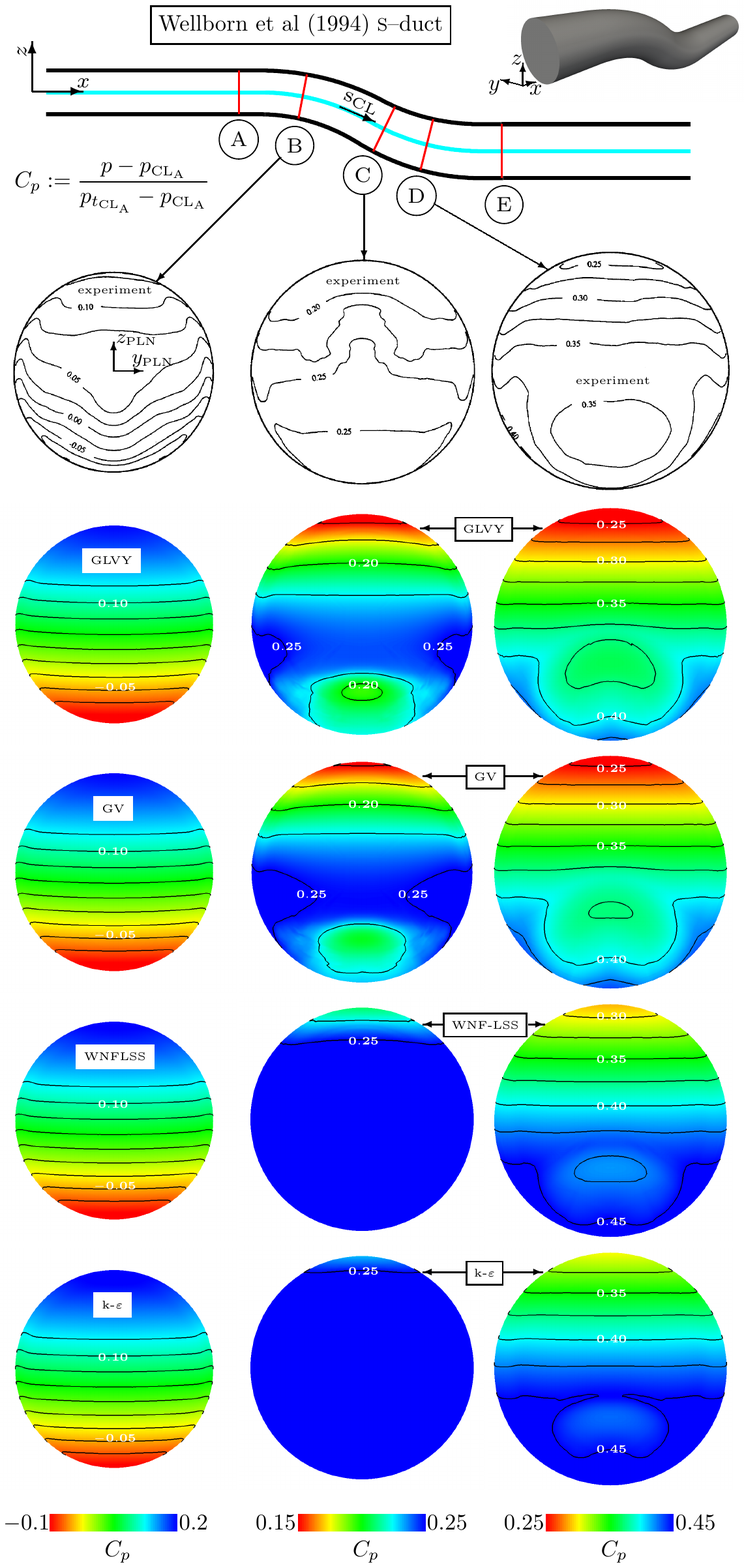}}
\end{picture}
\end{center}
\caption{Comparison, at 3 measurement planes (planes $\perp$ to the centerline define stations of circular cross-section),
of experimental \cite{Wellborn_Reichert_Okiishi_1994a} contours of pressure coefficient $C_p$ (based on centerline quantities at plane \tsn{A}; contour step 0.025),
with computations ($2\times10^6$ points grid discretizing the entire duct; \tabrefnp{Tab_RSMP3DDF_s_A_001})
using \parref{RSMP3DDF_s_TCsFS_ss_TCs} the \tsn{GV} \cite{Gerolymos_Vallet_2001a}, the \tsn{WNF--LSS} \cite{Gerolymos_Sauret_Vallet_2004a}
and the \tsn{GLVY} \cite{Gerolymos_Lo_Vallet_Younis_2012a} \tsn{RSM}s, and the \tsn{LS} \cite{Launder_Sharma_1974a} linear $\mathrm{k}$--$\varepsilon$ model
($Re_{\tsn{CL}_\tsn{A}}=2.6\times10^6$, $\bar M_{\tsn{CL}_\tsn{A}}\approxeq0.6$; \tabrefnp{Tab_RSMP3DDF_s_A_002}).}
\label{Fig_RSMP3DDF_s_A_ss_D3DSDWRO1994_004}
\end{figure}

Field pneumatic-probe measurements of $C_p$ \figref{Fig_RSMP3DDF_s_A_ss_D3DSDWRO1994_004} at plane \tsn{B} indicate a slight static-pressure distortion which is not predicted by the computations \figref{Fig_RSMP3DDF_s_A_ss_D3DSDWRO1994_004}
and is not observed in the wall-pressure measurements \figref{Fig_RSMP3DDF_s_A_ss_D3DSDWRO1994_003}. At plane \tsn{C}, the \tsn{GLVY} and \tsn{GV} \tsn{RSM}s are in reasonable agreement with measurements, correctly predicting the flow acceleration near the
ceiling \figref{Fig_RSMP3DDF_s_A_ss_D3DSDWRO1994_004} induced by the floor-separation blockage \figrefsab{Fig_RSMP3DDF_s_A_ss_D3DSDWRO1994_001}
                                                                                                         {Fig_RSMP3DDF_s_A_ss_D3DSDWRO1994_002}.
The \tsn{WNF--LSS RSM} which predicts separation downstream of experiment \figrefsab{Fig_RSMP3DDF_s_A_ss_D3DSDWRO1994_001}
                                                                                    {Fig_RSMP3DDF_s_A_ss_D3DSDWRO1994_002}
severely overestimates $C_p$ at plane \tsn{C} \figref{Fig_RSMP3DDF_s_A_ss_D3DSDWRO1994_004}, the linear \tsn{LS} $\mathrm{k}$--$\varepsilon$ model performing worse.
At the near-reattachment plane \tsn{D}, the \tsn{GLVY} and \tsn{GV} \tsn{RSM}s are in good agreement with measurements, substantially outperforming the \tsn{WNF--LSS RSM} 
and the \tsn{LS} $\mathrm{k}$--$\varepsilon$ closure \figref{Fig_RSMP3DDF_s_A_ss_D3DSDWRO1994_004}.
%
\begin{figure}[h!]
\begin{center}
\begin{picture}(340,490)
\put(15,-15){\includegraphics[angle=0,height=510pt,bb= 72 33 512 752]{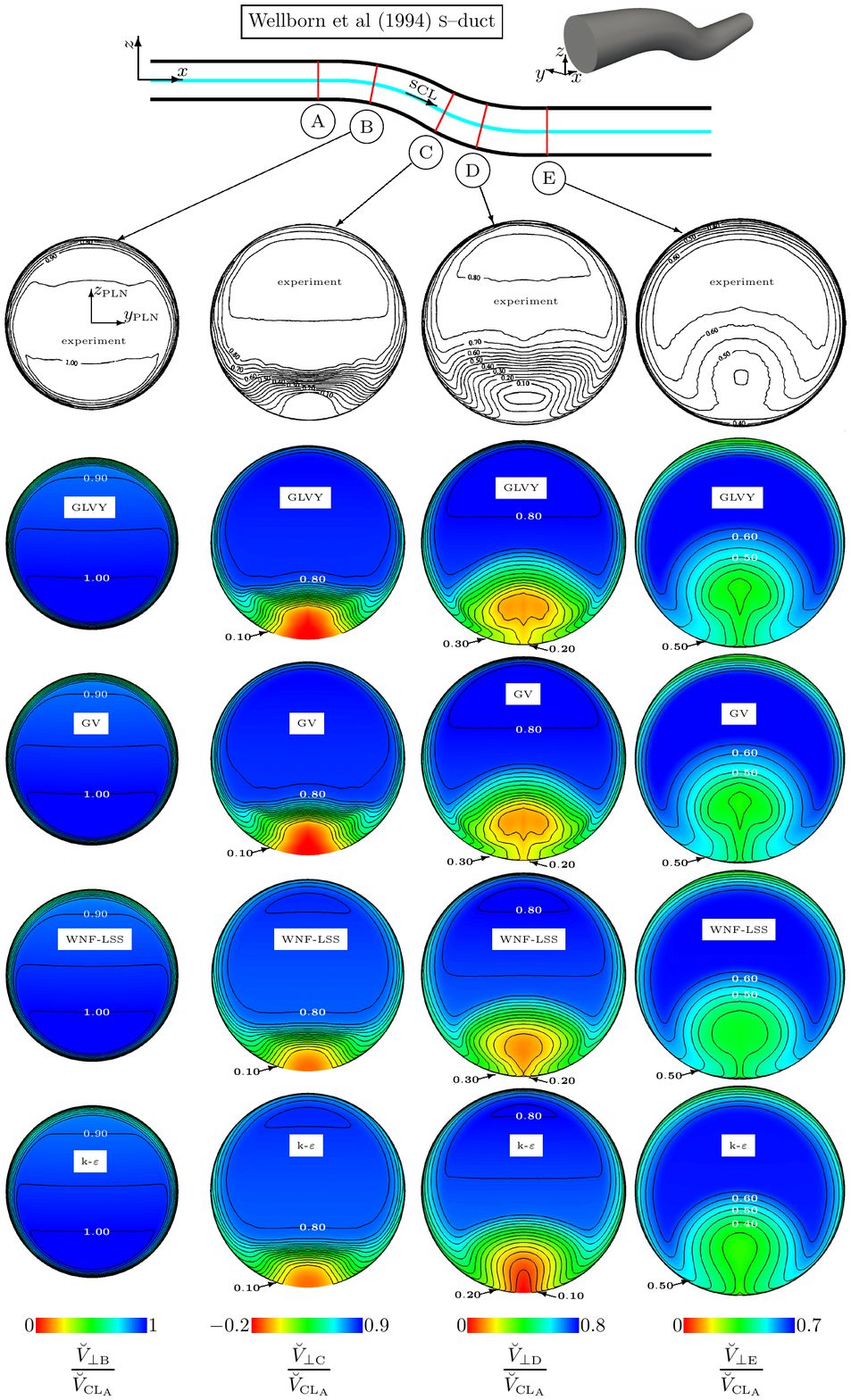}}
\end{picture}
\end{center}
\caption{Comparison, at 4 measurement planes (planes $\perp$ to the centerline define stations of circular cross-section),
of experimental \cite{Wellborn_Reichert_Okiishi_1994a} contours of normal-to-the-plane (streamwise) velocity $\breve V_\perp$ (made nondimensional
by the centerline velocity at plane \tsn{A}, $\breve V_{\tsn{CL}_\tsn{A}}$; contour step 0.05),
with computations ($2\times10^6$ points grid discretizing the entire duct; \tabrefnp{Tab_RSMP3DDF_s_A_001})
using \parref{RSMP3DDF_s_TCsFS_ss_TCs} the \tsn{GV} \cite{Gerolymos_Vallet_2001a}, the \tsn{WNF--LSS} \cite{Gerolymos_Sauret_Vallet_2004a}
and the \tsn{GLVY} \cite{Gerolymos_Lo_Vallet_Younis_2012a} \tsn{RSM}s, and the \tsn{LS} \cite{Launder_Sharma_1974a} linear $\mathrm{k}$--$\varepsilon$ model
($Re_{\tsn{CL}_\tsn{A}}=2.6\times10^6$, $\bar M_{\tsn{CL}_\tsn{A}}\approxeq0.6$; \tabrefnp{Tab_RSMP3DDF_s_A_002}).}
\label{Fig_RSMP3DDF_s_A_ss_D3DSDWRO1994_005}
\end{figure}
%
\begin{figure}[h!]
\begin{center}
\begin{picture}(340,490)
\put(15,-15){\includegraphics[angle=0,height=510pt,bb= 72 51 512 752]{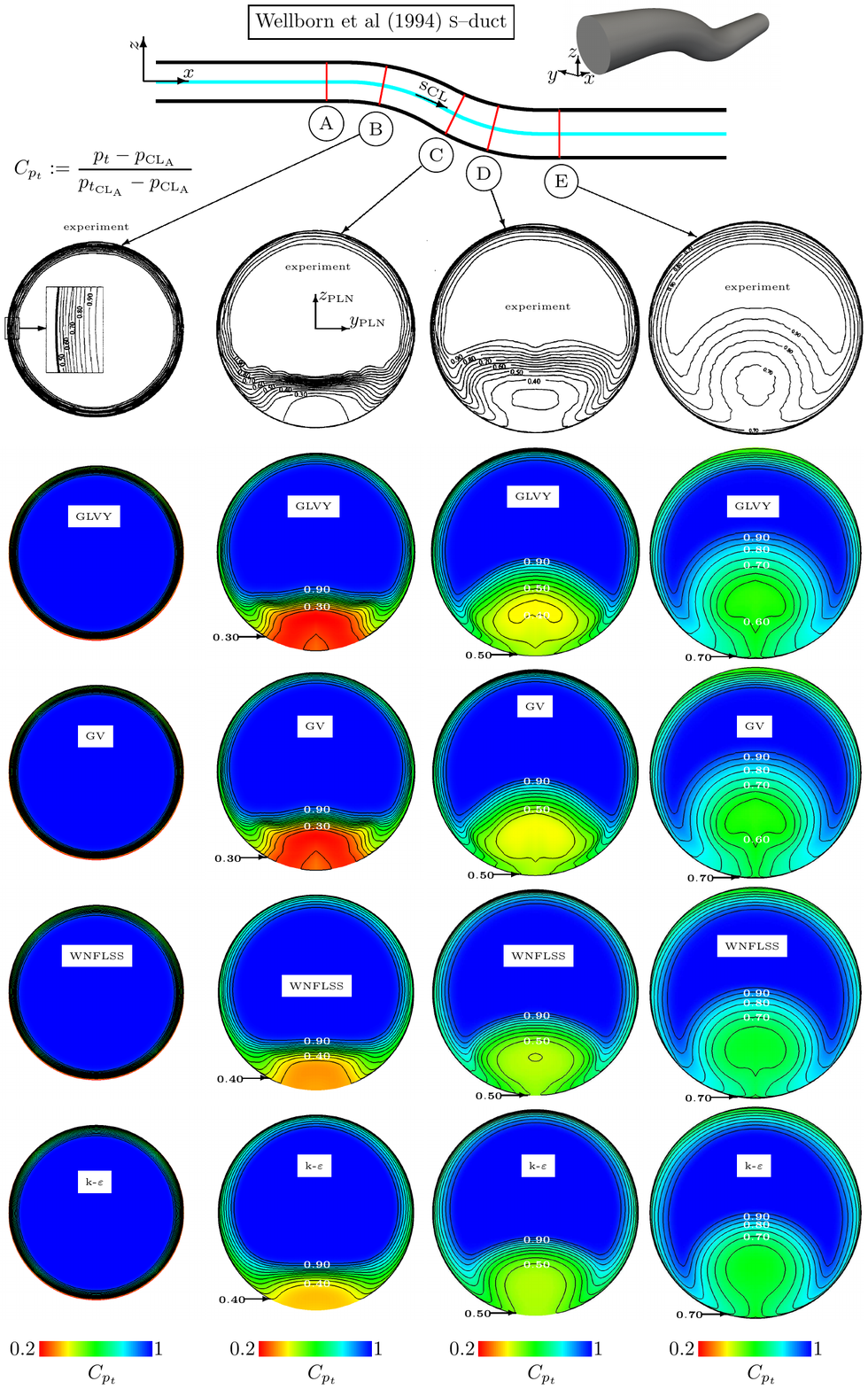}}
\end{picture}
\end{center}
\caption{Comparison, at 4 measurement planes (planes $\perp$ to the centerline define stations of circular cross-section),
of experimental \cite{Wellborn_Reichert_Okiishi_1994a} contours of total pressure coefficient $C_{p_t}$ (based on centerline quantities at plane \tsn{A}; contour step 0.05),
with computations ($2\times10^6$ points grid discretizing the entire duct; \tabrefnp{Tab_RSMP3DDF_s_A_001})
using \parref{RSMP3DDF_s_TCsFS_ss_TCs} the \tsn{GV} \cite{Gerolymos_Vallet_2001a}, the \tsn{WNF--LSS} \cite{Gerolymos_Sauret_Vallet_2004a}
and the \tsn{GLVY} \cite{Gerolymos_Lo_Vallet_Younis_2012a} \tsn{RSM}s, and the \tsn{LS} \cite{Launder_Sharma_1974a} linear $\mathrm{k}$--$\varepsilon$ model
($Re_{\tsn{CL}_\tsn{A}}=2.6\times10^6$, $\bar M_{\tsn{CL}_\tsn{A}}\approxeq0.6$; \tabrefnp{Tab_RSMP3DDF_s_A_002}).}
\label{Fig_RSMP3DDF_s_A_ss_D3DSDWRO1994_006}
\end{figure}
\clearpage

The velocity field, at each measurement plane, can be decomposed into a plane-normal component $\vec{V}_{\perp\tsn{PLN}}$ and an in-plane (parallel) component $\vec{V}_{\parallel\tsn{PLN}}$, $\vec{V}=\vec{V}_{\perp\tsn{PLN}}+\vec{V}_{\parallel\tsn{PLN}}$,
where \tsn{PLN} $\in \{\tsn{A},\tsn{B},\tsn{C},\tsn{D},\tsn{E}\}$. The plane-normal mean-velocities $\breve{V}_{\perp}$ \figref{Fig_RSMP3DDF_s_A_ss_D3DSDWRO1994_005} indicate the regions of separated and low-speed flow, which also
correspond to the high-loss regions (low $C_{p_t}$; \figrefnp{Fig_RSMP3DDF_s_A_ss_D3DSDWRO1994_006}).
\begin{figure}[h!]
\begin{center}
\begin{picture}(340,380)
\put(0,-5){\includegraphics[angle=0,width=340pt,bb= 71 74 558 623]{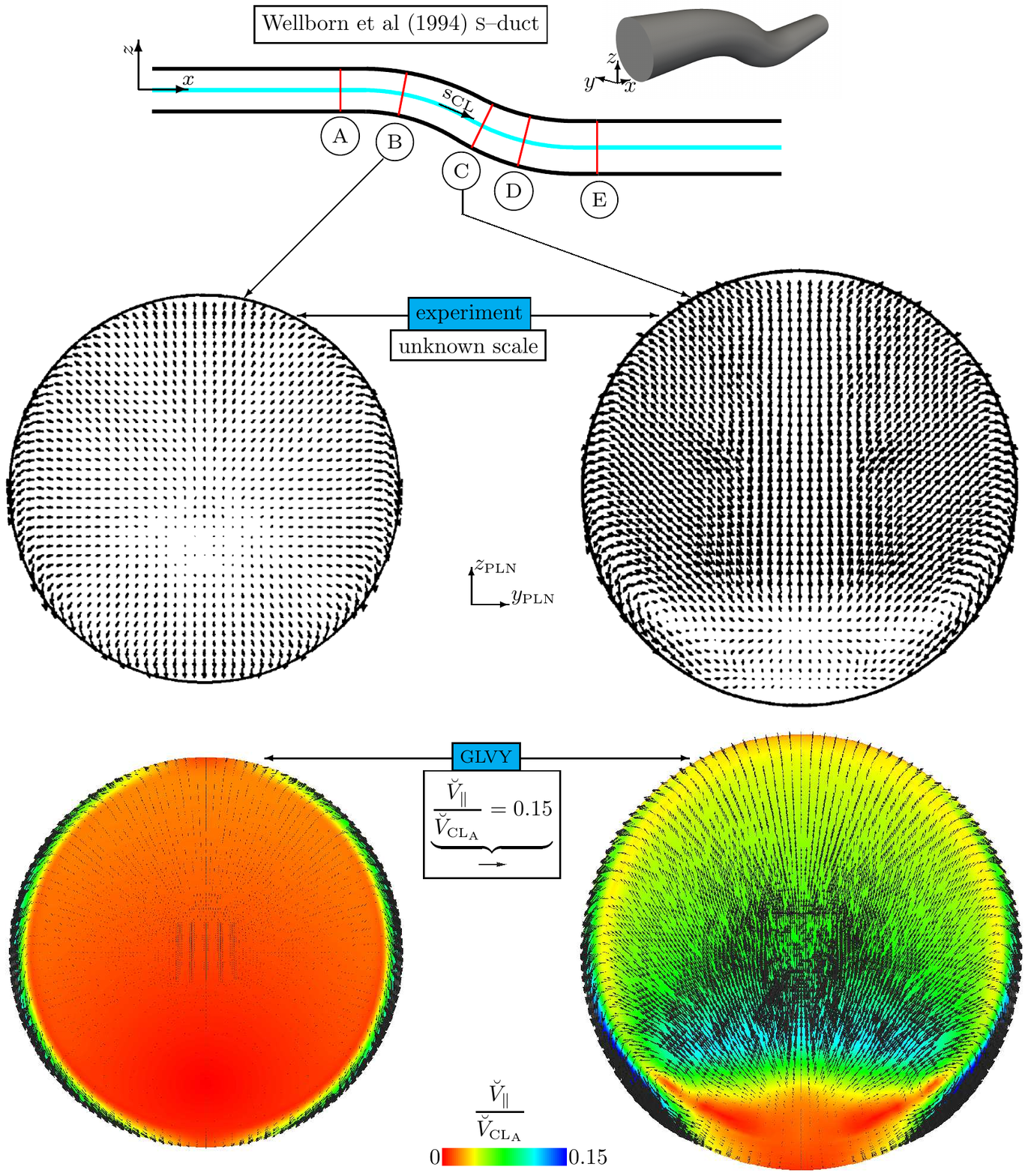}}
\end{picture}
\end{center}
\caption{Comparison, at 2 measurement planes (planes $\perp$ to the centerline define stations of circular cross-section),
of experimental \cite{Wellborn_Reichert_Okiishi_1994a} vectors (unknown scale) of in-plane (secondary) velocity $\bar{\vec{V}}_\parallel$ (made nondimensional
by the centerline velocity at plane \tsn{A}, $\breve V_{\tsn{CL}_\tsn{A}}$),
with computations ($2\times10^6$ points grid discretizing the entire duct; \tabrefnp{Tab_RSMP3DDF_s_A_001})
using \parref{RSMP3DDF_s_TCsFS_ss_TCs} the \tsn{GLVY} \cite{Gerolymos_Lo_Vallet_Younis_2012a} \tsn{RSM}
($Re_{\tsn{CL}_\tsn{A}}=2.6\times10^6$, $\bar M_{\tsn{CL}_\tsn{A}}\approxeq0.6$; \tabrefnp{Tab_RSMP3DDF_s_A_002}).}
\label{Fig_RSMP3DDF_s_A_ss_D3DSDWRO1994_007}
\end{figure}
%
\begin{figure}[h!]
\begin{center}
\begin{picture}(340,380)
\put(0,-5){\includegraphics[angle=0,width=340pt,bb= 65 74 564 623]{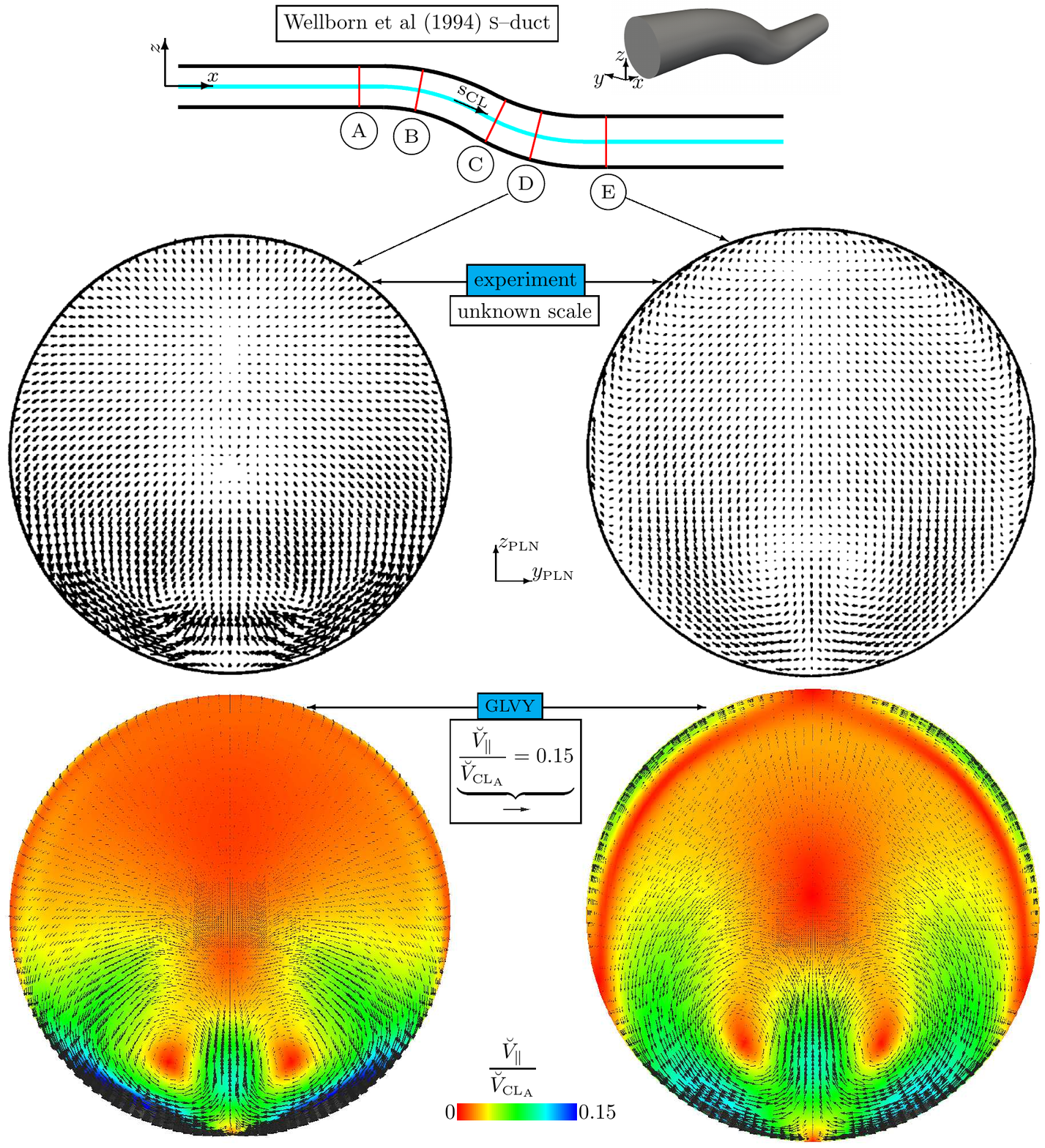}}
\end{picture}
\end{center}
\caption{Comparison, at 2 measurement planes (planes $\perp$ to the centerline define stations of circular cross-section),
of experimental \cite{Wellborn_Reichert_Okiishi_1994a} vectors (unknown scale) of in-plane (secondary) velocity $\bar{\vec{V}}_\parallel$ (made nondimensional
by the centerline velocity at plane \tsn{A}, $\breve V_{\tsn{CL}_\tsn{A}}$),
with computations ($2\times10^6$ points grid discretizing the entire duct; \tabrefnp{Tab_RSMP3DDF_s_A_001})
using \parref{RSMP3DDF_s_TCsFS_ss_TCs} the \tsn{GLVY} \cite{Gerolymos_Lo_Vallet_Younis_2012a} \tsn{RSM}
($Re_{\tsn{CL}_\tsn{A}}=2.6\times10^6$, $\bar M_{\tsn{CL}_\tsn{A}}\approxeq0.6$; \tabrefnp{Tab_RSMP3DDF_s_A_002}).}
\label{Fig_RSMP3DDF_s_A_ss_D3DSDWRO1994_008}
\end{figure}
The \tsn{GLVY} and \tsn{GV} \tsn{RSM}s are in overall satisfactory agreement with measurements \figrefsab{Fig_RSMP3DDF_s_A_ss_D3DSDWRO1994_005}
                                                                                                         {Fig_RSMP3DDF_s_A_ss_D3DSDWRO1994_006}
correctly predicting the backflow region at plane \tsn{C} and the flow blockage at the reattachment plane \tsn{D} and at the exit plane \tsn{E} \figref{Fig_RSMP3DDF_s_A_ss_D3DSDWRO1994_005}.
As a consequence, the \tsn{GLVY} and \tsn{GV} \tsn{RSM}s also predict correctly the high level of loss in the backflow region (low $C_{p_t}$; plane \tsn{C}; \figrefnp{Fig_RSMP3DDF_s_A_ss_D3DSDWRO1994_006})
and the subsequent streamwise evolution of the high-loss region (planes \tsn{D} and \tsn{E}; \figrefnp{Fig_RSMP3DDF_s_A_ss_D3DSDWRO1994_006}).
On the contrary, the linear \tsn{LS} $\mathrm{k}$--$\varepsilon$ model, and to a lesser extent the \tsn{WNF--LSS RSM}, underpredict both backflow \figref{Fig_RSMP3DDF_s_A_ss_D3DSDWRO1994_005}
and losses \figref{Fig_RSMP3DDF_s_A_ss_D3DSDWRO1994_006}, predicting a less thick low-speed high-loss region everywhere \figrefsab{Fig_RSMP3DDF_s_A_ss_D3DSDWRO1994_005}
                                                                                                                                  {Fig_RSMP3DDF_s_A_ss_D3DSDWRO1994_006}.
\begin{figure}[h!]
\begin{center}
\begin{picture}(340,460)
\put(0,-5){\includegraphics[angle=0,width=340pt,bb= 71 33 509 642]{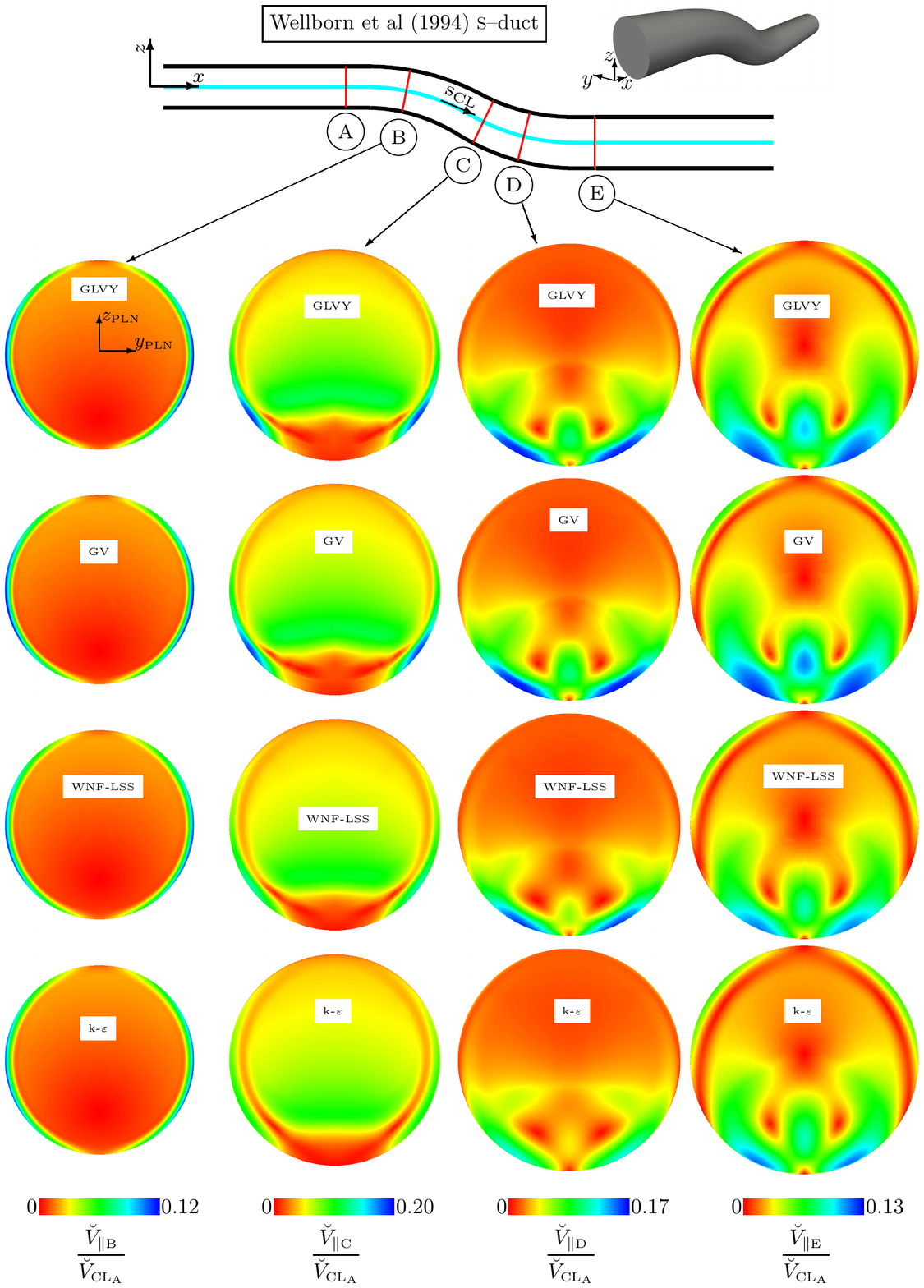}}
\end{picture}
\end{center}
\caption{Level plots of the module of in-plane (secondary) velocity $\bar{\vec{V}}_\parallel$ (made nondimensional by the centerline velocity at plane \tsn{A}, $\breve V_{\tsn{CL}_\tsn{A}}$),
at 4 measurement planes (planes $\perp$ to the centerline define stations of circular cross-section),
computed ($2\times10^6$ points grid discretizing the entire duct; \tabrefnp{Tab_RSMP3DDF_s_A_001})
using \parref{RSMP3DDF_s_TCsFS_ss_TCs} the \tsn{GV} \cite{Gerolymos_Vallet_2001a}, the \tsn{WNF--LSS} \cite{Gerolymos_Sauret_Vallet_2004a}
and the \tsn{GLVY} \cite{Gerolymos_Lo_Vallet_Younis_2012a} \tsn{RSM}s, and the \tsn{LS} \cite{Launder_Sharma_1974a} linear $\mathrm{k}$--$\varepsilon$ model
($Re_{\tsn{CL}_\tsn{A}}=2.6\times10^6$, $\bar M_{\tsn{CL}_\tsn{A}}\approxeq0.6$; \tabrefnp{Tab_RSMP3DDF_s_A_002}).}
\label{Fig_RSMP3DDF_s_A_ss_D3DSDWRO1994_009}
\end{figure}
\clearpage

The \tsn{GLVY RSM} (whose results are very close to those obtained with the  \tsn{GV RSM}; \figrefsnpatob{Fig_RSMP3DDF_s_A_ss_D3DSDWRO1994_001}
                                                                                                         {Fig_RSMP3DDF_s_A_ss_D3DSDWRO1994_006}),
predicts quite satisfactorily the structure of secondary (in-plane $\vec{V}_{\parallel}$) flows \figrefsab{Fig_RSMP3DDF_s_A_ss_D3DSDWRO1994_007}
                                                                                                          {Fig_RSMP3DDF_s_A_ss_D3DSDWRO1994_008}. 
At plane \tsn{B} \figref{Fig_RSMP3DDF_s_A_ss_D3DSDWRO1994_007}, the circumferential pressure-gradient \figref{Fig_RSMP3DDF_s_A_ss_D3DSDWRO1994_003} drives the boundary-layer flow
from ceiling to floor along the duct walls \figref{Fig_RSMP3DDF_s_A_ss_D3DSDWRO1994_007}.
At the separated-flow plane \tsn{C} \figref{Fig_RSMP3DDF_s_A_ss_D3DSDWRO1994_007} this downward flow interacts with the large separation at the duct's floor \figref{Fig_RSMP3DDF_s_A_ss_D3DSDWRO1994_001} forming 2 contrarotating vortices
\figref{Fig_RSMP3DDF_s_A_ss_D3DSDWRO1994_007}, which lift off the floor as they are convected downstream (planes \tsn{D} and \tsn{E}; \figrefnp{Fig_RSMP3DDF_s_A_ss_D3DSDWRO1994_008}).

The differences in predictive accuracy between the 4 turbulence models \figrefsatob{Fig_RSMP3DDF_s_A_ss_D3DSDWRO1994_001}
                                                                                   {Fig_RSMP3DDF_s_A_ss_D3DSDWRO1994_006}
is directly related to differences in the secondary-flow structure \figref{Fig_RSMP3DDF_s_A_ss_D3DSDWRO1994_009}.
At plane \tsn{B}, where the flow is still attached \figref{Fig_RSMP3DDF_s_A_ss_D3DSDWRO1994_005}, all 4 turbulence closures yield quite similar results \figref{Fig_RSMP3DDF_s_A_ss_D3DSDWRO1994_009}.
At the separated-flow plane \tsn{C}, \tsn{GLVY} and \tsn{GV RSM}s predict a thick low-$\vec{V}_{\parallel}$ region \figref{Fig_RSMP3DDF_s_A_ss_D3DSDWRO1994_009}, with distinct tails roughly marking the centers of 2 contrarotating vortices
\figref{Fig_RSMP3DDF_s_A_ss_D3DSDWRO1994_007}, in good agreement with measurements.
The \tsn{WNF--LSS RSM} predicts too thin a low-speed region \figref{Fig_RSMP3DDF_s_A_ss_D3DSDWRO1994_009} and the 2 tails are less sharp, these 2 defaults being even more pronounced for the linear \tsn{LS} $\mathrm{k}$--$\varepsilon$ model. The differences
between the \tsn{GLVY} and \tsn{GV} \tsn{RSM}s on the one hand and the \tsn{WNF--LSS RSM} and the linear \tsn{LS} $\mathrm{k}$--$\varepsilon$ closure on the other,
are much more pronounced at the reattachment plane \tsn{D} \figref{Fig_RSMP3DDF_s_A_ss_D3DSDWRO1994_009},
where the 2 vortices have lifted off the floor in the \tsn{GLVY} and \tsn{GV} \tsn{RSM}s predictions \figref{Fig_RSMP3DDF_s_A_ss_D3DSDWRO1994_009}, in quite satisfactory agreement with measurements \figref{Fig_RSMP3DDF_s_A_ss_D3DSDWRO1994_008},
whereas they are more diffuse and closer to the wall in the \tsn{WNF--LSS RSM} predictions, which also underestimate the 2 symmetric high-$\breve{V}_\parallel$ regions near the duct floor
(plane \tsn{D}; \figrefsnpab{Fig_RSMP3DDF_s_A_ss_D3DSDWRO1994_008}
                            {Fig_RSMP3DDF_s_A_ss_D3DSDWRO1994_009}).
These high-$V_{\parallel}$ regions are simply absent in the linear \tsn{LS} $\mathrm{k}$--$\varepsilon$ model predictions \figref{Fig_RSMP3DDF_s_A_ss_D3DSDWRO1994_009}.
At the exit plane \tsn{E}, the \tsn{GLVY} and \tsn{GV} \tsn{RSM}s predict sharp regions of low speed \figref{Fig_RSMP3DDF_s_A_ss_D3DSDWRO1994_009} which correspond to the centers of the vortices \figref{Fig_RSMP3DDF_s_A_ss_D3DSDWRO1994_008}, with
regions of high-$V_\parallel$ near the ducts floor \figrefsab{Fig_RSMP3DDF_s_A_ss_D3DSDWRO1994_008}
                                                             {Fig_RSMP3DDF_s_A_ss_D3DSDWRO1994_009}
and in the region between the 2 contrarotating vortices \figref{Fig_RSMP3DDF_s_A_ss_D3DSDWRO1994_009},
in good agreement with measurements. The vortices predicted by the \tsn{WNF--LSS RSM} and the linear $\mathrm{k}$--$\varepsilon$ model are closer to the duct floor and their centers are less sharp \figref{Fig_RSMP3DDF_s_A_ss_D3DSDWRO1994_009}.

For the Wellborn \etal\ \cite{Wellborn_Reichert_Okiishi_1994a} test-case as for the previous ones \parrefsab{RSMP3DDF_s_A_ss_DTFSDGE1981}
                                                                                             {RSMP3DDF_s_A_ss_CtoRTDDG1992},
the \tsn{GLVY} and \tsn{GV} \tsn{RSM}s yield very similar results, and are in quite satisfactory agreement with 
measurements, showing that properly calibrated \tsn{RSM--RANS} closures can predict flows with large separation and  wall-curvature effects. The \tsn{GLVY} and \tsn{GV} \tsn{RSM}s considerably outperform the
\tsn{WNF--LSS RSM}, and this is again attributed to the $C_\phi^{(\tsn{RH})}$ coefficient-function used \tabref{Tab_RSMP3DDF_s_TCsFS_ss_TCs_001},
because predictions of the Wellborn \etal\ \cite{Wellborn_Reichert_Okiishi_1994a} test-case using the \tsn{GV} and \tsn{GV--DH} (\cf\ \parrefnp{RSMP3DDF_s_A_ss_DTFSDGE1981}) \tsn{RSM}s are very similar one with another \cite[Figs. 11--12, pp. 1153--1154]{Vallet_2007a},
implying that the turbulent diffusion closure is less influential than pressure-strain redistribution in this flow.
On the other hand, the improvement of the \tsn{WNF--LSS RSM} over the linear \tsn{LS} $\mathrm{k}$--$\varepsilon$ model for this separation-dominated flow is weak.

%
%
%
%
%
%
%
%
%
\section{Conclusions}\label{RSMP3DDF_s_Cs}
%
%
%
%
%
%
%
%
%

In the present work, 3 wall-normal-free \tsn{RSM}s were assessed through comparison with experimental data for complex 3-D duct flows,
highlighting the impact of the closure used for the velocity/pressure-gradient tensor $\Pi_{ij}$ \eqref{Eq_RSMP3DDF_s_TCsFS_ss_TCs_007}
and for turbulent diffusion by the fluctuating velocities $d_{ij}^{(u)}$ \eqref{Eq_RSMP3DDF_s_TCsFS_ss_TCs_005} on the predictive accuracy of the models.

The Gessner and Emery \cite{Gessner_Emery_1981a} square duct flow is dominated by turbulence-anisotropy-driven secondary flows
whereas the Davis and Gessner \cite{Davis_Gessner_1992a} \tsn{C}-to-\tsn{R} transition duct flow combines pressure-driven secondary flows in the transition section
with turbulence-anisotropy-driven secondary flows in the straight constant cross-section exit part. Therefore, these test-cases are particularly useful in evaluating the predictive accuracy of turbulence closures for secondary flows where 
streamwise vorticity is important.
Finally, the Wellborn \etal\ \cite{Wellborn_Reichert_Okiishi_1994a} diffusing \tsn{S}-duct contains a large region of separated flow and tests the ability of the turbulence models to accurately predict 3-D separation and reattachment in presence of blockage due to
confinement and of secondary flows.

Results with the baseline \tsn{LS} \cite{Launder_Sharma_1974a} linear $\mathrm{k}$--$\varepsilon$ closure were included as a reference for comparison with the more advanced differential \tsn{RSM}s. The underlying Boussinesq's hypothesis
pathologically returns negligible levels of normal-stress anisotropy \cite[pp. 273--279]{Wilcox_1998a}
and for this reason the \tsn{LS} $\mathrm{k}$--$\varepsilon$ predicts negligibly weak ($\sim\!\!0$) secondary velocities both in the square-duct \cite{Gessner_Emery_1981a}
and in the straight exit part of the \tsn{C}-to-\tsn{R} duct \cite{Davis_Gessner_1992a}. Furthermore, in the \tsn{S}-duct \cite{Wellborn_Reichert_Okiishi_1994a} test-case,
the \tsn{LS} $\mathrm{k}$--$\varepsilon$ model, which has been calibrated for equilibrium shear flows, severely underestimates separation. For all of the 3 test-cases the 
\tsn{LS} $\mathrm{k}$--$\varepsilon$ closure compares very poorly with experimental data.

The \tsn{WNF--LSS RSM} \cite{Gerolymos_Sauret_Vallet_2004a} adopts the Launder-Shima \cite{Launder_Shima_1989a} closure for the homogeneous part of $\Pi_{ij}$ and is therefore calibrated in zero-pressure-gradient flat-plate boundary-layer flow. As a consequence,
it underestimates separation in the \tsn{S}-duct \cite{Wellborn_Reichert_Okiishi_1994a} test-case. On the other hand it has the differential \tsn{RSM}s' inherent ability to predict normal-stress anisotropy and performs quite well for the 
\tsn{C}-to-\tsn{R} duct \cite{Davis_Gessner_1992a} but underestimates the centerline velocity peak in the developing square-duct flow \cite{Gessner_Emery_1981a}; this inadequacy was traced to the cumulative influence of the homogeneous
rapid redistribution isotropisation-of-production closure \eqref{Eq_RSMP3DDF_s_TCsFS_ss_TCs_007} $C_\phi^{(\tsn{RH})}$ \tabref{Tab_RSMP3DDF_s_TCsFS_ss_TCs_001} and the Daly-Harlow turbulent-diffusion model \tabref{Tab_RSMP3DDF_s_TCsFS_ss_TCs_001}.

For all of the 3 test-cases that were examined \cite{Gessner_Emery_1981a,Davis_Gessner_1992a,Wellborn_Reichert_Okiishi_1994a},
the \tsn{GLVY} \cite{Gerolymos_Lo_Vallet_Younis_2012a} and \tsn{GV} \cite{Gerolymos_Vallet_2001a} \tsn{RSM}s yield very similar results in quite satisfactory agreement with measurements,
implying that the extra terms in the $\Pi_{ij}$ closure \eqref{Eq_RSMP3DDF_s_TCsFS_ss_TCs_007} used in the \tsn{GLVY RSM} \tabref{Tab_RSMP3DDF_s_TCsFS_ss_TCs_001}
have little influence for the secondary and/or separated 3-D flows studied in this paper; however, these extra terms were found to substantially improve the apparent transition behaviour of the model \cite{icp_Gerolymos_Vallet_2013a}.
The coefficient-function $C_\phi^{(\tsn{RH})}$ used in the \tsn{GLVY} and \tsn{GV}  \tsn{RSM}s \tabref{Tab_RSMP3DDF_s_TCsFS_ss_TCs_001} was calibrated
with reference to flows with large separation \cite{Gerolymos_Vallet_2001a,Gerolymos_Vallet_2002a,Gerolymos_Sauret_Vallet_2004b}.
As a result, the \tsn{GLVY} and \tsn{GV}  \tsn{RSM}s perform quite well in the \tsn{S}-duct \cite{Wellborn_Reichert_Okiishi_1994a} flow. They predict quite satisfactorily 
the other 2 test-cases \cite{Gessner_Emery_1981a,Davis_Gessner_1992a} as well, although they underpredict the strength of the secondary flow velocities and the level of the Reynolds-stress tensor anisotropy.

The results presented in the paper suggest that \tsn{RSM} \tsn{RANS} has the potential to predict complex 3-D flows with streamwise vorticity and separation. Further improvements in the prediction of the Reynolds-stress tensor anisotropy can be
achieved by the use of a differential model for the full Reynolds-stress-dissipation tensor $\varepsilon_{ij}$ \cite{Lumley_Yang_Shih_1999a,Gerolymos_Lo_Vallet_Younis_2012a}. Furthermore, the turbulence structure in separated 
and reattaching/relaxing flows exhibits strong hysteresis \cite{Gerolymos_Kallas_Papailiou_1989a} whose inclusion in the model should be investigated \cite{Olsen_Coakley_2001a}.

%
%
%
%
%
%
%
%
%
\section*{Acknowledgments}
%
%
%
%
%
%
%
%
%

The present work was initiated within the \tsn{DGA}-funded research project \tsn{CACV} with Dassault-Aviation.
The computations were performed using \tsn{HPC} ressources allocated at \tsn{GENCI--IDRIS} (Grants 2013-- and 2014--020218) from \tsn{ICS--UPMC} (\tsn{ANR--10--EQPX--29--01}).
The authors are listed alphabetically.

%
%
%
%
%
\bibliographystyle{spmpsci.bst}\footnotesize\bibliography{Aerodynamics,GV,GV_news}
%
%
%
%
%
\normalsize
\end{document}